\documentclass[a4paper,11pt]{article}
\pdfoutput=1 

\usepackage{jheppub} 
\usepackage{lmodern}
\usepackage[T1]{fontenc} 
\usepackage{lineno}

\newcommand{\msbar}{$\overline{\text{MS}}$}

\usepackage{xcolor}
\usepackage[normalem]{ulem}

\preprint{TTP-21-057}

\title{\boldmath Matching next-to-leading-order and high-energy-resummed calculations of heavy-quarkonium-hadroproduction cross sections}

\author[a]{Jean-Philippe Lansberg,}
\author[1,a,b]{Maxim Nefedov,\note{Corresponding author.}}
\author[a,c]{Melih A. Ozcelik}

\affiliation[a]{Universit\'{e} Paris-Saclay, CNRS, IJCLab,  91405 Orsay, France}
\affiliation[b]{National Centre for Nuclear Research (NCBJ), Pasteura 7, 02-093 Warsaw, Poland}
\affiliation[c]{Institute for Theoretical Particle Physics, KIT, 76128 Karlsruhe, Germany}

\emailAdd{Jean-Philippe.Lansberg@in2p3.fr}
\emailAdd{maxim.nefedov@desy.de}
\emailAdd{melih.oezcelik@kit.edu}

\abstract{The energy dependence of the total hadroproduction cross section of pseudo-scalar quarkonia is computed via matching Next-to-Leading Order (NLO) Collinear-Factorisation (CF) results with resummed higher-order corrections, proportional to $\alpha_s^{n}\ln^{n-1}(1/z)$, to the CF hard-scattering coefficient, where $z=M^2/\hat{s}$ with $M$ and $\hat{s}$ being the quarkonium mass and the partonic center-of-mass energy squared. The resummation is performed using  High-Energy Factorisation (HEF) in the Doubly-Logarithmic (DL) approximation, which is a subset of the leading logarithmic $\ln (1/z)$ approximation. Doing so, one remains strictly consistent with the NLO and NNLO DGLAP evolution of the PDFs. By improving the treatment of the small-$z$ asymptotics of the CF coefficient function, the resummation cures the unphysical results of the NLO CF calculation. The matching is directly performed  in the $z$-space and, for the first time, by  using the Inverse-Error Weighting (InEW) matching procedure. As a by-product of the calculation, the NNLO term of the CF hard-scattering coefficient proportional to $\alpha_s^2\ln(1/z)$ is predicted from HEF.}

\begin{document} 
\maketitle
\flushbottom

\section{Introduction: heavy-quarkonium cross-section instability at high energies}

Historically, the discovery of charm quarks and charmonia has been one of the most important milestones towards establishing QCD as the theory of strong interaction. Nowadays, charmonia and bottomonia attract a lot of attention as tools to study the proton structure, spin physics and/or to probe the quark-gluon plasma, see e.g. recent reviews~\cite{Chapon:2020heu, Arbuzov:2020cqg, Lansberg:2019adr, Brambilla:2014jmp, Brambilla:2010cs}. They have clean experimental signatures and the fact that the hard scale of their production process is limited from below by the heavy-quark mass justifies the application of perturbation theory. However, the usage of quarkonia as tools for precision studies is problematic because up to now there is no consensus in the community on what are the main mechanisms of quarkonium production. The Non-Relativistic QCD(NRQCD) factorisation approach~\cite{Bodwin:1994jh} at NLO in the $\alpha_s$ expansion and NNLO in the $v^2$ expansion, which seems to be the most systematic theoretical tool available so far, is able to describe the unpolarised $p_T$-differential hadroproduction cross sections of charmonia~\cite{Butenschoen:2010rq, Butenschoen:2011yh, Butenschoen:2012px, Butenschoen:2012qr, Chao:2012iv, Gong:2012ug, Bodwin:2014gia} and bottomonia~\cite{Gong:2010bk, Wang:2012is, Gong:2013qka}. However, this approach has essentially failed all other phenomenological tests. It is incapable~\cite{Butenschoen:2010rq, Butenschoen:2011yh, Butenschoen:2012px, Butenschoen:2012qr} of describing the unpolarised photo- and hadroproduction data together with the data on quarkonium polarisation observables using a single set of Long-Distance Matrix Elements (LDMEs). This problem is  known as the ``polarisation puzzle''. Moreover, the hadroproduction of $\eta_c$ at the LHC has unexpectedly turned out to be dominated by the color-singlet state $c\bar{c}[{}^1S_0^{[1]}]$, with no color-octet contribution needed to describe data~\cite{Butenschoen:2014dra}. This seems inconsistent with Heavy-quark spin symmetry relations between the LDMEs of $\eta_c$ and $J/\psi$. This inconsistency is often referred to as the ``heavy-quark-spin-symmetry puzzle''.  

Alternative theoretical approaches, such as the (Improved) Colour-Evaporation Model~\cite{Barger:1979js, Barger:1980mg, Gavai:1994in, Ma:2016exq} have their own phenomenological problems, e.g. being incapable to describe the $J/\psi$-pair hadroproduction~\cite{Lansberg:2020rft}, single $J/\psi$ hadroproduction at large $p_T$~\cite{Lansberg:2020rft} or $e^+e^-\to J/\psi + c +\bar{c}$ cross section~\cite{Lansberg:2019adr}. Moreover, recent theoretical developments, such as potential-NRQCD~\cite{Brambilla:2021abf} or soft-gluon factorisation~\cite{Ma:2017xno, Li:2019ncs, Chen:2020yeg} stay within the non-relativistic paradigm for the bound-state and try to either reduce number of free parameters in NRQCD or more accurately treat the kinematic effects of soft gluon emissions at hadronisation stage.

Given the situation described above, the logical way to proceed is to scrutinise all possible sources of numerically large higher-order corrections to the hard-scattering coefficient function in the NRQCD approach beyond fixed-order NLO in $\alpha_s$ computations and find possible ways to resum them to all orders in perturbation theory. For example, at high $p_T$, much larger than the quarkonium mass $M$, such a resummation has already being achieved via the fragmentation formalism which has been worked out up to the next-to-leading power in $M/p_T$~\cite{Kang:2011mg, Bodwin:2014gia}. 
In the present paper, we focus on a class of higher-order QCD corrections, which has been largely ignored so far in the CF heavy-quarkonium-production literature and which become important when the hadronic collision energy, $\sqrt{s}$, is much larger than any other scale of our process, i.e.  $M$ or $p_T$. 

For simplicity, we will focus on the case of the inclusive hadroproduction of the ${}^1S_0^{[1]}$, ${}^3P_{0}^{[1]}$ and ${}^3P_{2}^{[1]}$ NRQCD Fock states of a heavy-quark pair of mass $M$. These states can be produced at LO in CF via a $2\to 1$ process, namely by fusion of two on-shell gluons. The NLO CF corrections to the total hadroproduction cross section at most involves $2\to 2$ processes and can be computed in a closed form, which has been done long time ago~\cite{Kuhn:1992qw, Petrelli:1997ge}. In CF, the total hadroproduction cross section of a state $m={}^{2S+1}L_J^{[1,8]}$ in a $h_A+h_B$ hadronic collision at a center-of-mass energy $\sqrt{s}$ can be computed as a convolution:
\begin{equation}
\sigma^{[m], h_Ah_B}(\sqrt{s})=\sum\limits_{i,j=q,\bar{q},g}\hspace{1mm}\int\limits_{z_{\min}}^1\frac{dz}{z} {\cal L}^{h_Ah_B}_{ij}(z,\mu_F) \hat{\sigma}^{[m]}_{ij}(z,\mu_F,\mu_R),\label{eq:sigma-L-sighat}
\end{equation}
over the variable $z=M^2/\hat{s}$ which represents the squared fraction of the energy of the partons initiating the hard subprocess used to produce the observed final-state of mass $M$, with $z_{\min}=M^2/s$. The partonic luminosity for partons $i,j=q,\bar{q},g$, entering eqn.~(\ref{eq:sigma-L-sighat}) is:
\begin{equation}
{\cal L}^{h_Ah_B}_{ij}(z,\mu_F) =\int\limits_{-y_{\max}}^{+y_{\max}} dy\ \tilde{f}^{h_A}_i\left( \frac{M}{\sqrt{sz}} e^y,\mu_F \right) \tilde{f}^{h_B}_j\left(\frac{M}{\sqrt{sz}} e^{-y},\mu_F\right),\label{eq:L-def}
\end{equation}
where $y_{\max}=\ln \sqrt{sz}/M$ and\footnote{In what follows, we will omit $h_A$ and $h_B$ when referring to PDFs.} $\tilde{f}^{h}_i(x,\mu_F^2)=xf^{h}_i(x,\mu_F^2)$ are the collinear ``momentum-density'' PDFs of the parton $i$ in a hadron $h$, evaluated at the factorisation scale $\mu_F$. The quantity $\hat{\sigma}^{[m]}_{ij}$ in eqn.~(\ref{eq:sigma-L-sighat}) is the hard-scattering coefficient for the corresponding partonic channel $ij$ and state $m$. 

 The partonic luminosity is always decreasing when $z\to M^2/s$ corresponding to $y_{\max}\to 0$. 
 However, at $\mu_F\gg 1$ GeV this decrease is more rapid than at $\mu_F\sim 1$ GeV, characteristic of quarkonium physics, due to $\mu_F$-evolution of the $x$-dependence of PDFs in eqn.~(\ref{eq:L-def}). It turns out that, at high scales, the contribution of small values of $z$ to the integral (\ref{eq:sigma-L-sighat}) is suppressed by the partonic luminosity. However, at smaller scales $\mu_F\sim M\sim 1$ GeV most of the recent PDFs are relatively flat as functions of $x$, and the region of small $z$ starts to significantly contribute to the cross section at high energies $\sqrt{s}\gg M$, see the detailed discussion in ref.~\cite{Lansberg:2020ejc}. For quarkonium-production cross sections at NLO in CF, the effect of the small-$z$ region is dramatic, as it was understood already in refs.~\cite{Schuler:1994hy, Mangano:1996kg}. 

 For our forthcoming discussion, let us express the CF coefficient function from eqn.~(\ref{eq:sigma-L-sighat}) as follows:
 \begin{equation}
 \hat{\sigma}^{[m]}_{ij}=\sigma_{0}^{[m]}\left[ A_0^{[m]}\delta(1-z) + C_{ij} \frac{\alpha_s(\mu_R)}{\pi} \left(A_0^{[m]}\ln\frac{M^2}{\mu_F^2} + A_1^{[m]} \right) + {\cal O}(z\alpha_s ,\alpha_s^2) \right], \label{eq:sighat-exp-NLO}
 \end{equation}  
where the first term in square brackets contributes only for the states $Q\bar{Q}[m]$ which can be produced in fusion of two on-shell gluons at ${\cal O}(\alpha_s^2)$. For such states,  $A^{[m]}_0=1$. For all the other $Q\bar{Q}[m]$ states, $A^{[m]}_0=0$ so that $\hat{\sigma}_{ij}^{[m]}$ starts at ${\cal O}(\alpha_s^3)$. The second term in the square brackets of the eqn.~(\ref{eq:sighat-exp-NLO}) is the $z\to 0$ asymptotics of the ${\cal O}(\alpha_s^3)$ part of the coefficient function, while all the remaining terms are collected in the term ${\cal O}(\alpha_s z,\alpha_s^2)$. Overall factors $\sigma_0^{[m]}$ are defined in eqn.~(\ref{eq:sig0-m-defs}) below. 

 The Fock-state dependent constants $A_1^{[m]}$ in eqn.~(\ref{eq:sighat-exp-NLO})  are collected, for the cases when $A_0^{[m]}\neq 0$,  in table 1 of ref.~\cite{Lansberg:2020ejc} and they turn out to be negative. 
  The colour-factors $C_{ij}$ in eqn.~(\ref{eq:sighat-exp-NLO}) are: $C_{gg}=2C_A=2N_c$, $C_{qg}=C_{gq}=C_F=(N_c^2-1)/(2N_c)$ and $C_{q\bar{q}}=0$ (with $N_c=3$ being the number of colours). If the series (\ref{eq:sighat-exp-NLO}) is truncated at NLO in $\alpha_s$, then for scale choices satisfying $\mu_F\geq M$, the contribution of the region $z\ll 1$ to the integral (\ref{eq:sigma-L-sighat}) becomes purely negative and, at sufficiently high energies, it may outweigh the positive NLO contribution from the $z\to 1$ limit and even the LO contribution. 

  For charmonium production characterised by $M\simeq 3$ GeV, this already happens at $\sqrt{s}$ as modest as several hundreds of GeV~\cite{Mangano:1996kg,Lansberg:2020ejc}  making NLO calculation completely unpredictive at LHC energies. We find the problem described above very interesting because similar instabilities plague the energy-dependence of rapidity-differential cross sections of $J/\psi$ hadroproduction in NRQCD factorisation at NLO~\cite{Feng:2015cba} and could also affect the $p_T$-differential cross section at $p_T\lesssim M\ll \sqrt{s}$.  

As it was realised in ref.~\cite{Lansberg:2020ejc}, authored by two of us, these negative cross sections in CF at NLO stem from an {\it over}subtraction of the collinear divergences inside the renormalised PDFs within the \msbar\ scheme. In principle, such a subtraction should be compensated by the evolution of the PDFs which progressively become steeper when $\mu_F$ increases. However, the coefficients $A_1$, which are related to the internal structure of the $gg\to Q\bar Q[m]$ form factor, are process-dependent as opposed to the PDF evolution governed by the Dokshitzer-Gribov-Lipatov-Altarelli-Parisi (DGLAP)~\cite{DGLAP1,DGLAP2, DGLAP3} equations. As such, the negative numbers cannot be systematically compensated by the PDF evolution. This mismatch between the partonic cross section and the PDFs is most dramatic for low scale processes for which the PDFs are the flattest.

The small-$z$ behaviour of the NLO partonic cross section of the type~(\ref{eq:sighat-exp-NLO}) is characteristic of many hard processes, as it was shown for the first time in ref.~\cite{Ellis:1990hw} for the cases of  total open heavy-flavour hadro- and photoproduction as well as prompt-photon hadroproduction. In these cases, the coefficients $A_1$ turn out to be positive, whereas they are negative for the hadroproduction of ${}^{1}S_0$, ${}^3P_{0}$ and ${}^3P_2$-states of heavy quark--anti-quark pairs as mentioned above, as well as for the photoproduction of ${}^{3}S_1$~\cite{Serri:2021fhn}. For Higgs-boson hadroproduction, the sign of the coefficient $A_1$ depends on the ratio $M_{H^0}/m_t$ (see ref.~\cite{Lansberg:2020ejc}). These observations clearly point at the fact that it is impossible to absorb the impact of $A_1$ into the PDFs via the global scheme redefinition proposed in ref.~\cite{Ellis:1990hw}: the optimal subtraction scheme for the  open heavy-flavour hadroproduction will be worse than \msbar\ for heavy quarkonia at high energies, and vice-versa.

In ref.~\cite{Lansberg:2020ejc}, two of us have proposed a new scale prescription for the factorisation scale to cure the mismatch between the partonic cross section and the PDF evolution (for processes with $A_0=1$):
\begin{equation}
\hat{\mu}_F=M\exp\left[{{A_1^{[m]}}/{2} } \right], \label{eq:muF-hat}
\end{equation}  
which restores the positivity of quarkonium-production cross sections (see refs.~\cite{Lansberg:2020ejc,Serri:2021fhn}). As it will be explained in section~\ref{sec:muF-resumm}, this scale choice corresponds to an attempt to resum some higher-order QCD corrections proportional to:
\begin{equation}
\alpha_s^{n}\ln^{n-1}\frac{1}{z}, \label{eq:LLA-z}
\end{equation} 
at leading power in $z$ for $z\ll 1$ within $\hat{\sigma}_{ij}$.
In what follows, we will refer to the all order in $\alpha_s$ resummation of the contributions of the type (\ref{eq:LLA-z})  as the {\it Leading Logarithmic-$\ln(1/z)$ Approximation} or LL($\ln(1/z)$) for short. The advantage of the $\hat{\mu}_F$-prescription in comparison to the global scheme redefinition is that it is process-dependent, i.e. every process is evaluated with its own scale (\ref{eq:muF-hat}).

 However, beyond ${\cal O}(\alpha_s)$, the approximate resummation via the $\hat{\mu}_F$-prescription does not correctly capture the structure of QCD corrections of the type (\ref{eq:LLA-z}). For instance, as we will see, the $\alpha_s^2\ln 1/z$ coefficient of the expanded expression of the NLO cross section obtained with $\mu_F=\hat{\mu}_F$ does not correspond to the one obtained from the resummation formalism. 
 The systematic formalism for such a resummation requires the use of the Balitsky-Fadin-Kuraev-Lipatov (BFKL) evolution~\cite{BFKL1, BFKL2, BFKL3} of the partonic amplitude in the rapidity $Y\sim\ln(1/z)\gg 1$ with its non-trivial transverse-momentum dynamics to correctly capture higher-order QCD corrections even in the LL($\ln(1/z)$) approximation. Such a formalism is known as {\it High Energy Factorisation (HEF)} and has been developed in refs.~\cite{Catani:1990xk, Catani:1990eg, Collins:1991ty, Catani:1994sq} in the LL($\ln(1/z)$) approximation. In the past, this formalism has been successfully applied to study the high-energy structure of hard-scattering coefficients of CF for many processes, such as heavy flavor leptoproduction~\cite{Catani:1992rn}, Higgs~\cite{Hautmann:2002tu, Harlander:2009my}, Drell-Yan lepton-pair~\cite{Marzani:2008uh} and prompt-photon production~\cite{Diana:2009xv}. In particular, in ref.~\cite{Marzani:2008uh} the correctness of the resummation predictions up to NNLO in $\alpha_s$ has been verified. Recently, the NLL($\ln (1/z)$) resummation for lepton-hadron Deep-Inelastic Scattering has been shown to significantly improve the quality of NNLO PDF fits~\cite{Ball:2017otu, Abdolmaleki:2018jln}.  
 
  With the present paper, we fill the gap in the existing HEF literature, providing the first NLO CF + LL($\ln(1/z)$) matched calculation of the energy dependence of the total heavy-quarkonium-hadroproduction cross section. The aforementioned HEF calculations~\cite{Catani:1992rn,Hautmann:2002tu, Harlander:2009my,Marzani:2008uh,Diana:2009xv}  use the representation of the resummed cross section in Mellin space, which is convenient from the computational point of view, but is difficult to match with NLO CF results. For example, in the Ref.~\cite{Catani:1992rn} the matching between LL HEF and NLO CF calculations has been implemented via subtractive prescription in the Mellin space ($N$) conjugate to $z$. With such a technique, however, it would be quite difficult to come up with any alternative matching prescription, which is necessary to estimate the uncertainty on the obtained cross section results due to matching procedure. In contrast, we perform a direct $z$-space matching between the HEF-resummed hard-scattering coefficient, which is valid at $z\ll 1$, and the full NLO CF correction, which contains numerically important ${\cal O}(z)$ power corrections. Along the same lines as in ref.~\cite{Echevarria:2018qyi}, we propose a new matching procedure, described in section~\ref{sec:InEW}, which smoothly interpolates between $z\to 0$ and $z\to 1$ limits. 
  
  Let us also discuss now the relation of our work with other approaches to quarkonium production at high energies existing in the small-$x$ physics literature, such as calculations of the $J/\psi$-hadroproduction cross section within Color-Glass-Condensate (CGC) formalism in refs.~\cite{Ma:2014mri, Kang:2013hta}. In these papers, the all-order in $\alpha_s$ resummation of radiative corrections to the {\it observable cross section} $\sigma(\sqrt{s})$ enhanced by  $\ln(s/M^2)$ was the primary objective. The CGC formalism in principle includes all such corrections in the LLA w.r.t. $\ln(s/M^2)$ in the limit $\sqrt{s}\gg M$, yet disregarding\footnote{It is interesting to note, that the {\it colour-singlet}  $Q\bar{Q}[{}^3S_1^{(1)}]$ state is produced in refs.~\cite{Ma:2014mri, Kang:2013hta} via the {\it subleading twist} contribution, $g+(g+g)\to Q\bar{Q}[{}^3S_1^{(1)}]$, requiring at least {\it two} off-shell gluons to be taken from the ``target'' proton or nucleus. The leading-twist process $g+g\to Q\bar{Q}[{}^3S_1^{(1)}] + g$ appears at NLO in $\alpha_s$ in the CGC counting. } the ``twist'' expansion in powers of the hard scale $M^2$. In contrast, in the present study, we are resumming large logarithms (\ref{eq:LLA-z}) in the coefficient function of the CF $\hat{\sigma}(z)$, i.e. we are working in the leading power approximation w.r.t. the hard scale $M^2$. As such, our calculation includes at least partially the leading-twist part of the logarithmic corrections resummed in~\cite{Ma:2014mri, Kang:2013hta}. Indeed, at relatively small scales $\mu_F\sim M\sim 1$ GeV, the partonic luminosity function ${\cal L}_{ij}(z,\mu_F)$ in eqn.~(\ref{eq:sigma-L-sighat}) varies slowly for all values of $z$ except those very close to the kinematic endpoint $z_{\min}=M^2/s$, where it abruptly goes to zero, and therefore large logarithms (\ref{eq:LLA-z}) in $\hat{\sigma}$ lead to the contributions proportional to:
  \begin{equation}
        \int \limits_{z_{\min}}^1 \frac{dz}{z}\ \alpha_s^n \ln^{n-1}\frac{1}{z} \sim \alpha_s^n \ln^n \frac{1}{z_{\min}} = \alpha_s^n \ln^n \frac{s}{M^2}, \label{eq:small-x-logs}
  \end{equation}
  which is another way to explain why the corrections of the type (\ref{eq:LLA-z}) are especially important at $s\gg M^2$. The logarithmic contributions of the type (\ref{eq:small-x-logs}) can not be resummed completely in our calculation, because the HEF formalism is based on the linear BFKL equation, while corrections of the type (\ref{eq:small-x-logs})  lead to the non-linear effects already in the LLA($\ln(s/M^2)$). These are taken into account, in the planar limit, by Balitsky-Kovchegov~\cite{Balitsky:1995ub, Kovchegov:1999yj} or exactly in $N_c$, by (Balitsky-)Jalilian-Marian-Iancu-McLerran-Weigert-Leonidov-Kovner~\cite{Balitsky:1998kc,Jalilian-Marian:1997qno,Jalilian-Marian:1997jhx,Jalilian-Marian:1997ubg,Iancu:2001ad,Iancu:2000hn} evolution equations. However, we believe that for single-scale observables, such as the total cross section $\sigma(\sqrt{s})$, in the leading twist approximation, all LLA contributions of the type (\ref{eq:small-x-logs}) can be included into the LLA HEF calculation either via the perturbative resummation of logarithms (\ref{eq:LLA-z}) in $\hat{\sigma}$ or phenomenologically, via the modification of the $x$ dependence of the collinear PDFs. This point of view is supported by the fact that higher-twist corrections originating from the modification of the ${\bf q}_T$-dependence of the HEF resummation function in the region $|{\bf q}_T| \ll M$ are small, see appendix~\ref{sec:Gauss}. Our conclusion is that the resummation of logarithms~(\ref{eq:LLA-z}) as well as its smooth matching with the NLO CF corrections at $z\sim 1$ are the most numerically important effects for the single-scale observables such as $\sigma(\sqrt{s})$ at high energy and in the leading-twist approximation.   
  
  The paper has the following structure. In section~\ref{sec:HEF}, we describe the LL($\ln(1/z)$) HEF formalism, in particular the structure of the resummed partonic cross section in $z$-space (section~\ref{sec:sigma-z-space}), the corresponding process-dependent coefficient functions (section~\ref{sec:H-HEF}) and universal resummation factors (section~\ref{sec:C-factors}). In section~\ref{sec:NNLO-expansion}, we expand the resummed cross section up to NNLO in $\alpha_s$ to verify its consistency with NLO results~\cite{Kuhn:1992qw, Petrelli:1997ge} and provide predictions for future NNLO calculations of quarkonium production cross section. In  section~\ref{sec:muF-resumm}, we explain the relation of the HEF formalism with the $\hat \mu_F$ scale prescription. In section~\ref{sec:match}, we introduce two matching procedures in $z$-space, compare the corresponding numerical results and discuss the uncertainties of our calculation. Finally, in section~\ref{sec:concl}, we summarise our conclusions and prospects for future studies. The paper contains three appendices: in  appendix~\ref{sec:num}, we explain two methods which we have developed to calculate the resummed cross section in $z$-space in a numerically-stable way. In  appendix~\ref{sec:gamma-effects}, we show the numerical effects of the ${\cal O}(\alpha_s^2)$ corrections to the anomalous dimension~(\ref{eq:pert-gamma}) on the $\mu_F$ dependence of the resummed part of the cross section and, in  appendix~\ref{sec:Gauss}, we discuss the possible size of non-perturbative effects, due to the intrinsic transverse momentum of gluons in the proton, on the value of the total quarkonium production cross section in our approximation.

\section{High-energy factorisation in the leading-logarithmic approximation}
\label{sec:HEF}

\subsection{The resummed cross section in $z$-space}
\label{sec:sigma-z-space}

  Considering the $qg$ channel as an example, the corrections proportional to $\alpha_s^n \ln^{n-1} 1/z$ to the QCD hard subprocess for the production of the final-state of interest, $Q\bar{Q}[m]$, with a four-momentum $p$  come from processes with at most $n$ additional partons in the final state:
\begin{equation}
g(p_1)+q(p_2)\to g(k_1)+g(k_2)+\ldots+Q\bar{Q}[m](p) + \ldots +g(k_{n-1}) + q(k_n),
\end{equation}
where the four-momentum labels are given in parentheses.

\begin{figure}
\begin{center}
\includegraphics[width=0.9\textwidth]{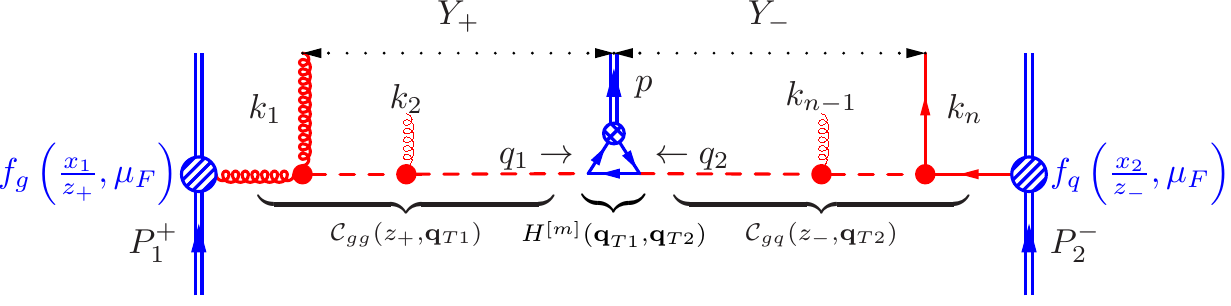}
\end{center}
\caption{Typical diagrams contributing to the high-energy factorisation amplitude in the $gq$-channel. The dashed lines denote Reggeised gluon exchanges, while solid circles denote Lipatov's vertices.}\label{fig:HEF}
\end{figure}

At leading-power in $z$ and in the LL($\ln(1/z)$) approximation, only the final states with  a strong ordering in the corresponding light-cone components of momenta contribute, that is when: 
\begin{eqnarray*}
p_1^+&\simeq& k_1^+\gg k_2^+ \gg \ldots \gg p^+ \gg \ldots \gg k_{n-1}^+ \gg k_n^+, \\
&&k_1^-\ll k_2^- \ll \ldots \ll p^- \ll \ldots \ll k_{n-1}^- \ll k_n^-\simeq p_2^-,
\end{eqnarray*}
 where $k_i^{\pm}=k_i^0\pm k_i^3$, so that, for the four-momenta of  the incoming protons, one has $P_1^+=\sqrt{s}$, $P_1^-=0$ and $P_2^+=0$, $P_2^-=\sqrt{s}$. This limit is referred to as the {\it Multi-Regge Kinematics (MRK)}. In this limit and at this accuracy, the QCD amplitudes are known to factorise into a product of gauge-invariant blocks: the Lipatov's scattering, the production vertices~\cite{BFKL1, BFKL2, BFKL3}, the $t$-channel Reggeised gluon propagators and the effective vertex for production of the $Q\bar{Q}[m]$ state, see the figure~\ref{fig:HEF}. For a review, the reader can consult the monographs~\cite{IFLQCD,kovchegov_levin_2012} or the lecture notes~\cite{RevDelDuca95}. This factorisation at the amplitude level allows one to factorise out the corresponding cross-section-level blocks: the universal {\it resummation factors} ${\cal C}_{gi}$, with $i=g,q,\bar{q}$, and  the process-dependent {\it coefficient function} $H^{[m]}$.   

The resummation factors are solutions of the BFKL equation, describing the rapidity-evolution of the cascade of emissions from the rapidities of the most backward/forward partons in the hard process to the rapidity of the observed final state. The corresponding rapidity intervals (figure~\ref{fig:HEF}) can be estimated in the MRK as:
\begin{eqnarray}
Y_+&=&\frac{1}{2}\left(\ln\frac{k_1^+}{k_1^-} - \ln \frac{p^+}{p^-} \right) \simeq \ln \left[ \frac{p_1^+}{p_+} \frac{M_T}{|{\bf k}_{T1}|} \right] = \ln \frac{1}{z_+} + \ln \frac{M_T}{|{\bf k}_{T1}|}, \label{eq:Y+} \\
Y_-&=&\frac{1}{2}\left(\ln \frac{p^+}{p^-} - \ln\frac{k_n^+}{k_n^-} \right) \simeq \ln \left[ \frac{p_2^-}{p_-} \frac{M_T}{|{\bf k}_{Tn}|} \right] = \ln \frac{1}{z_-} + \ln \frac{M_T}{|{\bf k}_{Tn}|}, \label{eq:Y-}
\end{eqnarray}
where $M_T^2=M^2+{\bf p}_T^2$, $z_+=p_+/p_1^+$ and $z_-=p_-/p_2^-$ are fractions of (+/-) light-cone components entering the hard process, which are used up to produce the final state $Q\bar{Q}[m]$. In terms of $z_+$ and $z_-$, the ``global'' $z$ variable introduced above is equal to:
\begin{equation}
z=\frac{M^2}{\hat{s}}=\frac{M^2}{p_1^+p_2^-}=\frac{M^2}{M_T^2}z_+z_-. \label{eq:z-appr}
\end{equation}

 The LL BFKL evolution of the resummation factors ${\cal C}_{gi}$ in rapidity resums higher-order corrections to the cross section proportional to $\alpha_s^n Y_{\pm}^{n-1}$. If one is only interested in the resummation of the $\ln 1/z$-corrections in the LL($\ln(1/z)$) approximation, as we do in the present paper, then one solves the BFKL equation for resummation factors, neglecting transverse-momentum logarithms in eqns.~(\ref{eq:Y+}) and (\ref{eq:Y-}), i.e. putting $Y_{\pm} \simeq \ln 1/z_{\pm}$, as it has been done in the seminal papers~\cite{Catani:1990xk, Catani:1990eg, Collins:1991ty, Catani:1994sq}. 
 
  If the transverse-momentum logarithms in eqns.~(\ref{eq:Y+}) and (\ref{eq:Y-}) are not neglected, then in addition to $\ln 1/z$-corrections the HEF actually resums the LL $\alpha_s^n \ln^{2n} (|{\bf p}_T|/M)$ and part of the NLL $\alpha_s^n \ln^{n} (|{\bf p}_T|/M)$ corrections, which are traditionally considered within the framework of Transverse-Momentum Dependent (TMD) factorisation~\cite{CollinsQCD}. This overlap between TMD and High-Energy factorisations has been explored in recent papers~\cite{Nefedov:2021vvy, Hentschinski:2021lsh}. 
  
  Postponing a more detailed discussion of the resummation factors until  section~\ref{sec:C-factors}, let us write down the factorisation formula for the rapidity-differential cross section in HEF:
  \begin{eqnarray}
  \frac{d\sigma}{dy}&=&\sum\limits_{i,j=g,q,\bar{q}}\int\limits_0^\infty d{\bf q}_{T1}^2 d{\bf q}_{T2}^2  \int\limits_{x_1}^1\frac{dz_+}{z_+} \tilde{f}_i\left( \frac{x_1}{z_+} ,\mu_F\right) {\cal C}_{gi}(z_+,{\bf q}_{T1}^2,\mu_F,\mu_R) \nonumber \\
  &\times & \int\limits_{x_2}^1\frac{dz_-}{z_-} \tilde{f}_j\left( \frac{x_2}{z_-} ,\mu_F\right) {\cal C}_{gj}(z_-,{\bf q}_{T2}^2,\mu_F,\mu_R)  \int\limits_0^{2\pi} \frac{d\phi}{2} \frac{H^{[m]}({\bf q}_{T1}^2,{\bf q}_{T2}^2,\phi)}{M_T^4},\label{eq:HEF-basic}
  \end{eqnarray}  
where $x_1=M_T e^y/\sqrt{s}$ and $x_2=M_Te^{-y}/\sqrt{s}$ with $M_T^2=M^2+({\bf q}_{T1}+{\bf q}_{T2})^2$ and $\phi$ being the azimuthal angle between ${\bf q}_{T1}$ and ${\bf q}_{T2}$. The HEF coefficient function $H^{[m]}$ non-trivially depends on the transverse momenta ${\bf q}_{T1,2}$ of the Reggeised gluons entering it (figure~\ref{fig:HEF}), and the procedure for its computation is described in section~\ref{sec:H-HEF}. As it will become clear below, this transverse-momentum dependence is the key to the factorisation of higher-order terms enhanced by $\ln 1/z$ in the QCD perturbative series in terms of the process-independent resummation functions ${\cal C}_{gj}$, encapsulating logarithms $\ln 1/z_{\pm}$ and the process dependent coefficients $H^{[m]}$, which are free from these large logarithms.  

  Integrating eqn.~(\ref{eq:HEF-basic}) over  the rapidity and introducing the variables $\eta=\ln(z_+/z_-)/2$ and $z$ via eqn.~(\ref{eq:z-appr}), one can recast the total hadronic cross section in HEF into the form of eqn.~(\ref{eq:sigma-L-sighat}) with the partonic coefficient:
  \begin{eqnarray}
  \hat{\sigma}_{ij}^{[m],{\rm \ HEF}}(z,\mu_F,\mu_R)&=& \int\limits_{-\infty}^{\infty} d\eta\int\limits_0^\infty d{\bf q}_{T1}^2 d{\bf q}_{T2}^2 \  {\cal C}_{gi}\left( \frac{M_T}{M}\sqrt{z} e^\eta, {\bf q}_{T1}^2,\mu_F,\mu_R \right) \nonumber \\ &\times& {\cal C}_{gj}\left( \frac{M_T}{M}\sqrt{z} e^{-\eta}, {\bf q}_{T2}^2,\mu_F,\mu_R \right) \int\limits_0^{2\pi} \frac{d\phi}{2} \frac{H^{[m]}({\bf q}_{T1}^2,{\bf q}_{T2}^2,\phi)}{M_T^4}, \label{eq:sig-part-rint-z} 
  \end{eqnarray}
  where the integration over $\eta$ actually proceeds only over $|\eta|\le \eta_{\rm max}=\ln\left[ M/(M_T\sqrt{z}) \right]$, because the arguments of the resummation functions $z_{\pm}$ should be smaller or equal than one. 
  
   Finally let us emphasise that, for the $gg$ channel, the transverse-momentum integrals in eqn.~(\ref{eq:sig-part-rint-z}) can not be dealt with numerically unless the LO contribution $\sigma_{0}^{[m]}\delta(1-z)$ is explicitly removed from  eqn.~(\ref{eq:sig-part-rint-z}):
  \begin{equation}
   \hat{\sigma}_{gg}^{[m],\ {\rm HEF}}(z,\mu_F,\mu_R)=\sigma_{0}^{[m]} \delta(1-z) + \check{\sigma}_{gg}^{[m],\ {\rm HEF}}(z,\mu_F,\mu_R), \label{eq:sig-check-def}
   \end{equation}
   where $\check{\sigma}_{gg}^{[m],\ {\rm HEF}}$ is the LO-{\it subtracted} resummed hard-scattering coefficient. The LO CF contribution is contained in the $\hat{\sigma}_{gg}^{[m],\ {\rm HEF}}$ because the perturbative expansion for resummation factors ${\cal C}_{gg}$ starts with the LO term $\delta(1-z)\delta({\bf q}_T^2)$, which corresponds to the case with no additional emissions. The procedure (\ref{eq:sig-check-def}) turns the integrals in eqn.~(\ref{eq:sig-part-rint-z}) into well-defined functions of $z$.  
   We describe two methods which we have used to isolate the LO contribution in the Appendix~\ref{sec:num}. The method of the section~\ref{sec:method-reg-C} relies on the form of resummation factor in the Doubly-Logarithmic approximation, which we will introduce in eqn.~(\ref{eq:Bluemlein}) in section~\ref{sec:C-factors}, while the method in section~\ref{sec:method-subtr} is more general and is applicable to the exact LL($\ln(1/z)$) resummation factor~(\ref{eq:C-full}) and its generalisations.

\subsection{HEF coefficient functions}
\label{sec:H-HEF}

For the following $2\to 1$ processes:
\begin{equation}
R(q_1)+R(q_2)\to c\bar{c}\left[ {}^1S_0^{[1,8]},{}^3S_1^{[8]},{}^3P_{0,1,2}^{[1,8]} \right],\label{eq:2-1-proc}
\end{equation}
the corresponding HEF coefficient functions are non-zero at LO in $\alpha_s$. In eqn.~(\ref{eq:2-1-proc}), the incoming Reggeised gluons are denoted by the symbol $R$ and their four-momenta can be parametrised as $q_1^\mu=x_1P_1^\mu+q_{T1}^\mu$ and $q_2^\mu=x_2P_2^\mu+q_{T2}^\mu$ (figure~\ref{fig:HEF}) so that $q_{1,2}^2=-{\bf q}_{T1,2}^2<0$. The Feynman rules of Lipatov's gauge-invariant EFT for Multi-Regge processes in QCD~\cite{Lipatov95} are used for the computation of HEF coefficient functions for general QCD processes, as it was done for quarkonia in Ref.~\cite{Kniehl:2006vm}. The use of the High-Energy EFT is particularly important for the gauge invariance\footnote{With respect to gauge choice of internal gluon propagator.} of the computation of the coefficient function for production of the ${}^3S_1^{[8]}$-state. Essentially, the Lipatov's vertex should be used instead of the usual three-gluon vertex.\footnote{Indeed, for the states listed in eqn.~(\ref{eq:2-1-proc}), except for ${}^3S_1^{[8]}$, the EFT computation is equivalent to the use of the following ``nonsense'' polarisation prescription for the initial-state gluons:
\begin{equation}
\varepsilon^\mu(q_{1,2})\to \frac{q_{T1,2}^\mu}{|{\bf q}_{T1,2}|}, \label{eq:non-sense-pol}
\end{equation}  
due to Slavnov-Taylor identities of QCD. However, for more complicated processes containing final-state gluons coupling to initial-state ones, or otherwise essentially non-Abelian, as e.g. double color-octet channels in heavy-quarkonium pair production~\cite{He:di-Jpsi}, the full set of Feynman rules of the EFT~\cite{Lipatov95} have to be used instead of the prescription (\ref{eq:non-sense-pol}). The coefficient functions for the processes (\ref{eq:2-1-proc}) have been computed for the first time in~\cite{Hagler:2000dda,Hagler:2000dd,Kniehl:2006sk} using the prescription (\ref{eq:non-sense-pol}) and later have been reproduced in the ref.~\cite{Kniehl:2006vm} using Lipatov's EFT.
}

 In the present paper we take the explicit expressions for HEF  coefficient functions, which can be directly plugged into our eqn.~(\ref{eq:sig-part-rint-z}),  from ref.~\cite{Kniehl:2006sk}. The HEF coefficient functions satisfy the on-shell-limit normalisation condition:
\begin{equation}
\int\limits_0^{2\pi} \frac{d\phi}{2\pi} \lim\limits_{{\bf q}_{T1,2}^2\to 0} H^{[m]}({\bf q}^2_{T1},{\bf q}_{T2}^2,\phi) = \frac{1}{4(N_c^2-1)^2}\sum\limits_{\lambda_{1,2}=\pm}\left\vert {\cal M} (g_{\lambda_1}g_{\lambda_2}\to Q\bar{Q}[m]) \right\vert^2, \label{eq:CF-limit}
\end{equation}
with the corresponding CF squared amplitude ${\cal M}$ averaged over colours and helicities $\lambda_{1,2}$ of the initial-state gluons. It is convenient to separate out the corresponding colour factors, the LDMEs of NRQCD $\langle {\cal O}[m] \rangle$ and the CF spin-orbital factors  from the coefficient functions as follows:
\begin{equation}
H^{[m]}({\bf q}_{T1}^2,{\bf q}_{T2}^2,\phi)= \frac{M^4 \sigma_0^{[m]}}{\pi}  F^{[m]} ({\bf q}_{T1}^2,{\bf q}_{T2}^2,\phi),   \label{eq:sig0-F-def}
\end{equation}
where the $\sigma_0^{[m]}$ factors are:
\begin{eqnarray}
  \sigma_0^{[{}^1S_0^{[1]}]} &=& \frac{8}{C_A^3C_F} \pi^3\alpha_s^2 \frac{\langle {\cal O} [ {}^1S_0^{[1])} ] \rangle}{M^5}, \label{eq:sig0-m-defs} \\
  \sigma_0^{[{}^1S_0^{[8]}]} &=& \frac{4(C_A^2-4)}{C_A^3C_F^2}\pi^3\alpha_s^2\frac{\langle{\cal O}[^1S_0^{[8]}]\rangle}{M^5}, \nonumber \\
  \sigma_0^{[{}^3S_1^{[8]}]} &=& \frac{8(C_A-2C_F)}{3C_F^2}\pi^3\alpha_s^2\frac{\langle{\cal O}[^3S_1^{[8]}]\rangle}{M^5}, \nonumber \\
\sigma_0^{[^3P_0^{[1]}]} &=&\frac{96}{C_A^3C_F}\pi^3 \alpha_s^2\frac{\langle{\cal O}[^3P_0^{[1]}]\rangle}{M^7}, \nonumber\\
\sigma_0^{[^3P_1^{[1]}]} &=&\frac{192}{C_A^3C_F}\pi^3 \alpha_s^2\frac{\langle{\cal O}[^3P_1^{[1]}]\rangle}{M^7}, \nonumber\\
\sigma_0^{[^3P_2^{[1]}]} &=&\frac{128}{5C_A^3C_F}\pi^3 \alpha_s^2\frac{\langle{\cal O}[^3P_2^{[1]}]\rangle}{M^7}, \nonumber\\
\sigma_0^{[^3P_0^{[8]}]} &=&\frac{48(C_A^2-4)}{C_A^3C_F^2} \pi^3 \alpha_s^2\frac{\langle{\cal O}[^3P_0^{[8]}]\rangle}{M^7}, \nonumber\\
\sigma_0^{[^3P_1^{[8]}]} &=&\frac{96(C_A^2-4)}{C_A^3C_F^2} \pi^3 \alpha_s^2\frac{\langle{\cal O}[^3P_1^{[8]}]\rangle}{M^7}, \nonumber\\
\sigma_0^{[^3P_2^{[8]}]} &=&\frac{64(C_A^2-4)}{5C_A^3C_F^2}\pi^3 \alpha_s^2\frac{\langle{\cal O}[^3P_2^{[8]}]\rangle}{M^7} \nonumber,   
\end{eqnarray}
where $C_A=N_c$, $C_F=(N_c^2-1)/(2N_c)$ and the averaging factors accounting for the number of polarisation ($2J+1$) and colour-singlet ($2N_c$) or colour-octet ($N_c^2-1$) states of the $Q\bar{Q}[m]$-pair have been included into the denominators of eqns.~(\ref{eq:sig0-m-defs}) according to the standard definition of LDMEs. The functions $F^{[m]}$ in the eqn.~(\ref{eq:sig0-F-def}) do not depend on the colour state of the $Q\bar{Q}$ pair and depend only on its spin and orbital momentum. For the $S$-states they are:
\begin{eqnarray}
F^{[^1S_0]}(t_1,t_2,\phi)
&=&2 \frac{\left(M^2+{\bf p}_T^2\right)^2}{(M^2+t_1+t_2)^2}
\sin^2{\phi},
\label{eq:F-1S0}\\
F^{[^3S_1]}(t_1,t_2,\phi)
&=&\frac{\left( M^2 + {\bf p}_T^2\right)
\left[ (t_1+t_2)^2 + M^2 \left(t_1+t_2-2\sqrt{t_1 t_2}
\cos\phi\right)\right]}{M^2 (M^2+t_1+t_2)^2},
\label{eq:F-3S1}
\end{eqnarray}
with ${\bf p}_T^2=({\bf q}_{T1}+{\bf q}_{T2})^2=t_1+t_2+2\sqrt{t_1t_2}\cos\phi$ with $t_{1,2}={\bf q}_{T1,2}^2$. For the $P$-wave states, one has:
\begin{eqnarray}
F^{[^3P_0]}(t_1,t_2,\phi)
&=&\frac{2}{9}\,\frac{
\left(M^2 + {\bf p}_T^2 \right)^2\left[(3 M^2 +t_1+t_2)
\cos{\phi} + 2\sqrt{t_1 t_2}\right]^2}{(M^2 + t_1 + t_2)^4},
\nonumber\\
F^{[^3P_1]}(t_1,t_2,\phi)
&=&\frac{2}{9}\,\frac{
\left(M^2 + {\bf p}_T^2 \right)^2
\left[(t_1+t_2)^2 \sin^2{\phi} + M^2 \left(t_1+t_2-2\sqrt{t_1
t_2}\cos{\phi}\right)\right]}{(M^2 + t_1 + t_2)^4},
\nonumber\\
F^{[^3P_2]}(t_1,t_2,\phi)
&=&\frac{1}{3}\,
\frac{\left(M^2+{\bf p}_T^2\right)^2}{(M^2+t_1+t_2)^4}  \left\{3 M^4 +
3M^2(t_1+t_2)+4t_1 t_2 \right.
\nonumber\\
&&{}+\left.
(t_1+t_2)^2\cos^2{\phi}+2 \sqrt{t_1 t_2}\left[3
M^2 +2 (t_1+t_2)\right]\cos{\phi}\right\}. \label{eq:H-F-functions_2-1}
\end{eqnarray}

 The on-shell limits (\ref{eq:CF-limit}) of the functions $F^{[m]}$ are equal to one or zero, depending on whether the corresponding state is allowed to be produced in a fusion of two on-shell gluons by the Landau-Yang theorem, i.e. they are equal to constants $A_0$ first introduced in eqn.~(\ref{eq:sighat-exp-NLO}). 
 
\subsection{Resummation factor in  the doubly-logarithmic approximation and beyond}
\label{sec:C-factors}
 Following the original refs.~\cite{Catani:1990xk, Catani:1990eg, Collins:1991ty, Catani:1994sq}, we introduce the Mellin transform with respect to the $z$-dependence of the resummation factor as follows: 
\begin{equation}
{\cal C}(N,{\bf q}_T,\mu_F,\mu_R)=\int\limits_0^1 dz\ z^{N-1} {\cal C}(z,{\bf q}_T,\mu_F,\mu_R),\label{eq:MT-def}
\end{equation}
so that the logarithms which we aim to resum are mapped to poles at $N=0$ in Mellin space:
\begin{equation}
\ln^{k-1} \frac{1}{z} \leftrightarrow \frac{(k-1)!}{N^{k}}.\label{eq:Log-Mellin}
\end{equation}

The emissions of additional partons at higher orders in $\alpha_s$ generate collinear divergences in the resummation factor which have been removed in ref.~\cite{Collins:1991ty} using a transverse-momentum cut off and using dimensional regularisation and the \msbar-subtraction prescription in ref.~\cite{Catani:1994sq}. After the subtraction of the collinear divergences, the dependence on factorisation scale $\mu_F$ arises in the resummation factor. The Mellin-space result~\cite{Catani:1990xk, Catani:1990eg, Collins:1991ty, Catani:1994sq} for the subtracted resummation factor is: 
\begin{equation}
{\cal C}_{gg}(N,{\bf q}_T^2,\mu_F,\mu_R)= R(\gamma_{gg}(N,\alpha_s)) \frac{\gamma_{gg}(N,\alpha_s)}{{\bf q}_T^2} \left(\frac{{\bf q}_T^2}{\mu_F^2} \right)^{\gamma_{gg}(N,\alpha_s)}, \label{eq:C-full}
\end{equation}
where $\hat{\alpha}_s=\alpha_s(\mu_R)C_A/\pi$. The choice of the scale at which the $\alpha_s$ is evaluated in the ${\cal C}_{gg}$ function is not strictly dictated by the LL($\ln(1/z)$) resummation or NLO CF results. However, since both ${\cal C}_{gg}$ functions and HEF coefficient function belong to the coeffcient function of CF $\hat{\sigma}_{ij}^{[m]}$ (see eqn.~(\ref{eq:sig-part-rint-z})), we opt to choose the same scale $\mu_R$ of the $\alpha_s$ in all these quantities. 
The anomalous dimension $\gamma_{gg}$ is the solution of the algebraic equation first derived in the ref.~\cite{Jaroszewicz:1982gr}:
\begin{equation}
\frac{\hat{\alpha}_s}{N} \chi(\gamma_{gg}(N,\alpha_s))=1, \label{eq:gamma-eqn}
\end{equation}
where $\chi(\gamma)=2\psi_0(1) - \psi_0(\gamma)-\psi_0(1-\gamma)$ is he Lipatov's LO characteristic function and $\psi_n(\gamma)=d^n \ln \Gamma(\gamma)/d\gamma^n$ is the Euler's $\psi$ function. The first few terms of the perturbative solution of eqn.~(\ref{eq:gamma-eqn}) are:
\begin{equation}
\gamma_{gg}(N,\alpha_s)=\gamma_N + 2\zeta(3)\frac{\hat{\alpha}_s^4(\mu_R)}{N^4} + 2\zeta(5)\frac{\hat{\alpha}_s^6(\mu_R)}{N^6} + {\cal O}(\alpha_s^7),{\rm \ where\ } \gamma_N=\frac{\hat{\alpha}_s(\mu_R)}{N}.\label{eq:pert-gamma}
\end{equation}

 Eqn.~(\ref{eq:C-full}) and the anomalous dimension (\ref{eq:pert-gamma}) resum the series of higher-order corrections proportional to $\alpha_s^n/N^n$. These corrections by virtue of the mapping (\ref{eq:Log-Mellin}) are equivalent to the LL($\ln(1/z)$)-approximation (\ref{eq:LLA-z}). While the LL($\ln(1/z)$) anomalous dimension (\ref{eq:pert-gamma}) is scheme-independent in a wide class of \msbar-like schemes~\cite{Catani:1994sq}, the factor $R(\gamma)$ encapsulates the subtraction-scheme dependence. For the \msbar{} scheme, its perturbative expansion starts at N$^3$LO~\cite{Catani:1994sq,Kirschner:2009qu}:
\begin{equation}
R(\gamma_{gg}(N,\alpha_s))=1+{\cal O}(\alpha_s^3).\label{eq:R-gamma-exp}
\end{equation}

 Let us now define the resummation factors for the quark-induced channels. As depicted in figure~\ref{fig:HEF}, in the LL($\ln(1/z)$)-approximation, only Reggeised gluons can propagate in the $t$-channel. Hence, the quarks can only participate in so-called  LO {\it partonic impact factors}. The corresponding expression for the squared amplitude in the LL($\ln(1/z)$) approximation differs from the similar squared amplitude with gluon substituted by the quark only by an overall colour factor. This picture leads to the simple relation between the quark and gluon-induced resummation factors in the LL($\ln(1/z)$) approximation:
 \begin{equation}
 {\cal C}_{gq}(z,{\bf q}_T^2,\mu_F,\mu_R)=\frac{C_F}{C_A}\left[ {\cal C}_{gg}(z,{\bf q}_T^2,\mu_F,\mu_R) - \delta(1-z)\delta({\bf q}_T^2) \right], \label{eq:Cgq-ren}
 \end{equation}
 where the subtraction term in square brackets removes the contribution of the direct $g\to R$ transition in the on-shell limit. The latter is absent in case of quarks due to the fermion-number conservation. 

  To clarify the physical meaning of the anomalous dimension $\gamma_{gg}$ in eqn.~(\ref{eq:C-full}), let us observe that, since the Mellin transform turns $z_{\pm}-$convolutions in eqn.~(\ref{eq:HEF-basic}) into products, the $\mu_F$ independence of the cross section order by order in $\alpha_s$  is achieved provided that:
  \begin{equation}
 \sum\limits_{i=q,\bar{q},g}  \frac{\partial}{\partial\ln\mu_F^2}\left[ \tilde{f}_i(N,\mu_F) {\cal C}_{gi}(N,{\bf q}_T^2,\mu_F,\mu_R)  \right] =0, \label{eq:muF-ind}
  \end{equation}
  at any ${\bf q}_T$. Substituting eqns.~(\ref{eq:C-full}) and (\ref{eq:Cgq-ren}) one finds that eqn.~(\ref{eq:muF-ind}) is indeed satisfied due to the following form of DGLAP equations at LL($\ln 1/z$) accuracy:
  \begin{eqnarray}
  \frac{\partial \tilde{f}_g(N,\mu_F)}{\partial \ln\mu_F^2} &=& \gamma_{gg}(N,\alpha_s) \left[ \tilde{f}_g(N,\mu_F) + \frac{C_F}{C_A} \sum\limits_{i=q,\bar{q}} \tilde{f}_i(N,\mu_F) \right], \label{eqn:DGLAPg-LLz}\\
  \frac{\partial \tilde{f}_q(N,\mu_F)}{\partial \ln\mu_F^2} &=& 0. \label{eqn:DGLAPq-LLz}
  \end{eqnarray}
  In other words, the $\gamma_{gg}(N,\alpha_s)$ is the LL($\ln(1/z)$) approximation to the anomalous dimension of the DGLAP equations, governing the scale dependence of the  gluon momentum density distribution, while $C_F\gamma_{gg}/C_A$ determines the feed down from quarks to gluon in this approximation. Indeed, the corresponding LO DGLAP splitting functions have the following $z\to 0$ asymptotics:
  \begin{eqnarray}
  \frac{\alpha_s}{2\pi}zP_{gg}(z) &=& \hat{\alpha}_s + {\cal O}(z), \label{eq:Pgg-zasy-LO} \\   \frac{\alpha_s}{2\pi}zP_{gq}(z) &=& \frac{C_F}{C_A}\hat{\alpha}_s + {\cal O}(z). \label{eq:Pgq-zasy-LO}
  \end{eqnarray}
  Eqn.~(\ref{eq:Pgg-zasy-LO}) is equivalent to the first term of the solution~(\ref{eq:pert-gamma}) via the mapping (\ref{eq:Log-Mellin}), while eqn.~(\ref{eq:Pgq-zasy-LO}) is related in the same way to the first term of the expansion of $C_F\gamma_{gg}/C_A$ in $\alpha_s$. The coefficients in front of $\alpha_s\ln(1/z)$ and $\alpha_s^2\ln^2(1/z)$ terms in the NLO and NNLO contributions to $zP_{gg}(z)$ are zero in QCD, consistently with eqn.~(\ref{eq:pert-gamma}). The first non-zero contributions at leading power in $z$ beyond LO in $\alpha_s$ in QCD are ${\cal O}(\alpha_s^2)$ and ${\cal O}(\alpha_s^3 \ln(1/z))$. They belong to the NLL($\ln(1/z)$) approximation and their consistency with NLL($\ln(1/z)$) BFKL predictions has been verified in ref.~\cite{Vogt:2004mw}, while the consistency of the LL($\ln(1/z)$) series (\ref{eq:pert-gamma}) with the DGLAP anomalous dimension in ${\cal N}=4$-supersymmetric Yang-Mills theory in the large-$N_c$ limit has been checked in ref.~\cite{Lukowski:2009ce} up to five loops using integrability-based methods.   
  
 The connection with the DGLAP anomalous dimension, outlined above, is very important for the phenomenological strategy of the usage of the LL($\ln(1/z)$) resummation. Most of the common PDF sets\footnote{With the notable exception of the sets obtained in refs.~\cite{Ball:2017otu, Abdolmaleki:2018jln}.} use the fixed-order approximation for the DGLAP splittings/anomalous dimensions, i.e. no $\ln(1/z)$-resummation is performed in the evolution. In particular, it means that existing LO, NLO and NNLO PDFs contain information only about the $\gamma_N$ term in the series (\ref{eq:pert-gamma}). Including further terms of this series into the CF coefficient function via the resummation will only increase the mismatch between the $\mu_F$ dependence of PDFs and the resummation factors and, hence, blow up our $\mu_F$-scale uncertainty at high energy. 
 
 The goal of the present paper is to use HEF to cure the unphysical behaviour of the CF coefficient function at small $z$ in a way consistent with the NLO DGLAP evolution of PDFs. Therefore, we will use the following {\it Doubly-Logarithmic approximation (DLA)} for resummation factor, which is obtained by neglecting all terms of the series (\ref{eq:pert-gamma}) except the first one:    
 \begin{equation}
 {\cal C}_{gg}^{\rm DLA}(N,{\bf q}_T,\mu_F,\mu_R)=\frac{ \gamma_N}{ {\bf q}_T^2} \left( \frac{{\bf q}_T^2}{\mu_F^2} \right)^{\gamma_N}.  \label{eq:qT-UPDF}
 \end{equation}
In the DLA, only terms $\propto\left(\frac{\alpha_s}{N} \ln\frac{{\bf q}_T^2}{\mu_F^2}\right)^n \leftrightarrow \alpha_s^n \ln^{n-1}\frac{1}{z} \ln^n\frac{{\bf q}_T^2}{\mu_F^2}$ are included in the resummation factor. One also takes $R(\gamma)=1$ in the eqn.~(\ref{eq:qT-UPDF}) for the same reason: the corresponding scheme-dependence starts at N$^3$LO, eqn.~(\ref{eq:R-gamma-exp}), and is not taken into account in the NLO or even NNLO PDFs. The inclusion of this factor to the resummation function ${\cal C}$ would only lead to an unphysical subtraction-scheme mismatch between the PDFs and the CF coefficient function.

The inverse Mellin transform:
 \begin{equation}
 {\cal C}(z,{\bf q}_T^2,\mu_F,\mu_R)= \int\limits_{-i\infty}^{+i\infty}\frac{dN}{2\pi i} z^{-N} {\cal C}(N,{\bf q}_T^2,\mu_F,\mu_R), \label{eq:inverse-Mellin}
 \end{equation}
 of the resummation factor in the DLA can be computed straightforwardly, e.g. order-by-order in $\alpha_s$ with the use of relation (\ref{eq:Log-Mellin}), leading to the following result:   
\begin{equation}
{\cal C}_{gg}^{\text{DLA}}(z,{\bf q}_T^2,\mu_F,\mu_R)=\frac{\hat{\alpha}_s(\mu_R)}{{\bf q}_T^2} \left\{ \begin{matrix}
J_0\left( 2\sqrt{\hat{\alpha}_s(\mu_R) \ln\left(\frac{1}{z}\right) \ln \left(\frac{\mu_F^2}{{\bf q}_T^2} \right) } \right) & \text{if }{\bf q}_T^2<\mu_F^2, \\
I_0\left( 2\sqrt{\hat{\alpha}_s(\mu_R) \ln\left(\frac{1}{z}\right) \ln \left(\frac{{\bf q}_T^2}{\mu_F^2} \right) } \right) & \text{if }{\bf q}_T^2>\mu_F^2,
\end{matrix} \right. \label{eq:Bluemlein}
\end{equation}
 where $J_0/I_0$ are Bessel functions of first/second kind.  Eqn.~(\ref{eq:Bluemlein}) is known in the HEF community\footnote{See e.g. Ref.~\cite{Kniehl:2011hc} for its application to the inclusive jet production at the LHC.} as the {\it (Collins-Ellis-)Bl\"umlein formula,} and first appears in  ref.~\cite{Blumlein:1995eu} where the solution of the evolution equation written in ref.~\cite{Collins:1991ty} has been studied.

The DLA resummation function (\ref{eq:qT-UPDF}), or equivalently (\ref{eq:Bluemlein}), has a very important normalisation property:
 \begin{eqnarray}
 \int\limits_0^{\mu_F^2}d{\bf q}_T^2\ {\cal C}_{gg}^{\rm DLA}(N,{\bf q}_T^2,\mu_F,\mu_R) = 1 \leftrightarrow \int\limits_0^{\mu_F^2}d{\bf q}_T^2\ {\cal C}_{gg}^{\rm DLA} (z,{\bf q}_T^2,\mu_F,\mu_R) = \delta(1-z), \label{eq:Blu-norm}
 \end{eqnarray}
 which will be used extensively in the rest of the paper. The similar normalisation condition for ${\cal C}_{gq}$ in the DLA follows form eqn.~(\ref{eq:Cgq-ren}):
 \begin{equation}
 \int\limits_0^{\mu_F^2} d{\bf q}_T^2\ {\cal C}^{\rm DLA}_{gq}(z,{\bf q}_T^2,\mu_F,\mu_R)=0. \label{eq:Cgq-norm}
 \end{equation}

\subsection{Small-$z$ hard-scattering coefficient at NLO and NNLO from resummation}
\label{sec:NNLO-expansion}

As a consistency check, the LL($\ln(1/z)$) resummation described above should reproduce the small-$z$ behaviour (\ref{eq:sighat-exp-NLO}) of the known NLO result for the coefficient function~\cite{Kuhn:1992qw, Petrelli:1997ge, Lansberg:2020ejc}. To this end, one has to expand the resummed CF coefficient functions (\ref{eq:sig-part-rint-z}) at least up to NLO in $\alpha_s$. In this section, we perform such an expansion up to NNLO in order to provide predictions for the $\alpha_s^2 \ln(1/z)$ NNLO terms in the CF coefficient functions. The results of this section will also be employed in sec.~\ref{sec:InEW} to construct the weight functions needed for our matching procedure.  The DLA resummation factor (\ref{eq:qT-UPDF}) is appropriate for the expansion up to NNLO because the exact LL($\ln(1/z)$)-resummation factor (\ref{eq:C-full}) differs from it only starting at ${\cal O}(\alpha_s^3)$. 

 To perform such an expansion, it is convenient to work in Mellin space, plugging in the inverse Mellin transform (\ref{eq:inverse-Mellin}) of eqns.~(\ref{eq:qT-UPDF}) into eqn.~(\ref{eq:sig-part-rint-z}). Passing to the dimensionless transverse momenta ${\bf n}_{T1}={\bf q}_{T1}/M$ and ${\bf n}_{T2}={\bf q}_{T2}/M$, one can rewrite the resummed CF coefficient functions as:
 \begin{equation}
 \hat{\sigma}^{[m],\text{\ HEF}}_{ij}=\sigma_0^{[m]} \int\limits_{-\infty}^\infty d\eta \int\limits_{-i\infty}^{+i\infty}\frac{dN_+ dN_-}{(2\pi i)^2}\ z^{-\frac{N_+ +N_-}{2}} e^{\eta (N_+-N_-)}  G^{[m]}_{ij}\left(\gamma_{N_+},\gamma_{N_-},\frac{N_++N_-}{2},\frac{M^2}{\mu_F^2}\right), \label{eq:dsighat-for-Exp} 
 \end{equation}
 where
 \begin{equation}
G^{[m]}_{gg}\left(\gamma_1,\gamma_2,\nu,\frac{M^2}{\mu_F^2}\right)= \left( \frac{M^2}{\mu_F^2} \right)^{\gamma_{1}+\gamma_{2}} \int\limits_0^{\infty}\frac{\gamma_1\gamma_2 d{\bf n}_{T1}^2 d{\bf n}_{T2}^2}{({\bf n}_{T1}^2)^{1-\gamma_1}({\bf n}_{T2}^2)^{1-\gamma_2}} \int\limits_0^{2\pi} \frac{d\phi}{2\pi} f^{[m]}({\bf n}_{T1}^2,{\bf n}_{T2}^2,\phi,\nu), \label{eq:F-gamma} 
\end{equation}
for the gluon channel, while for quark-induced channels, one takes into account eqn.~(\ref{eq:Cgq-ren}) to obtain
\begin{eqnarray}
 G^{[m]}_{qg}&=& \left(\frac{M^2}{\mu_F^2} \right)^{\gamma_2} \frac{C_F}{C_A} \int\limits_0^{\infty} \frac{\gamma_2 d{\bf n}_{T2}^2}{({\bf n}_{T2}^2)^{1-\gamma_2}} \int\limits_0^{\infty} d{\bf n}_{T1}^2 \left[ \left(\frac{M^2}{\mu_F^2}\right)^{\gamma_1} \gamma_1({\bf n}_{T1}^2)^{-1+\gamma_1} - \delta({\bf n}_{T1}^2) \right] \nonumber \\
 &\times& \int\limits_0^{2\pi} \frac{d\phi}{2\pi} f^{[m]}({\bf n}_{T1}^2,{\bf n}_{T2}^2,\phi), \label{eq:Fqg-gamma} 
 \end{eqnarray}
and 
 \begin{eqnarray}
 G^{[m]}_{q\bar{q}}&=& \left(\frac{C_F}{C_A}\right)^2 \int\limits_0^{\infty} d{\bf n}_{T1}^2 d{\bf n}_{T2}^2 \left[ \left(\frac{M^2}{\mu_F^2}\right)^{\gamma_1} \gamma_1({\bf n}_{T1}^2)^{-1+\gamma_1} - \delta({\bf n}_{T1}^2) \right] \nonumber \\
 &\times& \left[ \left(\frac{M^2}{\mu_F^2}\right)^{\gamma_2} \gamma_2({\bf n}_{T2}^2)^{-1+\gamma_2} - \delta({\bf n}_{T2}^2) \right] \int\limits_0^{2\pi} \frac{d\phi}{2\pi} f^{[m]}({\bf n}_{T1}^2,{\bf n}_{T2}^2,\phi), \label{eq:Fqq-gamma}
 \end{eqnarray}
where, in eqns.~(\ref{eq:F-gamma}), (\ref{eq:Fqg-gamma}) and (\ref{eq:Fqq-gamma}), the dimensionless function
 \begin{equation}
 f^{[m]}({\bf n}_{T1}^2,{\bf n}_{T2}^2,\phi,\nu)=\frac{\pi H^{[m]}(M^2{\bf n}_{T1}^2,M^2{\bf n}_{T2}^2,\phi)}{\sigma_0^{[m]} M^4(1+({\bf n}_{T1}+{\bf n}_{T2})^2)^{2+\nu}} = \frac{F^{[m]}(M^2{\bf n}_{T1}^2,M^2{\bf n}_{T2}^2,\phi)}{(1+({\bf n}_{T1}+{\bf n}_{T2})^2)^{2+\nu}}, \label{eq:fm-def}
 \end{equation}
 encapsulates the HEF coefficient function $H^{[m]}$ (or $F^{[m]}$ of eqn.~(\ref{eq:sig0-F-def})) and the kinematic factor $1/M_T^4$ from eqn.~(\ref{eq:sig-part-rint-z}).
 
 The only quantity depending on $\alpha_s$ in eqn.~(\ref{eq:dsighat-for-Exp}) is $\gamma_N=\hat{\alpha}_s/N$. One thus only has to Taylor-expand the function $G_{ij}^{[m]}$ with respect to its arguments $\gamma_{N_+}$ and $\gamma_{N_-}$. The corresponding poles in $N_+$ and $N_-$  map to the following functions of $z$ via  eqn.~(\ref{eq:dsighat-for-Exp}):
 \begin{eqnarray}
 1&\to& \delta(1-z), \nonumber \\
 \frac{1}{N_+} \text{\ and\ } \frac{1}{N_-} &\to& \theta(1-z), \nonumber \\
 \frac{1}{N_+^2},\ \frac{1}{N_-^2}, \text{\ and\ }\frac{1}{N_+N_-} &\to& \theta(1-z)\ln\frac{1}{z}. \label{eq:Npm-z} 
 \end{eqnarray}
 
  The expansion of the ${\bf n}_{T1,2}^2$ integrand of the function $G_{ij}^{[m]}$ in $\gamma_1$ and $\gamma_2$ however has to be done in a distributional sense because otherwise the order-by-order integrals will just diverge at ${\bf n}_{T1,2}^2\to 0$. To this end, one isolates the ${\bf n}_{T1,2}^2\to 0$ behaviour of the $f^{[m]}$ function\footnote{Which can be understood as a test function, on which functionals (\ref{eq:F-gamma}), (\ref{eq:Fqg-gamma}) and (\ref{eq:Fqq-gamma}) depend.} in the integrands of eqns.~(\ref{eq:F-gamma}), (\ref{eq:Fqg-gamma}) and (\ref{eq:Fqq-gamma}) as follows:
   \begin{eqnarray}
&&  f^{[m]}({\bf n}_{T1}^2,{\bf n}_{T2}^2,\phi,\nu) =\nonumber \\
&&  {\left\{  f^{[m]}({\bf n}_{T1}^2,{\bf n}_{T2}^2,\phi,\nu) -  f^{[m]}(0,{\bf n}_{T2}^2,\phi,\nu)\theta(1-{\bf n}_{T1}^2) \right. } \nonumber \\
&&  {\left. - f^{[m]}({\bf n}_{T1}^2,0,\phi,\nu)\theta(1-{\bf n}_{T2}^2) + f^{[m]}(0,0,\phi,\nu)\theta(1-{\bf n}_{T1}^2) \theta(1-{\bf n}_{T2}^2)  \right\}} \nonumber \\
&&+  {\left[ f^{[m]}({\bf n}_{T1}^2,0,\phi,\nu) - f^{[m]}(0,0,\phi,\nu)\theta(1-{\bf n}_{T1}^2) \right]\theta(1-{\bf n}_{T2}^2)} \nonumber \\
&& +{\left[ f^{[m]}(0,{\bf n}_{T2}^2,\phi,\nu) - f^{[m]}(0,0,\phi,\nu)\theta(1-{\bf n}_{T2}^2) \right]\theta(1-{\bf n}_{T1}^2) } \nonumber \\ 
&& {+ f^{[m]}(0,0,\phi,\nu)\theta(1-{\bf n}_{T1}^2) \theta(1-{\bf n}_{T2}^2). } \label{eq:f-nT-exp}
  \end{eqnarray}
  
  The expression in curly brackets in eqn.~(\ref{eq:f-nT-exp}) tends to zero when ${\bf n}^2_{T1}\to 0$ and/or ${\bf n}_{T2}^2\to 0$, so the factors $({\bf n}_{T1}^2)^{-1+\gamma_{1}}$ and $({\bf n}_{T2}^2)^{-1+\gamma_{2}}$ in front of it in  eqns.~(\ref{eq:F-gamma}) -- (\ref{eq:Fqq-gamma}) can be safely Taylor-expanded in $\gamma_{1,2}$. The expression in the fourth line of eqn.~(\ref{eq:f-nT-exp}) tends to zero when ${\bf n}_{T1}^2\to 0$ while its dependence on ${\bf n}_{T2}^2$ is just a step function. As a consequence, when this expression is substituted to eqns.~(\ref{eq:F-gamma}) -- (\ref{eq:Fqq-gamma}), the dependence on ${\bf n}_{T2}^2$ can be integrated out, while  the factor $({\bf n}_{T1}^2)^{-1+\gamma_{1}}$ can be Taylor-expanded in $\gamma_1$. For the term in the fifth line of eqn.~(\ref{eq:f-nT-exp}), the situation is symmetric, up to  the replacement ${\bf n}_{T1}\leftrightarrow {\bf n}_{T2}$. Finally, with the last term in eqn.~(\ref{eq:f-nT-exp}), the integrations over ${\bf n}_{T1}^2$ and ${\bf n}_{T2}^2$ can be performed straightforwardly. The resulting expansions for  the functions $G_{ij}^{[m]}$ in $\gamma_1$ and $\gamma_2$ at $\nu=0$ are:    
 \begin{eqnarray}
 G^{[m]}_{gg} &=& \left( \frac{M^2}{\mu_F^2} \right)^{\gamma_{1}+\gamma_{2}} \left[ A^{[m]}_0+(\gamma_{1} + \gamma_{2})A^{[m]}_1 + (\gamma_{1}^2+\gamma_{2}^2)A^{[m]}_2 + \gamma_{1}\gamma_{2} B^{[m]}_2 + {\cal O}(\gamma^3) \right] . \label{eq:F-gamma-exp}  \\
 G^{[m]}_{qg}&=&  \frac{C_F}{C_A}\left(\frac{M^2}{\mu_F^2} \right)^{\gamma_2}\left\{ \left[\left(\frac{M^2}{\mu_F^2} \right)^{\gamma_1}-1 \right] A^{[m]}_0 + \left(\frac{M^2}{\mu_F^2} \right)^{\gamma_1} (\gamma_1A^{[m]}_1+\gamma_1^2 A^{[m]}_2)   \right. \label{eq:Fqg-gamma-exp} \\
 &+&\left. \left[\left(\frac{M^2}{\mu_F^2} \right)^{\gamma_1}-1 \right] (\gamma_2A^{[m]}_1+\gamma_2^2 A^{[m]}_2) + \left(\frac{M^2}{\mu_F^2} \right)^{\gamma_1} \gamma_1\gamma_2B^{[m]}_2 + {\cal O}(\gamma^3) \right\}, \nonumber \\
 G^{[m]}_{q\bar{q}}&=&  \left(\frac{C_F}{C_A}\right)^2 \left\{ \left[\left(\frac{M^2}{\mu_F^2} \right)^{\gamma_1}-1 \right] \left[\left(\frac{M^2}{\mu_F^2} \right)^{\gamma_2}-1 \right]A^{[m]}_0 \right. \label{eq:Fqq-gamma-exp} \\
 &+&\left[\left(\frac{M^2}{\mu_F^2} \right)^{\gamma_2}-1 \right] \left(\frac{M^2}{\mu_F^2} \right)^{\gamma_1}(\gamma_1A^{[m]}_1 + \gamma_1^2 A^{[m]}_2) \nonumber \\
 &+&\left.\left[\left(\frac{M^2}{\mu_F^2} \right)^{\gamma_1}-1 \right] \left(\frac{M^2}{\mu_F^2} \right)^{\gamma_2}(\gamma_2A^{[m]}_1 + \gamma_2^2 A^{[m]}_2) + \left(\frac{M^2}{\mu_F^2} \right)^{\gamma_1+\gamma_2} \gamma_1\gamma_2B^{[m]}_2 + {\cal O}(\gamma^3)\right\}, \nonumber 
 \end{eqnarray}
 where the Taylor expansion of the factors $(M^2/\mu_F^2)^{\gamma_{1,2}}$ up to the second order have to be done as well. The coefficients $A_{0,1,2}^{[m]}$ entering this expansion are:
 \begin{eqnarray}
A_0^{[m]}&=&\int\limits_0^{2\pi} \frac{d\phi}{2\pi} f^{[m]}(0,0,\phi,0), \label{eq:A0-formula}\\
A_1^{[m]}&=&\int\limits_0^\infty \frac{d{\bf n}_{T1}^2}{{\bf n}_{T1}^2} \int\limits_0^{2\pi} \frac{d\phi}{2\pi} \left[ f^{[m]}({\bf n}_{T1}^2,0,\phi,0) - f^{[m]}(0,0,\phi,0)\theta(1-{\bf n}_{T1}^2) \right],\label{eq:A1-formula}\\ 
A_2^{[m]}&=&\int\limits_0^\infty \frac{d{\bf n}_{T1}^2}{{\bf n}_{T1}^2} \ln {\bf n}_{T1}^2 \int\limits_0^{2\pi} \frac{d\phi}{2\pi} \left[ f^{[m]}({\bf n}_{T1}^2,0,\phi,0) - f^{[m]}(0,0,\phi,0)\theta(1-{\bf n}_{T1}^2) \right],  \label{eq:A2-formula}
 \end{eqnarray}
 while, for the coefficient $B_2^{[m]}$, one has
  \begin{eqnarray}
 B^{[m]}_2&=&\int\limits_0^\infty \frac{d{\bf n}_{T1}^2 d{\bf n}_{T2}^2}{{\bf n}_{T1}^2 {\bf n}_{T2}^2} \int\limits_0^{2\pi}\frac{d\phi}{2\pi} \left\{  f^{[m]}({\bf n}_{T1}^2,{\bf n}_{T2}^2,\phi,0) -  f^{[m]}(0,{\bf n}_{T2}^2,\phi,0)\theta(1-{\bf n}_{T1}^2) \right. \nonumber \\
&& \left. - f^{[m]}({\bf n}_{T1}^2,0,\phi,0)\theta(1-{\bf n}_{T2}^2) + f^{[m]}(0,0,\phi,0)\theta(1-{\bf n}_{T1}^2) \theta(1-{\bf n}_{T2}^2)  \right\}. \label{eq:B2-formula}
 \end{eqnarray}
 
 Substituting the expansion (\ref{eq:F-gamma-exp}) to eqn.~(\ref{eq:dsighat-for-Exp}) due to the mappings (\ref{eq:Npm-z}), one obtains the following result for the $z$-dependent CF coefficient function in $gg$ channel:
   \begin{eqnarray}
   \hat{\sigma}^{[m],\text{\ HEF}}_{gg}&=&\sigma^{[m]}_{0} \left\{ A^{[m]}_0\delta(1-z) + \frac{\alpha_s}{\pi}2C_A\left[ A^{[m]}_1+A^{[m]}_0\ln\frac{M^2}{\mu_F^2} \right] \right.  \label{eq:sighat-gg-NNLO} \\ &+& \left.\left(\frac{\alpha_s}{\pi}\right)^2 C_A^2 \ln\frac{1}{z} \left[ 2A^{[m]}_2 + B^{[m]}_2 + 4A^{[m]}_1\ln\frac{M^2}{\mu_F^2} + 2A^{[m]}_0\ln^2\frac{M^2}{\mu_F^2} \right] + {\cal O}(\alpha_s^3) \right\}, \nonumber
   \end{eqnarray}
   where $\sigma^{[m]}_{0}$ has been defined in eqn.~(\ref{eq:sig0-m-defs}). The coefficients $A_{0,1,2}^{[m]}$ and $B_2^{[m]}$ can be computed for various spin-orbital states $m={}^{2S+1}L_J$ by substituting the HEF coefficient functions from eqn.~(\ref{eq:H-F-functions_2-1}) to eqns.~(\ref{eq:A0-formula}), (\ref{eq:A1-formula}), (\ref{eq:A2-formula}) and (\ref{eq:B2-formula}). The results of this computation are listed in the table~\ref{tab:pert-coefs}.
   
 \begin{table}
  \begin{center}\renewcommand{\arraystretch}{1.3}
  \begin{tabular}{|ccccc|}
  \hline
  State     &  $A_0^{[m]}$   & $A_1^{[m]}$                        & $A_2^{[m]}$ & $B_2^{[m]}$ \\
  \hline
  ${}^1S_0$ &  $1$         & \boldmath{$-1$}            & $\frac{\pi^2}{6} $        &    $\frac{\pi^2}{6}$       \\  
  ${}^3S_1$ &  $0$         & $1$                              & $0$          &  $\frac{\pi^2}{6}$         \\ 
  ${}^3P_0$ &  $1$         & \boldmath{$-\frac{43}{27}$}& $\frac{\pi^2}{6}+\frac{2}{3}$          & $\frac{\pi^2}{6}+\frac{40}{27}$          \\  
  ${}^3P_1$ &  $0$         & $\frac{5}{54}$                   & $-\frac{1}{9}$          &    $-\frac{2}{9}$       \\  
  ${}^3P_2$ &  $1$         & \boldmath{$-\frac{53}{36}$}& $\frac{\pi^2}{6}+\frac{1}{2}$          & $\frac{\pi^2}{6}+\frac{11}{9}$          \\    
  \hline
    \end{tabular}
  \end{center}
  \caption{Coefficients entering into eqns.~(\ref{eq:sighat-gg-NNLO}) -- (\ref{eq:sighat-qq-NNLO}) for several states $m={}^{2S+1}L_J$ of a $Q\bar{Q}$ pair. The highlighted numbers coincide with the corresponding results from the table 1 in ref.~\cite{Lansberg:2020ejc}. \label{tab:pert-coefs}}
 \end{table}
 
 Similarly, from the expansions (\ref{eq:Fqg-gamma-exp}) and (\ref{eq:Fqq-gamma-exp}), one obtains the following predictions for the small-$z$ behaviour of  the quark-induced channels at NLO and NNLO:
  \begin{eqnarray}
 \hat{\sigma}^{[m],\text{\ HEF}}_{qg}&=&\sigma_{0}^{[m]}\left\{ \frac{\alpha_s}{\pi} C_F \left[ A^{[m]}_1+A^{[m]}_0\ln\frac{M^2}{\mu_F^2} \right] + \left(\frac{\alpha_s}{\pi}\right)^2 C_AC_F \ln\frac{1}{z}  \right. \nonumber \label{eq:sighat-qg-NNLO}\\
 &\times& \left.\left[ A^{[m]}_2+B^{[m]}_2+3A^{[m]}_1\ln\frac{M^2}{\mu_F^2} + \frac{3}{2}A^{[m]}_0\ln^2\frac{M^2}{\mu_F^2} \right]\right\}, \\
  \hat{\sigma}^{[m],\text{\ HEF}}_{q\bar{q}}&=& \sigma_{0}^{[m]} \left(\frac{\alpha_s}{\pi}\right)^2 C_F^2 \ln\frac{1}{z} \left[B^{[m]}_2+2A^{[m]}_1\ln\frac{M^2}{\mu_F^2} + A^{[m]}_0 \ln^2\frac{M^2}{\mu_F^2} \right].  \label{eq:sighat-qq-NNLO}
 \end{eqnarray}
 Where the same coefficients $A_{0,1,2}^{[m]}$ and $B_2^{[m]}$ appear as in the $gg$ case. The NLO parts of eqns.~(\ref{eq:sighat-gg-NNLO}) and (\ref{eq:sighat-qg-NNLO}) coincide with the $z\to 0$ asymptotics of the full NLO results, obtained in refs.~\cite{Mangano:1996kg,Lansberg:2020ejc}, which is yet another non-trivial cross-check of the HEF formalism. 
 
  \begin{figure}
 \begin{center}
 \includegraphics[width=0.6\textwidth]{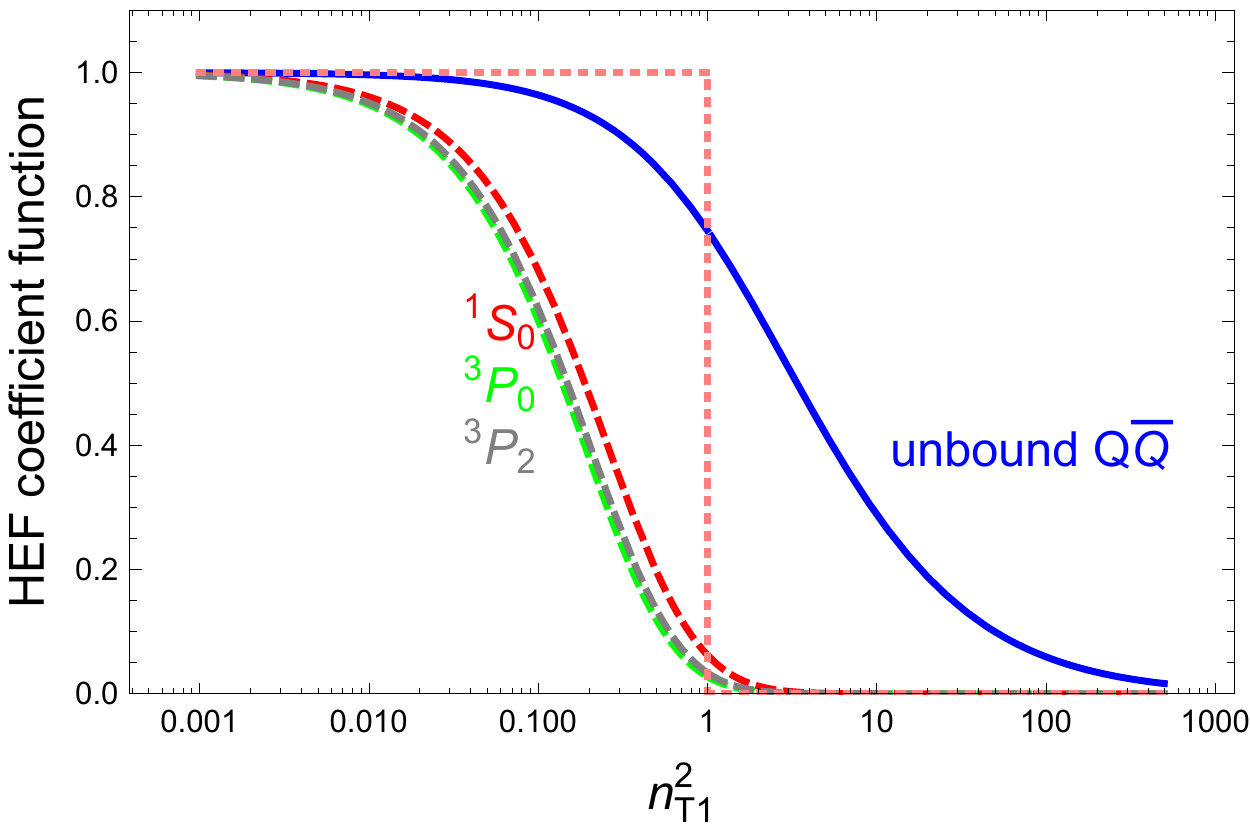}
 \end{center}
 \caption{ Dimensionless integrand function (\ref{eq:fm-def}) averaged over $\phi$, at ${\bf n}_{T2}^2=0$, as a function of ${\bf n}_{T1}^2={\bf q}_{T1}^2/(4m_Q^2)$, normalised on its value at ${\bf n}_{T1}^2=0$. The three dashed curves correspond to the coefficient functions for production of $Q\bar{Q}[m]$-states with $m={}^1S_0,$ ${}^3P_0$ and ${}^3P_2$ (section~\ref{sec:H-HEF}). The solid curve depicts the coefficient function for the unbound $Q\bar{Q}$-pair production~\cite{Ellis:1990hw}, integrated over the phase space of the $Q\bar{Q}$ pair. The short-dashed line represents the \msbar\ subtraction term $\theta(1-{\bf n}_{T1}^2)$.}\label{fig:IF-plot}
 \end{figure} 
 
 Let us emphasise that the specific transverse-momentum dependence of the HEF coefficient functions is crucial for the consistency of the HEF results with the high-energy limit of the QCD scattering amplitudes.  This is very different from the case of TMD factorisation, which is applicable only at ${\bf p}_T^2\ll M^2$. The NLO coefficient $A_1^{[m]}$ is determined  by the  behaviour of the coefficient function for ${\bf p}_T^2\simeq {\bf q}_{T1,2}^2\sim M^2$ as follows. In the integrand of eqn.~(\ref{eq:A1-formula}), the step-function $\theta(1-{\bf n}_{T1}^2)$ is subtracted from the dimensionless function  $f^{[m]}({\bf n}_{T1}^2,0,\phi,0)$, which describes the coefficient function in the case when one gluon is off-shell with momenta ${\bf q}_{T,1}$ and the second one is on-shell ${\bf q}_{T,2}^2=0$. In the case of quarkonia, the latter function starts to decrease already for ${\bf n}_{T1}^2={\bf q}_{T1}^2/M^2<1$, so the subtraction leads to a large negative contribution, as illustrated by the plot in the figure~\ref{fig:IF-plot}. For comparison, the coefficient function for open heavy flavour production, integrated over the phase-space of the final-state heavy quarks, given by eqns.~(4.8) -- (4.10) in ref.~\cite{Ellis:1990hw}, is also plotted in  figure~\ref{fig:IF-plot}. From this figure, one immediately realises that the bound state is easily broken by the transverse-momentum imbalance between the incoming off-shell gluons, which leads to a quickly decreasing HEF coefficient function. This dependence is quite different from the ``expectation'' of the \msbar\ scheme\footnote{See e.g. section 3.4 of ref.~\cite{CollinsQCD} for the relation of the \msbar{}  subtraction of $1/\epsilon$ singularity with the transverse momentum cut off.}, according to which the coefficient function should behave roughly like $\theta(1-{\bf q}_{T1,2}^2/M^2)$ thus leading to the consequent oversubtraction of the collinear behaviour. For the case of open heavy flavour, integrated over its invariant mass, the situation is opposite and the coefficient function has a substantial tail at ${\bf q}_{T1,2}^2> M^2$, which leads to large positive NLO corrections at high energy.

\subsection{Resummation by a $\mu_F$ choice}
\label{sec:muF-resumm}

The connection between the HEF resummation and the $\mu_F$-scale optimisation approach (the $\hat \mu_F$ prescription) proposed in ref.~\cite{Lansberg:2020ejc} is most easily illustrated by eqn.~(\ref{eq:F-gamma}). Substituting the scale choice (\ref{eq:muF-hat}), one obtains: 
 \begin{eqnarray}
  G^{[m]}_{gg}\left(\gamma_{N_+},\gamma_{N_-},\nu,\frac{M^2}{\hat{\mu}_F^2}\right)&=& \exp\left[-A_1^{[m]} (\gamma_{N_+}+\gamma_{N_-}) \right] \label{eq:F-gamma_muF-hat} \\&\times& \int\limits_0^{\infty}\frac{\gamma_{N_+}\gamma_{N_-} d{\bf n}_{T1}^2 d{\bf n}_{T2}^2}{({\bf n}_{T1}^2)^{1-\gamma_{N_+}}({\bf n}_{T2}^2)^{1-\gamma_{N_-}}}  \int\limits_0^{2\pi} \frac{d\phi}{2\pi} f^{[m]}({\bf n}_{T1}^2,{\bf n}_{T2}^2,\phi,\nu). \nonumber
 \end{eqnarray}
 
 The exponent in the first line of eqn.~(\ref{eq:F-gamma_muF-hat}), which arose from the scale choice (\ref{eq:muF-hat}), resums a series of corrections proportional to $\gamma_{N_{\pm}}^n=\hat{\alpha}_s^n/N_{\pm}^n$, which belong to the LL($\ln(1/z)$)-approximation. This resummation would be equivalent to that performed by the HEF only if this exponent is able to cancel the $\gamma_{N_{\pm}}$ dependence of the integral in the second line of eqn.~(\ref{eq:F-gamma_muF-hat}). If this was the case then all LL($\ln(1/z)$) corrections would be removed from the CF coefficient function and absorbed into the scale evolution of PDFs. However, such a perfect cancellation is possible only if the dimensionless function $f^{[m]}$ complies to the relation:
 \begin{equation}
 \int\limits_0^{2\pi} \frac{d\phi}{2\pi} f^{[m]}({\bf n}_{T1}^2,{\bf n}_{T2}^2,\phi,\nu)=\theta\left( e^{A_1^{[m]}}- {\bf n}_{T1}^2\right)\theta\left( e^{A_1^{[m]}}- {\bf n}_{T2}^2\right), \label{eq:f-ex-muF-hat}
 \end{equation}
 which when substituted to the eqn.~(\ref{eq:F-gamma_muF-hat}) leads to $G_{gg}^{[m]}\left(\gamma_{N_+},\gamma_{N_-},\nu,{M^2}/{\hat{\mu}_F^2}\right)=1$.  Neither the coefficient functions for quarkonium production listed in the section~\ref{sec:H-HEF}, nor any other coefficient functions for physical processes known to the authors provide a sharp cut-off in the transverse momentum as in eqn.~(\ref{eq:f-ex-muF-hat}). Instead, they always smoothly depend on transverse momenta, as e.g. in the figure~\ref{fig:IF-plot}. Therefore, the HEF is not equivalent to the $\hat{\mu}_F$ prescription. By construction, the $\hat{\mu}_F$ prescription  takes into account the coefficient $A_{1}^{[m]}$ and, hence, it is correct up to NLO in $\alpha_s$. However, already at NNLO in $\alpha_s$, the results of HEF and the $\hat{\mu}_F$ prescription differ. This is clear because the NNLO coefficients obtained from HEF (table~\ref{tab:pert-coefs}) contain the transcendental number $\pi^2/6$, which can not arise from the expansion of the exponent in the first line of eqn.~(\ref{eq:F-gamma_muF-hat}).  
 
\section{NLO+DLA matching in $z$-space}
\label{sec:match}

\begin{figure}[hbt!]
\begin{center}
\includegraphics[width=0.49\textwidth]{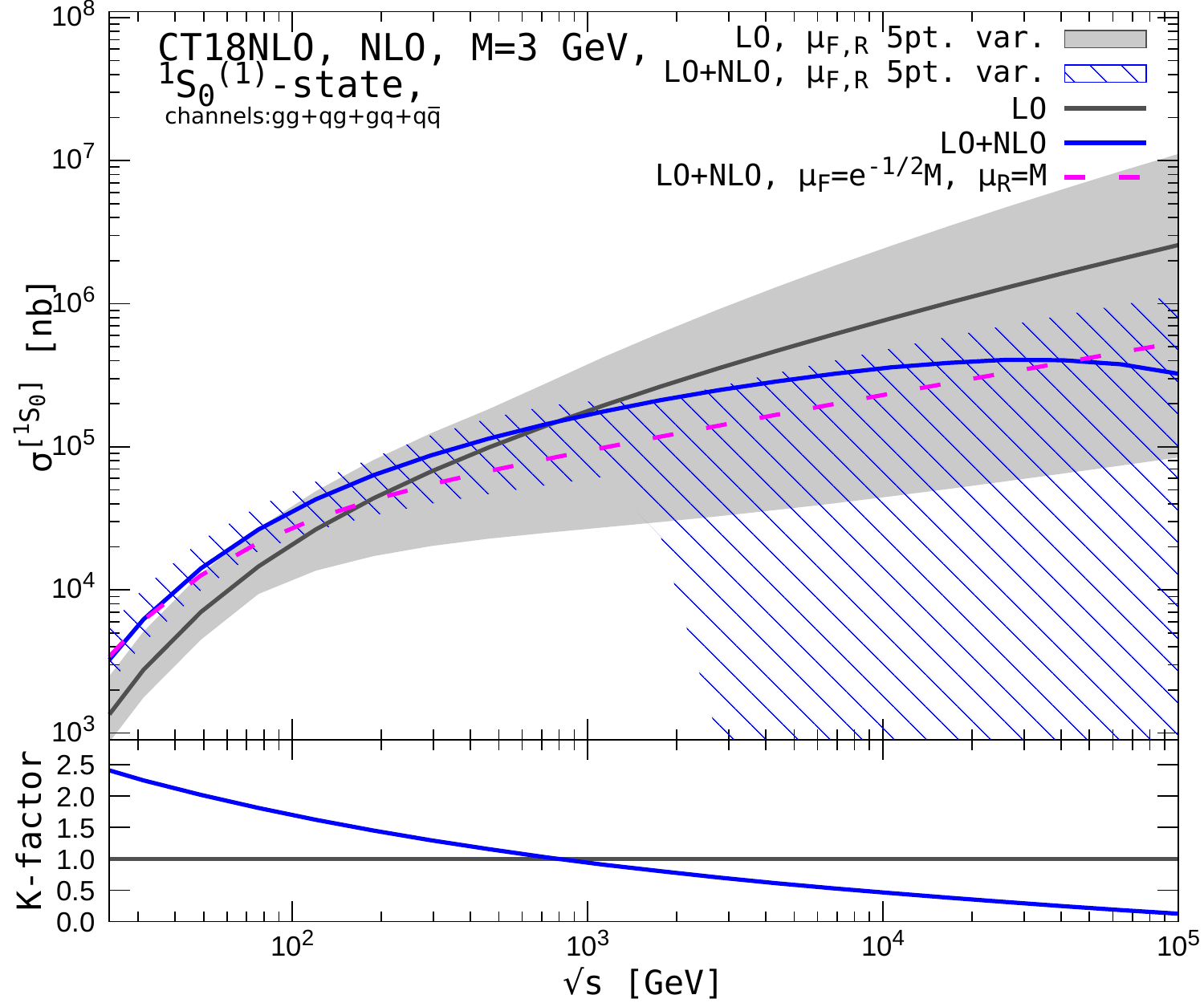}\includegraphics[width=0.49\textwidth]{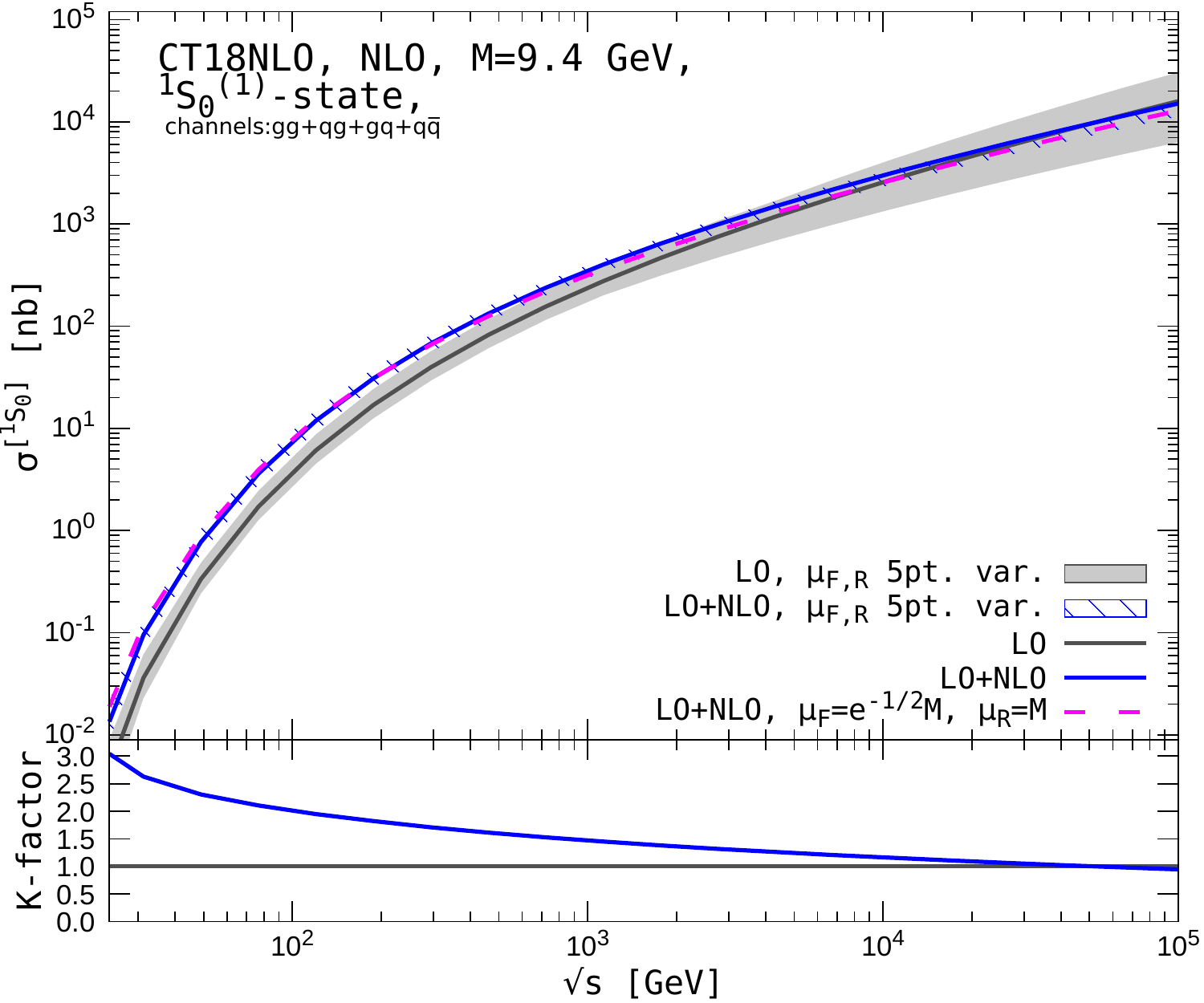}
\end{center}
\caption{Comparison of the energy dependence of the hadroproduction cross sections of a $Q\bar{Q}[{}^1S_0^{(1)} ]$ state with  $M=3$ GeV ({\bf left panel}) and $M=9.4$ GeV ({\bf right panel}) in CF at LO (grey curve) and NLO(blue curve). The central member of the \texttt{CT18NLO} PDF set~\cite{Hou:2019efy} has been used. The dashed line depicts the central prediction of the NLO calculation using the $\hat\mu_F$ prescription (Equation \ref{eq:muF-hat})~\cite{Lansberg:2020ejc}. The LDME $\langle {\cal O} [ {}^1S_0^{(1)} ] \rangle$ was set to 1~GeV$^3$ in both cases for illustration purposes.}\label{fig:LO-NLO1}\label{fig:LO-NLO2}
\end{figure}

Before discussing various approaches to match  the DLA HEF and NLO CF contributions into a single prediction for the total cross section, let us set the baseline by showing how the NLO CF cross section with the ``canonical'' scale choice $\mu_{F}=\mu_{R}=M$ depends on the energy and how large the associated scale uncertainty is. The plots of figure~\ref{fig:LO-NLO1} show the LO and NLO CF predictions for the total cross section of the production of ${}^1S_0^{[1]}$ states with masses of 3 and 9.4 GeV. The scale uncertainties, depicted as solid or shaded bands in all these plots, has been estimated using the five-point scale-variation procedure. The latter procedure consists in evaluating the cross section for each value of the energy with the scale choice $\mu_F=2^{\zeta_1} M$ and $\mu_R=2^{\zeta_2} M$ for $(\zeta_1,\zeta_2)\in \{ (0,\pm 1), (\pm 1,0) \}$ and taking, as an uncertainty estimate, the largest positive or negative deviation of the obtained value form the cross section with the default scale choice. 

  From the left panel of figure~\ref{fig:LO-NLO1}, one can see that, for $c\bar{c}$ states, the instability of the cross section at high energy is dramatic. Above $\sqrt{s}=1$ TeV, such a computation does not show any more predictive power and, above $\sqrt{s}=100$~GeV, the NLO $K$ factor, defined as the ratio of the NLO to the LO cross section, decreases with energy, which is very different from the behaviour of the matched result as we will see. For bottomonia (see the plot in the right panel of figure~\ref{fig:LO-NLO2}), the situation is much better: the scale uncertainty of the NLO calculation is significantly smaller than the LO one all the way up to $\sqrt{s}\sim 100$ TeV. However, the decrease of the NLO $K$ factor is evident even at this scales. The dependence of this behaviour on the PDF choice has been analysed in detail in ref.~\cite{Lansberg:2020ejc}. 
  
   A specific factorisation scale choice (\ref{eq:muF-hat}) has been proposed in ref.~\cite{Lansberg:2020ejc} as a possible resolution of the problem of high-energy instability of quarkonium hadroproduction cross sections. As we have shown in section~\ref{sec:muF-resumm}, this prescription is equivalent to the resummation of some LL($\ln 1/z$)-terms, which is correct at NLO but fails at higher orders. In practice, for $\eta_Q$ hadroproduction, this procedure leads to significantly smaller cross section at high energies, shown by dashed lines in figure~\ref{fig:LO-NLO1}, than that predicted by LO CF at the default scale. Indeed the PDFs are less evolved thus smaller. In this regards, we stress that, for other processes like $H^0$ production, $\hat{\mu}_F$ can be larger than the mass of the produced particle~\cite{Lansberg:2020ejc} and will lead to larger cross sections with further evolved gluon PDFs.
  
   The DL HEF resummation proposed in section~\ref{sec:C-factors} is compatible with the factorisation scheme and the evolution of the usual NLO and NNLO PDFs and correctly reproduces the LL($\ln 1/z$) terms in the CF coefficient function up to NNLO in $\alpha_s$, as it was shown in sections~\ref{sec:C-factors} and~\ref{sec:NNLO-expansion}. However this resummation is applicable only in the region of $z\ll 1$. Power corrections ${\cal O}(z)$ to the CF coefficient function are missing in the HEF calculation with any N${}^{k}$LL accuracy, while they are equally importantas the logarithmic terms  at $z\sim 1$. For the value of the total cross section, the whole range of $z$ from 0 to 1 contributes due to the integration over $z$ in eqn.~(\ref{eq:sigma-L-sighat}). In the following subsections, we will propose two approaches to match these DL HEF and NLO CF results in $z$ space and will compare the corresponding numerical results. 

\subsection{The subtractive-matching prescription}
\label{sec:Subtr-match}

The simplest matching prescription between NLO CF and DLA HEF calculations consists in the subtraction of the $z\to 0$ asymptotics of the NLO CF coefficient function, to avoid double-counting it with the HEF contribution:
\begin{equation}
\sigma^{[m]}_{\rm NLO+HEF}= \sigma^{[m]}_{\text{LO CF}} + \sum\limits_{i,j=q,\bar{q},g} \int\limits_{z_{\min}}^1 \frac{dz}{z}\  \left[ \check{\sigma}^{[m],ij}_{\text{HEF}}(z)  +  \hat{\sigma}^{[m],ij}_{\text{NLO CF}}(z) -  \hat{\sigma}_{\text{NLO CF}}^{[m],ij}(0)   \right] {\cal L}_{ij}(z),\label{eq:subtr-match}
\end{equation}
where $\check{\sigma}^{[m],gg}_{\text{HEF}}$ is defined in eqn.~(\ref{eq:sig-check-def}) while, for other partonic channels, $\check{\sigma}^{[m],ij}_{\text{HEF}}=\hat{\sigma}^{[m],ij}_{\text{HEF}}$ and $\hat{\sigma}^{[m],ij}_{\text{NLO CF}}(z)$ includes only the ${\cal O}(\alpha_s)$ CF terms from the original papers~\cite{Kuhn:1992qw,Petrelli:1997ge}. 

\begin{figure}[hbt!]
\begin{center}
\includegraphics[width=0.49\textwidth]{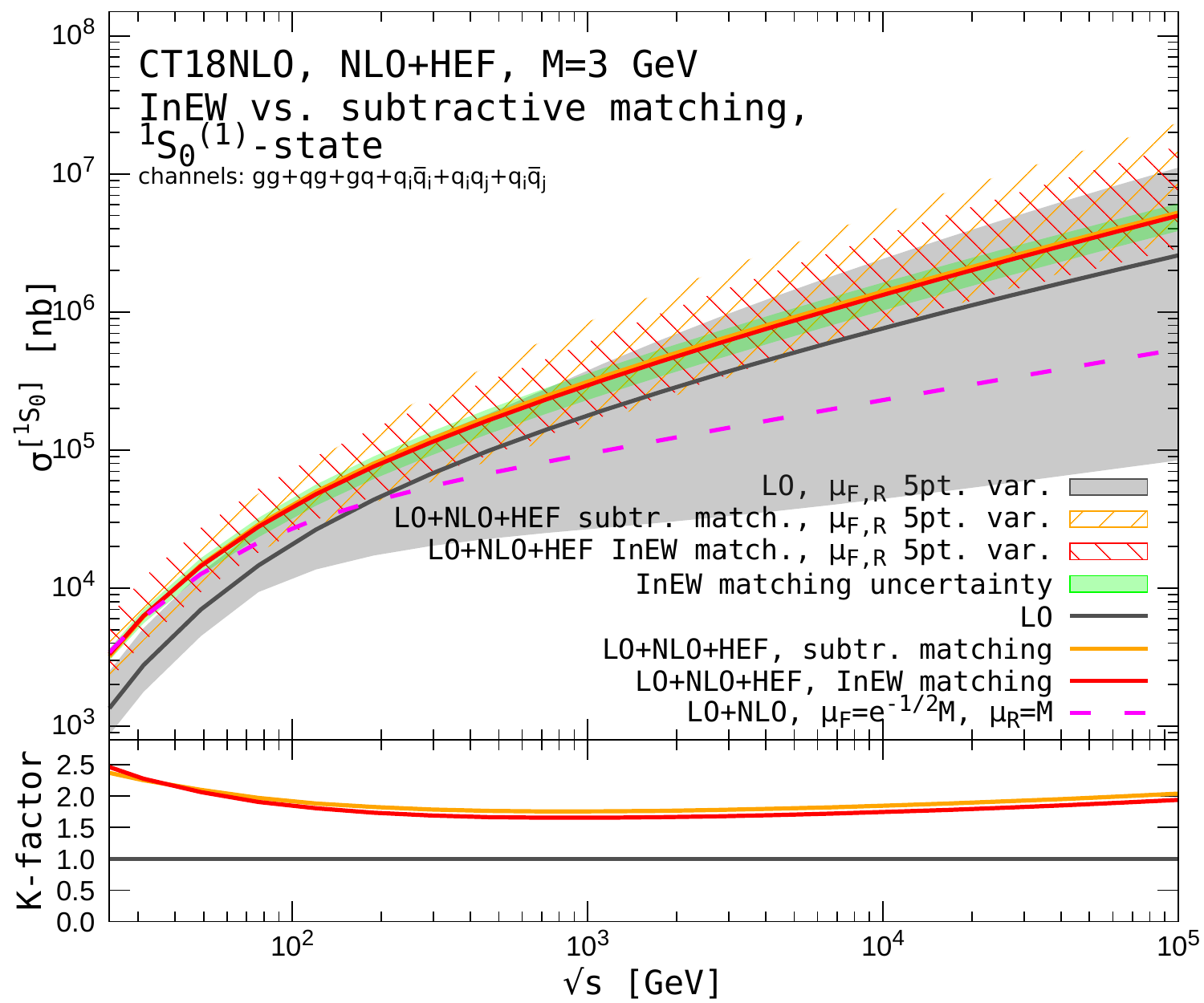}\includegraphics[width=0.49\textwidth]{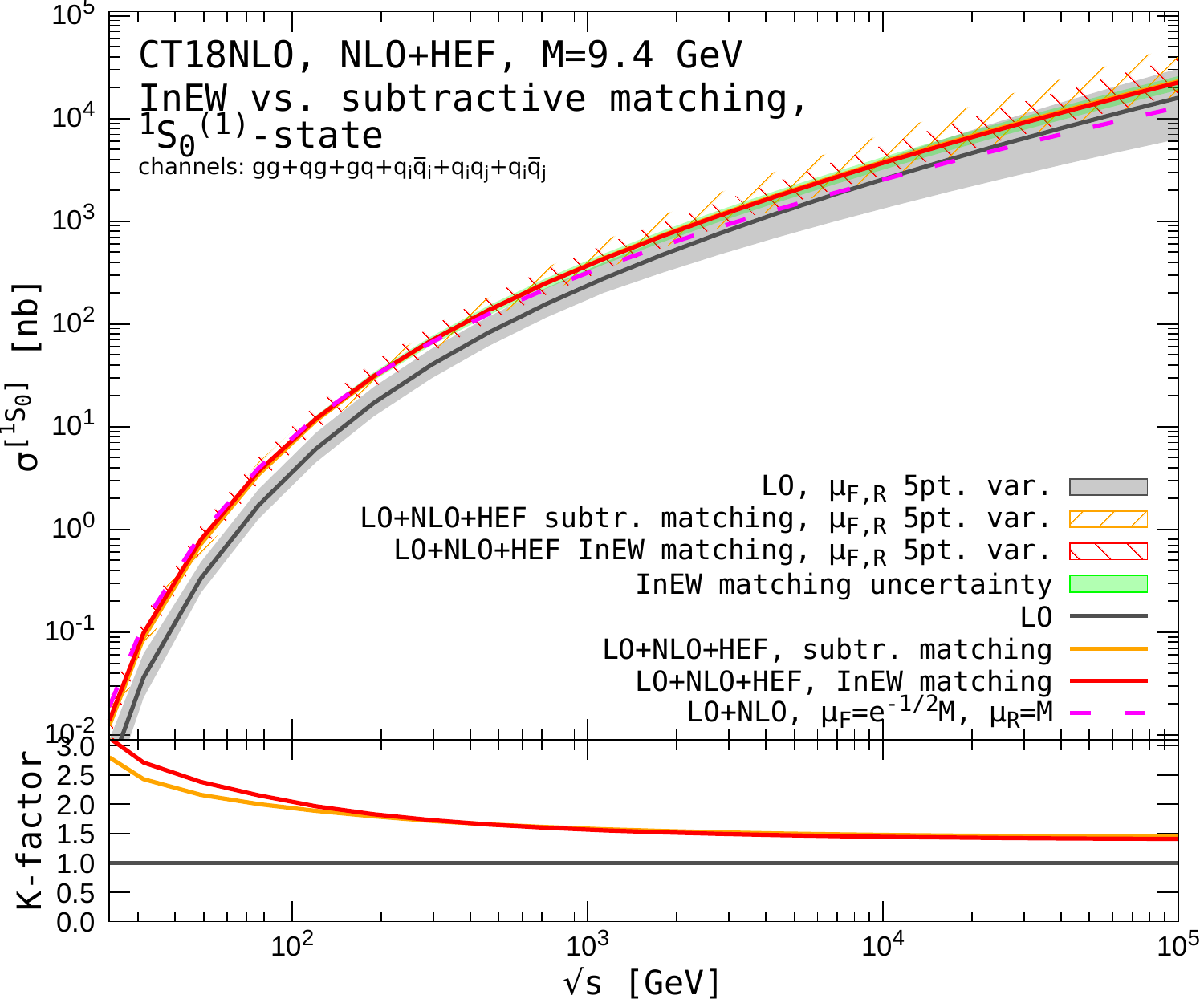}
\end{center}
\caption{Comparison of the energy dependence of the hadroproduction cross sections of a $Q\bar{Q}[{}^1S_0^{[1]} ]$ state with $M=3$ GeV ({\bf left panel}) and $M=9.4$ GeV ({\bf right panel}) for the matched NLO CF + DLA HEF calculations using the subtractive (orange curve and right-shaded band) and InEW matchings(red curve and left-shaded band).  The dashed line depicts the central prediction of the NLO calculation using the $\hat\mu_F$ prescription (Equation \ref{eq:muF-hat})~\cite{Lansberg:2020ejc}. The LDME $\langle {\cal O} [ {}^1S_0^{[1]} ] \rangle$ was set to 1~GeV$^3$.}\label{fig:InEW-match1}\label{fig:InEW-match2}
\end{figure}

  The corresponding numerical results for $Q\bar{Q}[{}^1S_0^{(1)}]$ states with masses of 3 and 9.4~GeV are shown in the figure~\ref{fig:InEW-match1}. One can see that the high-energy instability of the NLO cross section has gone away and the ratio of the matched cross section to the LO one becomes almost constant at high energy, which signal that energy dependence is now mostly driven by the PDFs, not by the hard-scattering coefficient.   
  
\subsection{The inverse-error-weighting matching}
\label{sec:InEW}

 The InEW matching method has been proposed in ref.~\cite{Echevarria:2018qyi} to match the $|{\bf p}_{Tl^+l^-}|$ spectrum of Drell-Yan lepton pairs predicted by the TMD factorisation  for $|{\bf p}_{Tl^+l^-}|\ll M_{l^+l^-}$ with the NLO CF calculation of the same spectrum, which is free from large logarithms of $|{\bf p}_{Tl^+l^-}|/M_{l^+l^-}$ for $|{\bf p}_{Tl^+l^-}|\gtrsim M_{l^+l^-}$.  Following the same ideas, we introduce smooth weight functions $w_{\rm HEF}(z)$ and $w_{\rm CF}(z)$ into our cross section formula as follows:
 \begin{eqnarray}
\sigma^{[m]}_{\rm NLO+HEF}= \sigma^{[m]}_{\text{LO CF}} +&& \label{eq:InEW-match} \\ + \sum\limits_{i,j=q,\bar{q},g} \int\limits_{z_{\min}}^1 dz\ &&\left\{ \left[ \check{\sigma}^{[m],ij}_{\text{HEF}}(z) \frac{{\cal L}_{ij}(z)}{z} \right] w_{\rm HEF}^{ij}(z)   + \left[ \hat{\sigma}^{[m],ij}_{\text{NLO CF}}(z)\frac{{\cal L}_{ij}(z)}{z} \right] w_{\rm CF}^{ij}(z)   \right\}, 
\nonumber \end{eqnarray}
 where we emphasise that the weights are introduced on the level of the $z$ integrand of the cross section, so that the distributions $1/(1-z)_+$ and $\ln(1-z)/(1-z)_+$, contained in the $\hat{\sigma}^{[m]}_{\text{NLO CF},ij}$, have already ``acted'' on the function ${\cal L}_{ij}(z)/z$. The weights should suppress one of the contributions in the region where it is unreliable, so naturally they are chosen to be proportional to the inverse squared errors of each contribution and to add up to one:
 \begin{equation}
 w^{ij}_{\rm HEF}(z) = \frac{[\Delta\sigma_{\rm HEF}^{ij}(z)]^{-2}}{ [\Delta\sigma_{\rm HEF}^{ij}(z)]^{-2} + [\Delta\sigma_{\rm CF}^{ij}(z)]^{-2} },\ w^{ij}_{\rm CF}(z) = 1-w^{ij}_{\rm HEF}(z), \label{eq:InEw-w-def}
 \end{equation}
where, by $\Delta \sigma^{ij}_{\rm HEF/CF}$, we refer to an estimate of the theoretical uncertainty of the corresponding contributions to the $z$ integrand of the total cross section, which should correctly capture the fact that the HEF or the CF contributions respectively become unreliable in the $z\to 1$ or $z\to 0$ limits. The integrand uncertainty due to the matching procedure then follows from:
 \begin{equation}
 \Delta \sigma^{ij}_{\rm Match.}(z) =\left([\Delta\sigma_{\rm HEF}^{ij}(z)]^{-2}+[\Delta\sigma_{\rm CF}^{ij}(z)]^{-2}\right)^{-1/2}, \label{eq:match-uns-InEW}
 \end{equation}
 which, upon integration over $z$, gives the total cross section uncertainty due to matching procedure.
 
 The errors entering eqns.~(\ref{eq:InEw-w-def}) and (\ref{eq:match-uns-InEW}) can be estimated as follows. The NLO CF cross section contains no information about the small-$z$ behaviour of $\hat{\sigma}^{[m],ij}(z)$ beyond ${\cal O}(\alpha_s)$, while the LL HEF calculation provides the $\alpha_s^2 \ln 1/z$ and higher-order LLA terms. Hence one can take the ${\cal O}(\alpha_s^2)$ expansion of the HEF cross section, obtained in Sec.~\ref{sec:NNLO-expansion} as an estimate for the error of the NLO CF integrand. We should also take into account the fact, that the constant ${\cal O}(\alpha_s^2)$ term of the $z\to 0$ asymptotics of the CF coefficient function can not be obtained from the LL HEF calculation, because it belongs to the NLL approximation of the HEF. We estimate the coefficient in front of this term as equal to $1$ and treat it as an independent source of uncertainty. Adding the constant term to the $\alpha_s^2 \log 1/z$ contribution in quadrature we obtain the following uncertainty estimate for the NLO CF contribution:
  \begin{equation}
\Delta\sigma_{\rm CF}^{ij}(z)= \alpha_s^2 \sigma_0^{[m]}   \frac{{\cal L}_{ij}(z)}{z} \sqrt{\left[ \frac{C_{\rm LL}^{ij}}{\pi}\ln\frac{1}{z} \right]^2 + 1 },\label{eq:D-CF-an}
 \end{equation}
 where
 \begin{eqnarray}
 C^{gg}_{LL}&=& C_A^2\left[ 2A_2 + B_2 + 4A_1\ln\frac{M^2}{\mu_F^2} + 2A_0\ln^2\frac{M^2}{\mu_F^2} \right], \\
 C^{qg}_{LL}&=& C_AC_F\left[ A_2 + B_2 + 3A_1\ln\frac{M^2}{\mu_F^2} + \frac{3}{2} A_0\ln^2\frac{M^2}{\mu_F^2} \right], \\
 C^{q\bar{q}}_{LL}&=& C_F^2\left[ B_2 + 2A_1\ln\frac{M^2}{\mu_F^2} + A_0\ln^2\frac{M^2}{\mu_F^2} \right],
 \end{eqnarray}
 and the coefficients $A_0$, $A_{1,2}$, $B_2$ can be found in table~\ref{tab:pert-coefs}.
 
 Conversely, the HEF result is valid only at $z\ll 1$ and contains no information about the ${\cal O}(z)$ power corrections to the CF coefficient function. On the other hand, the NLO CF coefficient function is an exact function of $z$ at NLO in $\alpha_s$. Hence one can take the NLO CF hard-scattering coefficient with its $z\to 0$ asymptotics subtracted, as an estimate of the missing ${\cal O}(z)$ power corrections in HEF, i.e. of the error of the latter. We also take into account the unknown $z\to 1$ asymptotics of the NNLO CF coefficient function as ${\cal O}(\alpha_s^2)$ term and add this estimate to the total error of the HEF contribution in quadrature as follows:
 \begin{equation}
  \Delta\sigma_{\rm HEF}^{ij}(z)= \sqrt{ \left[  \left( \hat{\sigma}^{[m],ij}_{\text{NLO CF}}(z) -  \hat{\sigma}_{\text{NLO CF}}^{[m],ij}(0) \right) \frac{{\cal L}_{ij}(z)}{z} \right]^2   +  \left[ \alpha_s^2 \sigma^{[m]}_0  \frac{{\cal L}_{ij}(z)}{z} \right]^2 } . \label{eq:D-HEF-an}
 \end{equation}

\begin{figure}
\begin{center}
\includegraphics[width=0.5\textwidth]{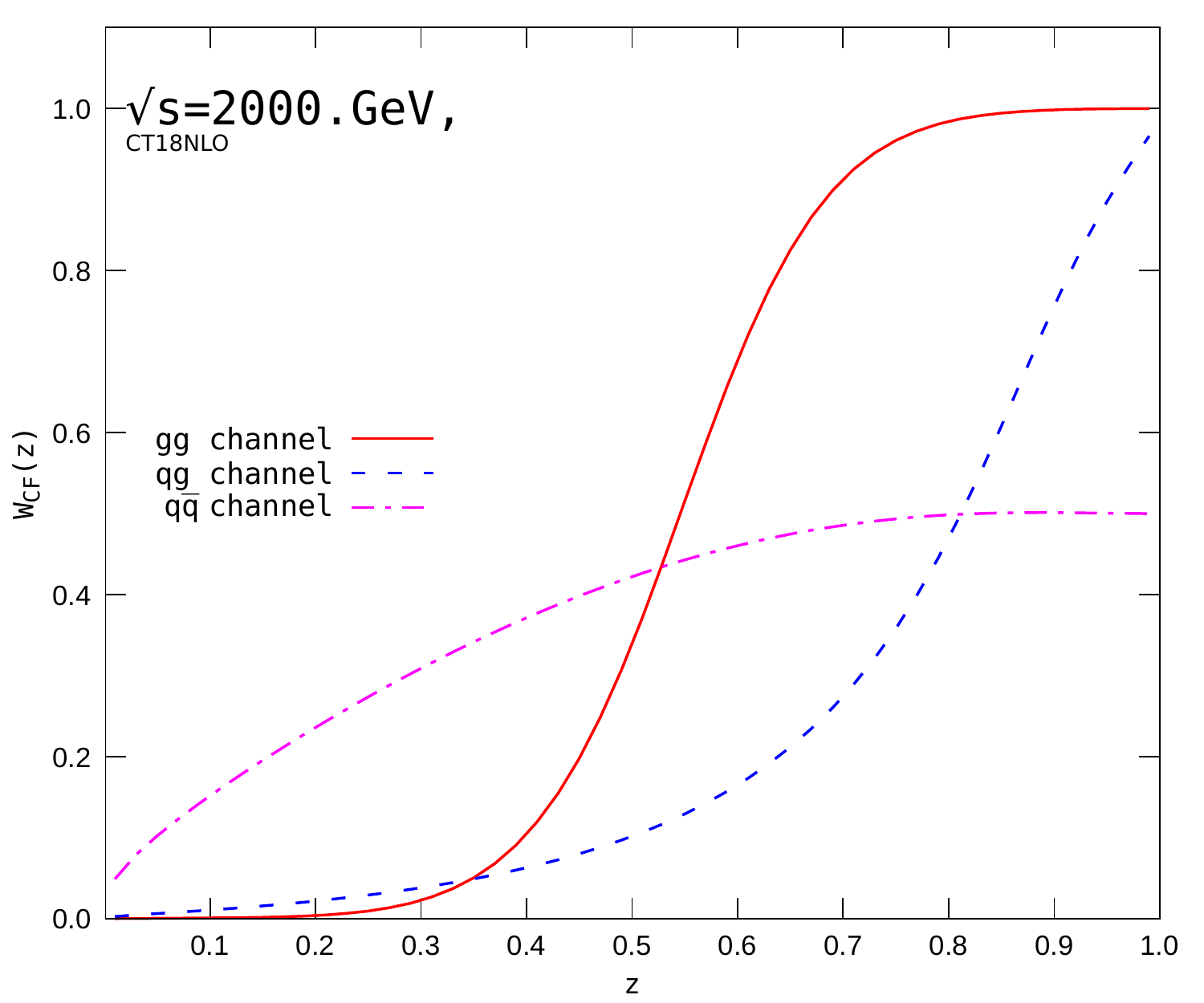}
\end{center}
\caption{Plots of the InEW weights of the CF contribution for different channels as function of $z$ with $\mu_F=\mu_R=M$.}\label{fig:kappa-var}
\end{figure}

Using formulas (\ref{eq:D-CF-an}) and (\ref{eq:D-HEF-an}), one can construct weights $w_{\rm CF}$ and $w_{\rm HEF}$ with expected behaviour: e.g. $w^{(gg)}_{\rm CF}(z)$ increases between 0 and 1 when $z\to 1$, figure~\ref{fig:kappa-var}. The latter property is ensured by the  behaviour of errors (\ref{eq:D-CF-an}) and (\ref{eq:D-HEF-an}). The CF error (\ref{eq:D-CF-an})  increases towards $z\to 0$, while HEF error (\ref{eq:D-HEF-an}) increases towards $z=1$. Among three weights shown in the figure~\ref{fig:kappa-var}, only the weight for $q\bar{q}$ channel does not reach $1$ when $z\to 1$. This behaviour is expectable, because the NLO coefficient function, and hence the error (\ref{eq:D-HEF-an}) has no $\ln(1-z)$ enhancement at $z\to 1$ for this channel, compared to other channels. But this feature of the weight function of the $q\bar{q}$ channel is not important numerically, because the $q\bar{q}$ contribution amounts to ${\cal O}(1\%)$ of the total cross section at high energy.

\begin{figure}
\begin{center}
\includegraphics[width=0.49\textwidth]{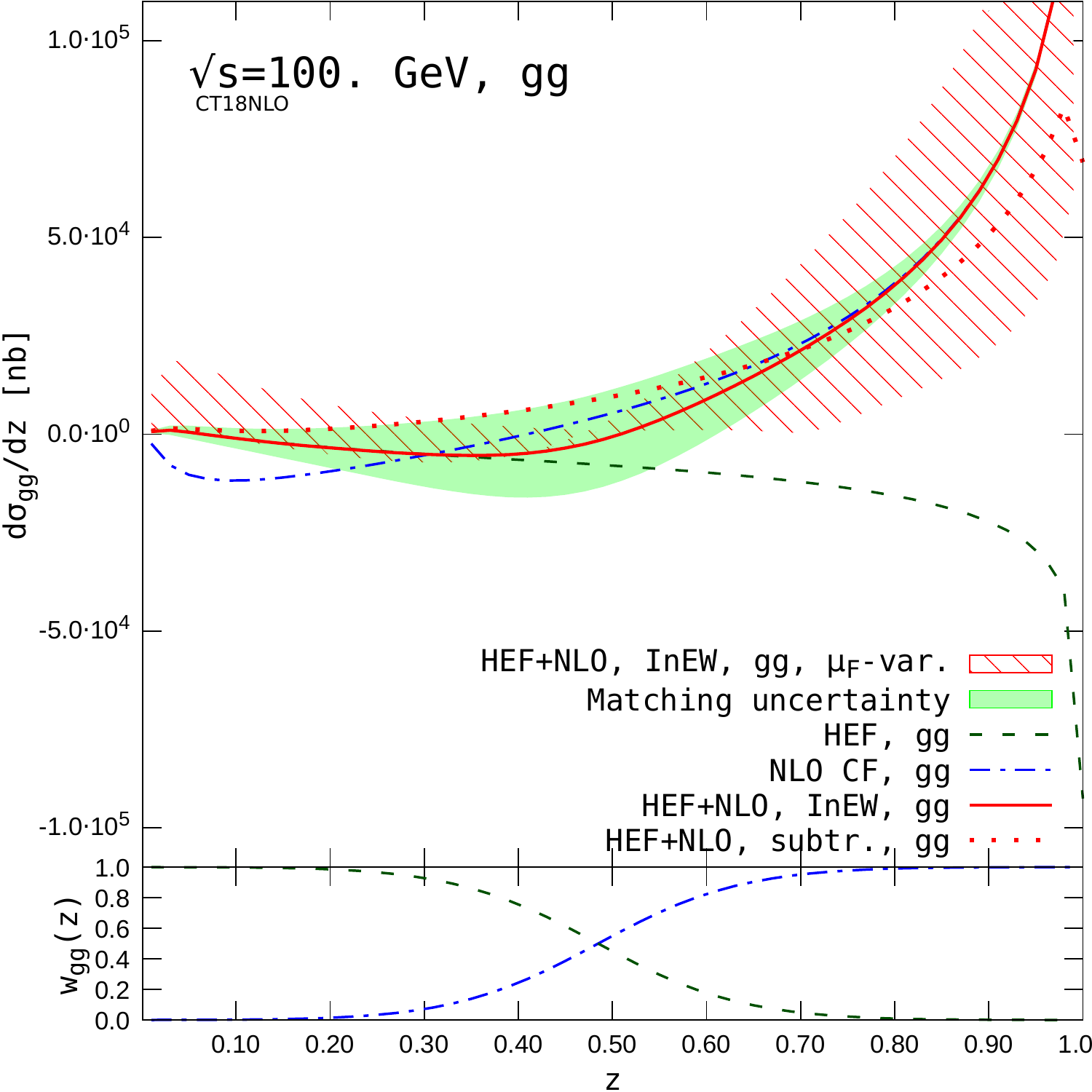}
\includegraphics[width=0.49\textwidth]{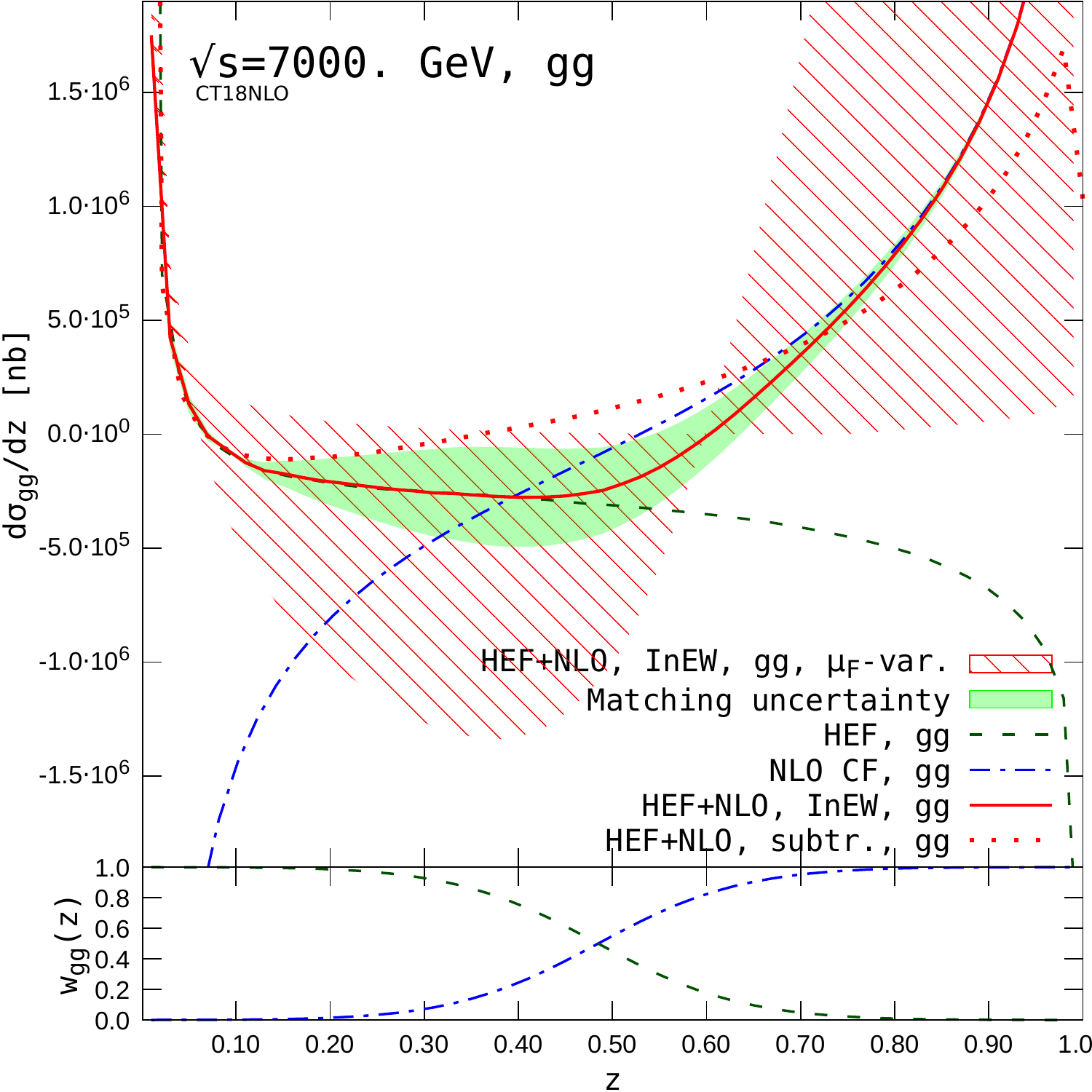}
\end{center}
\caption{Matching plots for the $gg$-channel contribution to the ${}^1S_0^{[1]}$-state hadroproduction with $M=3$ GeV. The solid curve depicts the $z$-integrand of eqn.~(\ref{eq:InEW-match}), the dashed curve the HEF contribution (without the InEW weight),  the dash-dotted curve the NLO CF contribution (without the InEW weight), and the dotted red line the integrand of eqn.~(\ref{eq:subtr-match}), i.e. the result of the subtractive matching prescription for comparison. The plots of the InEW weights are shown in the bottom inset, while the matching uncertainty (\ref{eq:match-uns-InEW}) is shown as the solid band. The LDME $\langle {\cal O} [ {}^1S_0^{[1]} ] \rangle$ was set to 1~GeV$^3$. }\label{fig:match-z-gg}
\end{figure}

\begin{figure}
\begin{center}
\includegraphics[width=0.49\textwidth]{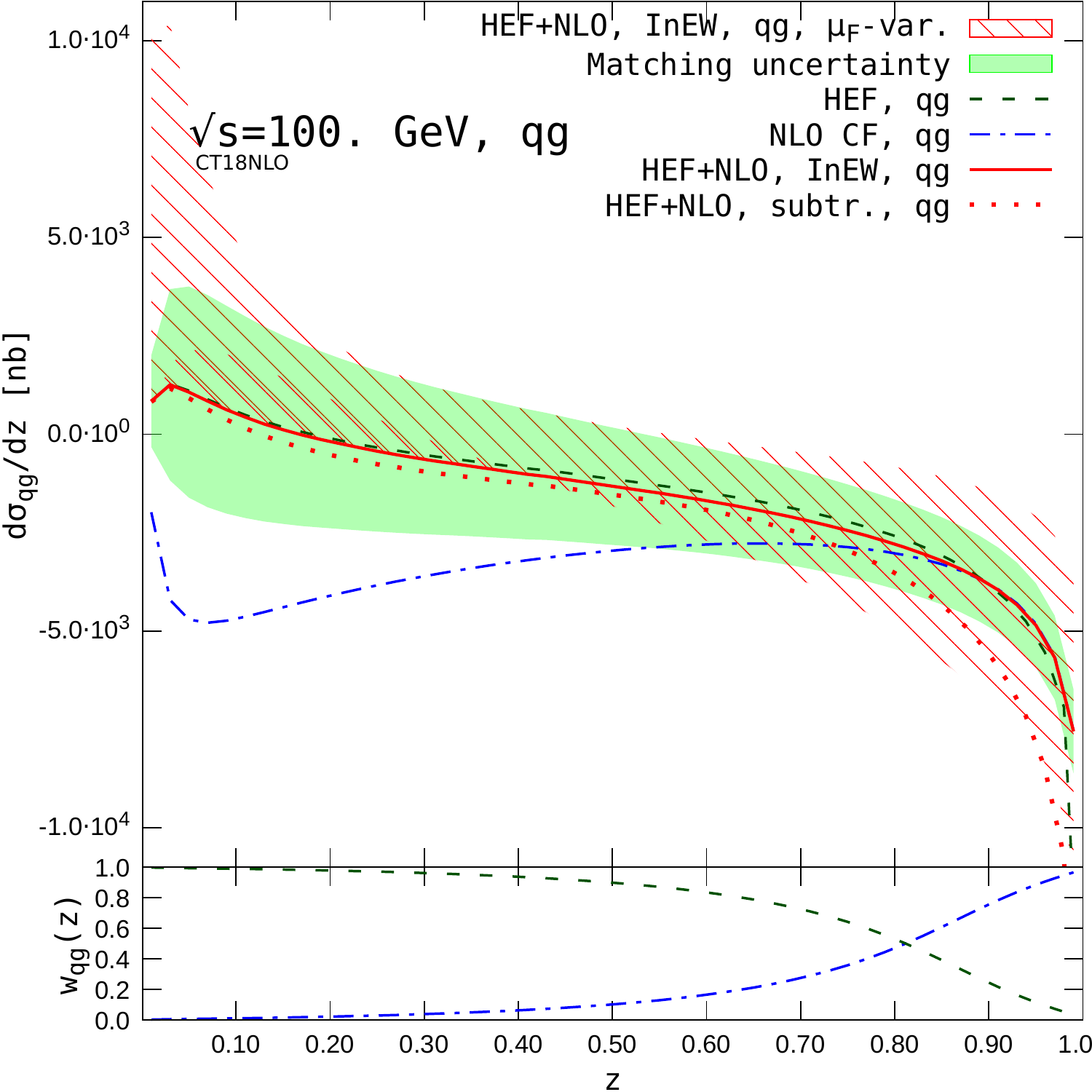}
\includegraphics[width=0.49\textwidth]{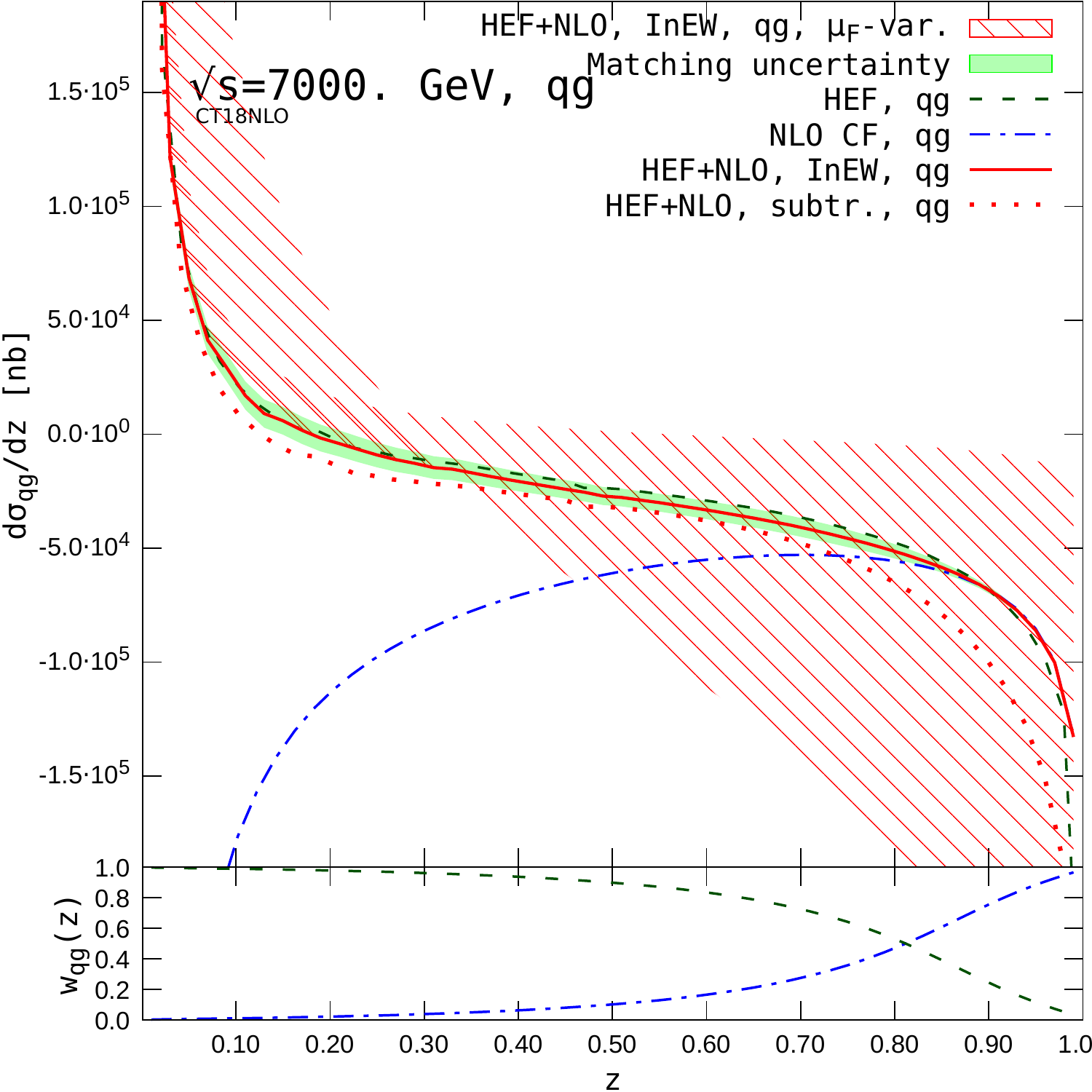}
\end{center}
\caption{Matching plots for $qg$-channel to the ${}^1S_0^{[1]}$-state hadroproduction with $M=3$ GeV. The integrand of eqn.~(\ref{eq:subtr-match}), i.e. subtractive matching prescription is shown for comparison by the dotted red line. The plots of InEW weights are shown in the bottom inset, while the matching uncertainty (\ref{eq:match-uns-InEW}) is shown as the solid band. The LDME $\langle {\cal O} [ {}^1S_0^{[1]} ] \rangle$ was set to 1~GeV$^3$. }\label{fig:match-z-qg}
\end{figure}

\begin{figure}
\begin{center}
\includegraphics[width=0.49\textwidth]{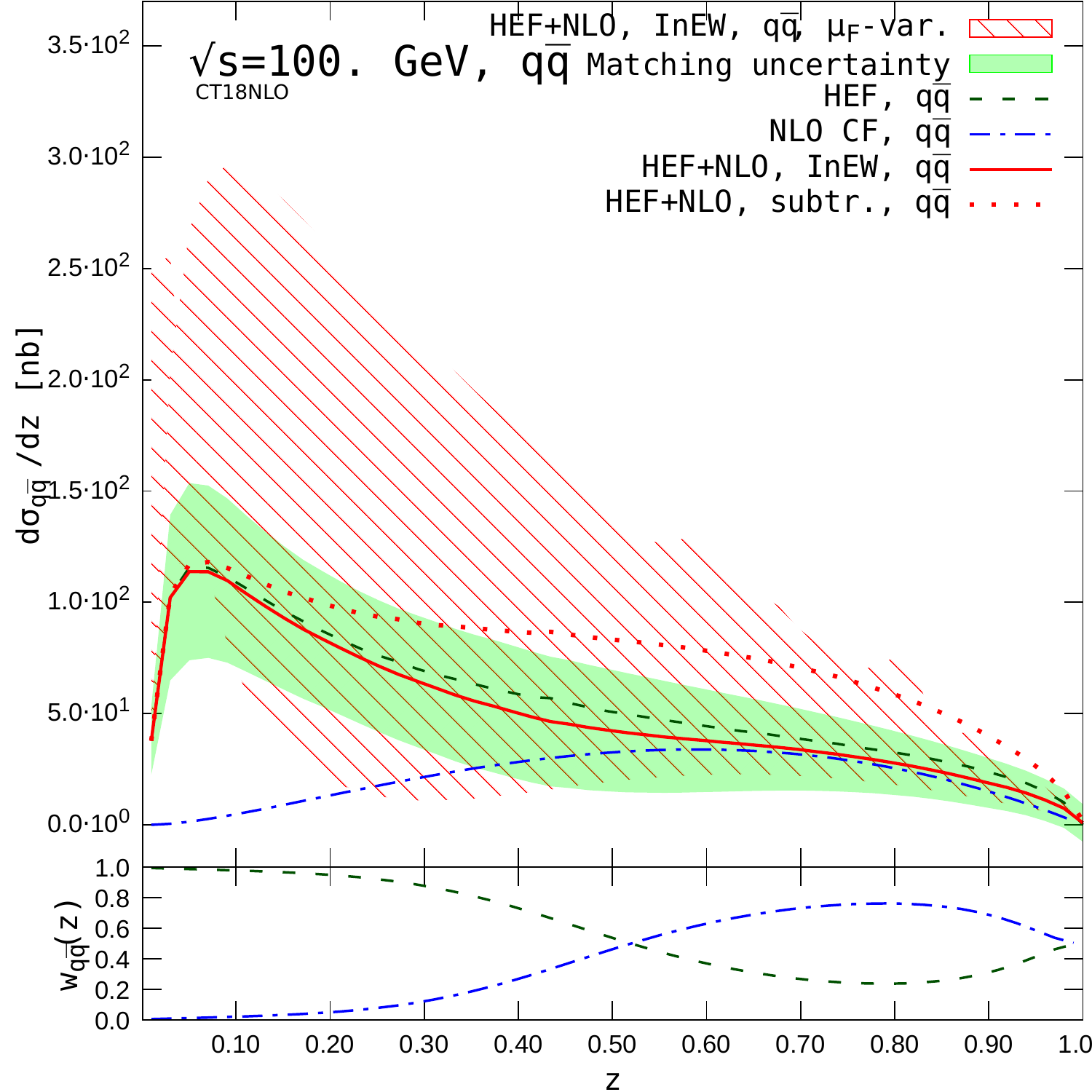}
\includegraphics[width=0.49\textwidth]{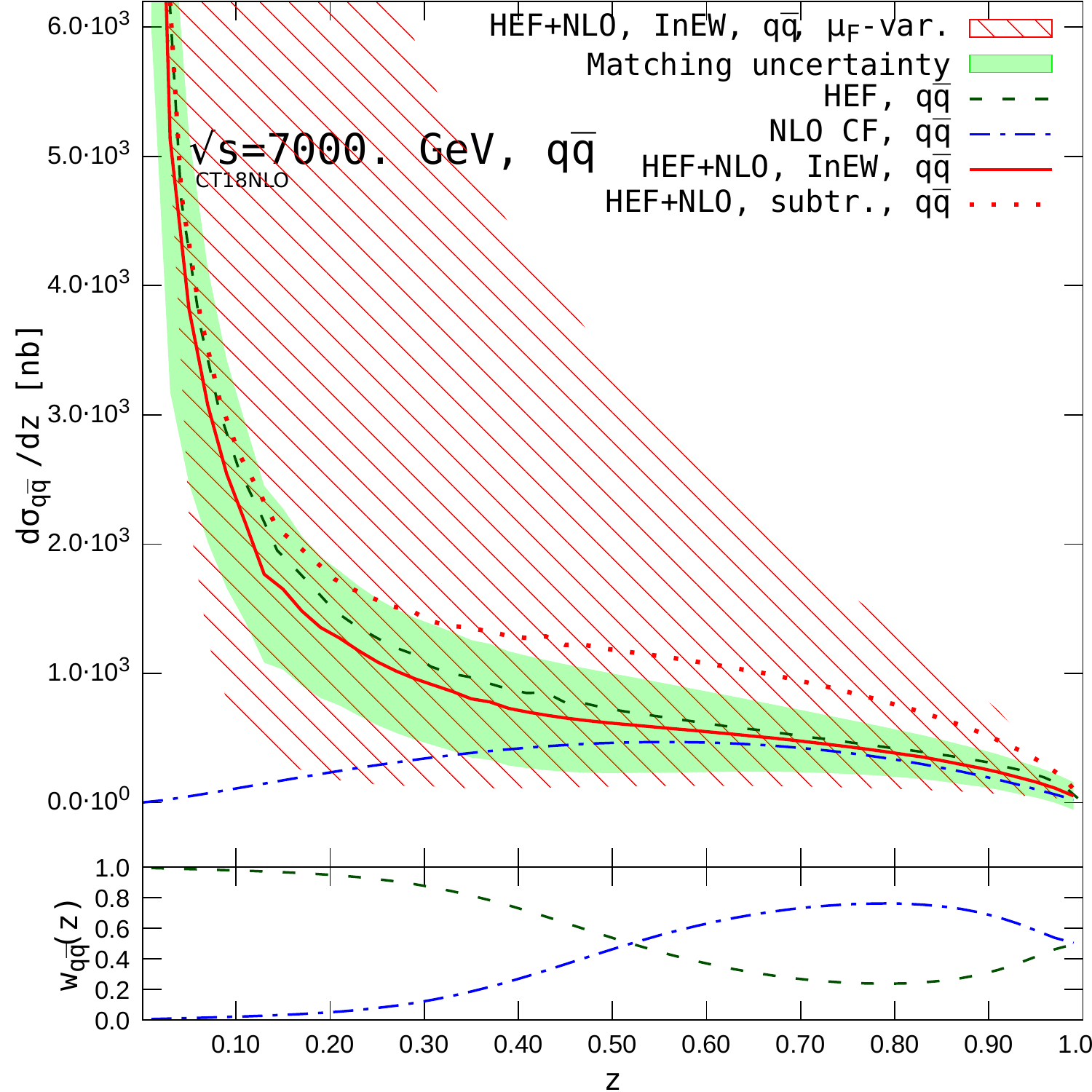}
\end{center}
\caption{Matching plots for $q\bar{q}$-channel to the ${}^1S_0^{[1]}$-state hadroproduction with $M=3$ GeV. The integrand of eqn.~(\ref{eq:subtr-match}), i.e. subtractive matching prescription is shown for comparison by the dotted red line. The plots of InEW weights are shown in the bottom inset, while the matching uncertainty (\ref{eq:match-uns-InEW}) is shown as the solid band. The LDME $\langle {\cal O} [ {}^1S_0^{[1]} ] \rangle$ was set to 1~GeV$^3$. }\label{fig:match-z-qq}
\end{figure}

 In figures~\ref{fig:match-z-gg}, \ref{fig:match-z-qg} and \ref{fig:match-z-qq} several plots of the $z$ integrand of  eqn.~(\ref{eq:InEW-match}) are shown as red solid lines, using weights obtained above. One can see that the InEW procedure provides a smooth interpolation between the HEF curves at $z\ll 1$ and the CF curves at $z$ closer to 1. 
 
 The corresponding curves for the subtractive matching are shown for comparison by the red dotted lines. From figures~\ref{fig:match-z-gg}, \ref{fig:match-z-qg} and \ref{fig:match-z-qq}, the problem of subtractive matching is evident: it subtracts the $z\to 0$ asymptotics of the NLO CF coefficient function at all values of $z$, thus introducing the unphysical modification of the $z\to 1$ behaviour. The InEW matching, on the other hand, smoothly connects the asymptotic regions.    

The results for the matched cross section in the InEW procedure are shown by red solid curves on figure~\ref{fig:InEW-match1}. The matching uncertainty, estimated by eqn.~(\ref{eq:match-uns-InEW}) is shown separately as the green solid band in the  figure~\ref{fig:InEW-match1} and it is significantly smaller than the residual scale uncertainty, shown as the shaded band. The scale uncertainty band is slightly smaller in the InEW matching case. 

\begin{figure}[hbt!]
    \centering
    \includegraphics[width=0.49\textwidth]{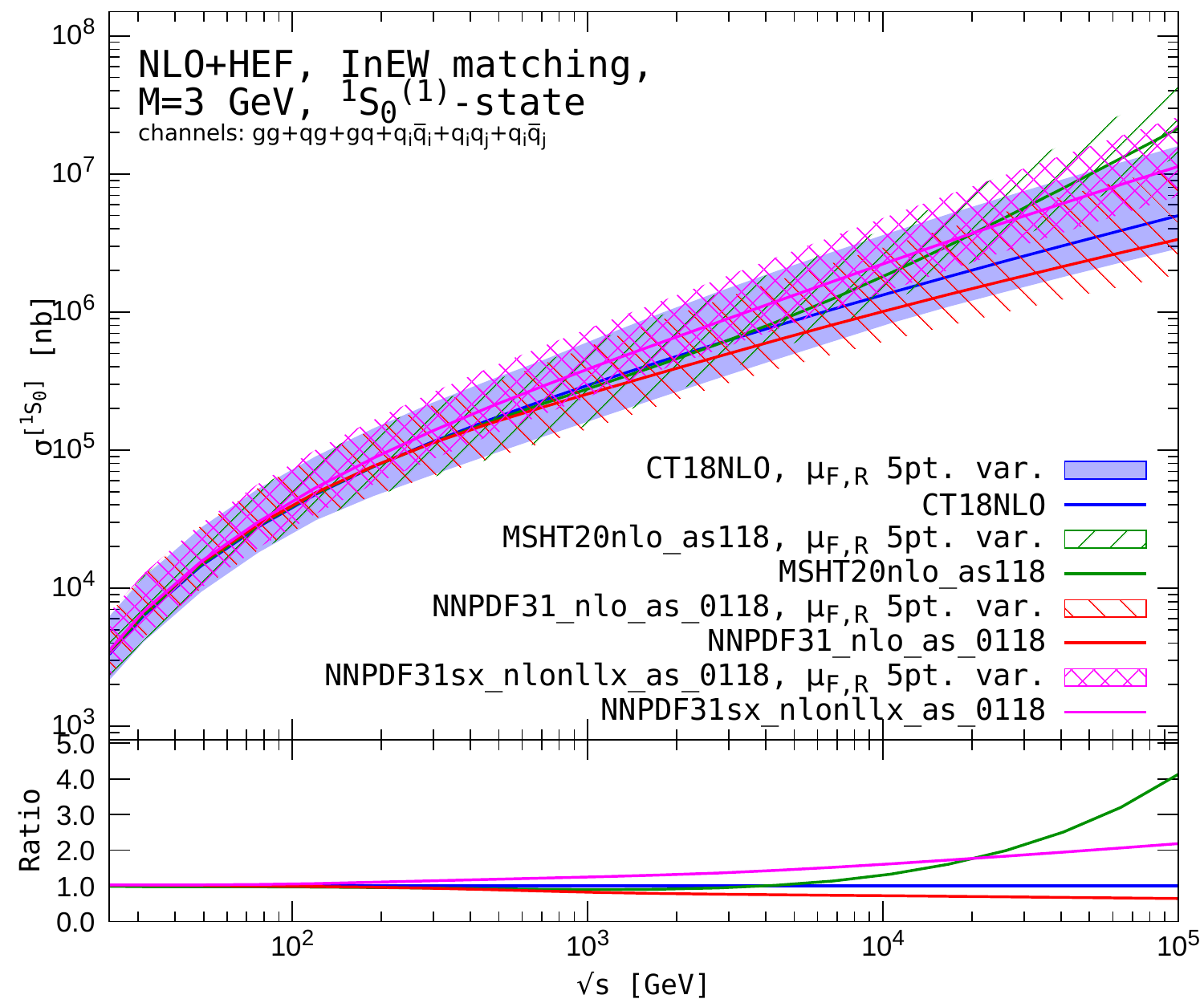}
    \includegraphics[width=0.49\textwidth]{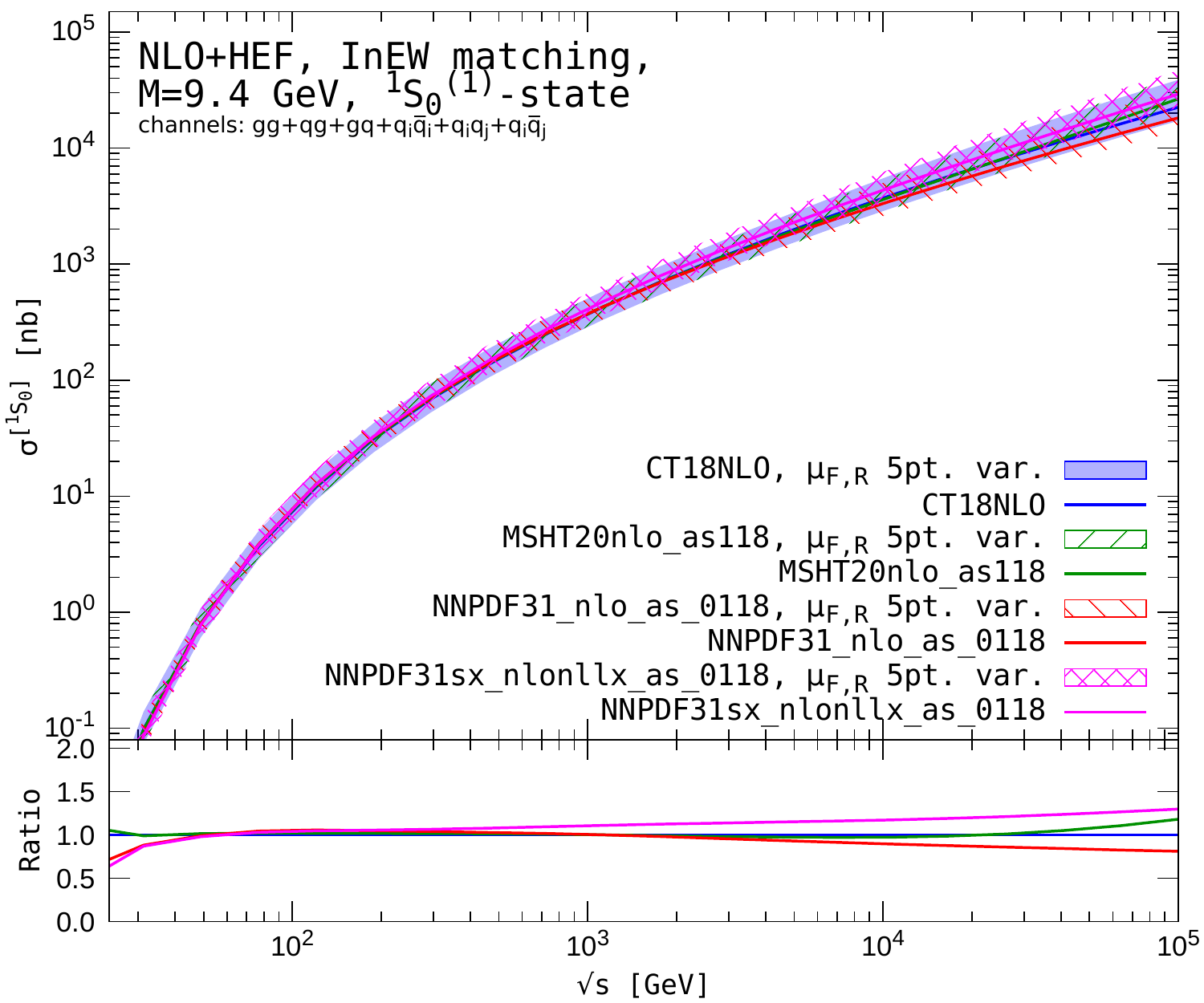}
    \caption{Dependence of the matched cross section on the choice of collinear PDF. The LDME $\langle {\cal O} [ {}^1S_0^{(1)} ] \rangle$ was set to 1~GeV$^3$.}
    \label{fig:PDF-comp}
\end{figure}

Finally, figure~\ref{fig:PDF-comp} illustrates how the matched cross section varies depending on the choice of the collinear PDF. Since our computations are rather computer intensive, only the central members of the \texttt{CT18NLO}~\cite{Hou:2019efy}, \texttt{MSHT20nlo\_as118}~\cite{Bailey:2020ooq}, \texttt{NNPDF31\_nlo\_as\_0118}~\cite{NNPDF:2017mvq} and \texttt{NNPDF31sx\_nlonllx\_as\_0118}~\cite{Ball:2017otu} PDF sets, as implemented in the LHAPDF library~\cite{Buckley:2014ana} has been used for the comparison. The first three PDFs evolve according to the usual NLO DGLAP splitting functions and thus fit seamlessly into our NLO+DLA matching scheme. The last PDF set~\cite{Ball:2017otu} contains the NLL($\ln(1/z)$) resummation and is included in  figure~\ref{fig:PDF-comp} to show that the corresponding cross section, obtained in the NLO+DLA $\ln(1/z)$-resummation scheme, is consistent with the other results within uncertainties.

On one hand, figure~\ref{fig:PDF-comp} shows that there is a significant spread in the charmonium production cross sections at high energy, due to the PDF uncertainties at small values of $x$. At $\sqrt{s}=13$ TeV, the central values of NLO+LL($\ln(1/z)$) predictions obtained with different PDFs vary as much as by factor 1.5 up and down from our default \texttt{CT18NLO}-based estimations. This demonstrates a potential usefulness of this observable to constrain PDFs at small $x$. On the other hand, the theoretical uncertainties of our predictions, dominated by the scale-variation uncertainties shown in  figure~\ref{fig:PDF-comp}, are still larger than the spread between central curves, so to reliably discriminate between them, the scale uncertainties have to be significantly reduced which we expected to be realised once a NLL computation is available.    

\section{Conclusions and outlook}
\label{sec:concl}

In the present paper, we have performed an exploratory study of the effects of High-Energy resummation on heavy-quarkonium hadroproduction cross sections. We have introduced the DLA for the resummation contributions, which is consistent with the fixed-order DGLAP evolution of PDFs and have shown that matching the NLO CF and the DL resummed HEF results solves the problem of the negative NLO CF cross sections at high energy. This opens up the possibility to use this observable to constrain PDFs at small-$x$, however the scale uncertainties of our predictions have to be significantly reduced. To this end, advancing such calculation to a complete NLO+NLL($\ln(1/z)$) accuracy is required, which is a work in progress. We also emphasize, that performing the matching between NLO CF and HEF contributions in $z$-space (as opposed to matching in Mellin-conjugate $N$-space, as in Ref.~\cite{Catani:1992rn}) is a simple and flexible procedure, which allows one to estimate the uncertainty of the cross section due to the matching. When the NLL($\ln(1/z)$) calculation in HEF will become available, it will be easy to match it with the NLO or possibly even NNLO CF results using the $z$-space technique.

\acknowledgments

We thank Renaud Boussarie, Francesco Hautmann and Samuel Wallon for useful discussions as well as Kate Lynch and Yelyzaveta Yedelkina for comments on the manuscript. 
This project has received funding from the European Union’s Horizon 2020 research and innovation programme under grant agreement STRONG–2020 No 824093 in order to contribute to the EU Virtual Access {\sc NLOAccess} (VA1-WG10) \& to the Joint Research activity ``Fixed Target Experiments at the LHC'' (JRA2).
This project has also received funding from the Agence Nationale de la Recherche (ANR) via the grant ANR-20-CE31-0015 (``PrecisOnium'') and via the IDEX Paris-Saclay ``Investissements d’Avenir'' (ANR-11-IDEX-0003-01) through the GLUODYNAMICS project funded by the ``P2IO LabEx (ANR-10-LABX-0038)''.
This work  was also partly supported by the French CNRS via the GDR QCD, via the IN2P3 project GLUE@NLO, via the Franco-Polish EIA (Gluegraph).
M.A.O.’s work was partly supported by the ERC grant 637019 ``MathAm''.

\appendix
\section{Numerics in $z$-space}
\label{sec:num}  

\subsection{Method 1: regularised resummation factor}
\label{sec:method-reg-C}

  This method is based on the following regularised form of the resummation factor, where the small-${\bf q}_T$ oscillations are replaced by a constant in ${\bf q}_T$-behaviour for ${\bf q}_T^2<{\bf q}_{T0}^2$ in such a way that the normalisation condition (\ref{eq:Blu-norm}) still holds exactly:
  \begin{equation}
  \widetilde{\cal C}^{\rm \ DLA}_{gg}(z,{\bf q}_T^2,\mu_F^2)=\left\{ \begin{matrix}
  \frac{1}{{\bf q}_{T0}^2}\left[ \delta(1-z) - {\cal F}\left( \frac{{\bf q}^2_{T0}}{\mu_F^2},z \right) \right], & \text{ if }{\bf q}_{T}^2<{\bf q}_{T0}^2,\\
  {\cal C}_{gg}^{\rm DLA}(z,{\bf q}_T^2,\mu_F^2), & \text{ if }{\bf q}_T^2>{\bf q}_{T0}^2,
  \end{matrix} \right. \label{eq:C-reg-cut}
  \end{equation}
  where 
  \begin{equation}
 {\cal F}\left(\frac{{\bf q}_{T0}^2}{\mu_F^2},z\right)=\int\limits_0^{L_{\max}} dL\ J_0\left(2\sqrt{L \ln\frac{1}{z}} \right) = L_{\max}\ {}_0\tilde{F}_1\left(2,-L_{\max} \ln\frac{1}{z}\right), \label{eq:F-def}
 \end{equation}
with $L_{\max}=\hat{\alpha}_s \ln({{\bf q}_{T0}^2}/{\mu_F^2})$ and ${}_0\tilde{F}_1$, which is the  regularised confluent hypergeometric function. In the ${\bf q}_{T0}^2\to 0$ limit, the function ${\cal F}$ turns into $\delta(1-z)$ but this limit is approached very slowly. So, for all reasonable values of $|{\bf q}_{T0}|$, even as small as $10^{-3}$ GeV, the function (\ref{eq:F-def}) is still rather smoothly dependent on $z$ and does not cause any numerical problems. For the quark-induced channel, the regularised resummation factor:
  \begin{equation}
  \widetilde{\cal C}^{\rm \ DLA}_{gq}(z,{\bf q}_T^2,\mu_F^2)= \frac{C_F}{C_A}\left\{ \begin{matrix}
  -\frac{1}{{\bf q}_{T0}^2} {\cal F}\left( \frac{{\bf q}^2_{T0}}{\mu_F^2},z \right), & \text{ if }{\bf q}_{T}^2<{\bf q}_{T0}^2,\\
  {\cal C}^{\rm DLA}_{gg}(z,{\bf q}_T^2,\mu_F^2), & \text{ if }{\bf q}_T^2>{\bf q}_{T0}^2,
  \end{matrix} \right. \label{eq:Cgq-reg-cut}
  \end{equation}
  satisfies the normalisation condition (\ref{eq:Cgq-norm}).
  
   After substituting the regularised resummation factors (\ref{eq:C-reg-cut}) or (\ref{eq:Cgq-reg-cut}) into  eqn.~(\ref{eq:sig-part-rint-z}), one splits the ${\bf q}_{T1,2}^2$-integrations into regions with ${\bf q}_{T1,2}^2<{\bf q}_{T0}^2$ and ${\bf q}_{T1,2}^2>{\bf q}_{T0}^2$. If, for example ${\bf q}_{T1}^2<{\bf q}_{T0}^2$, then this transverse momentum can be neglected in all the rest of the integrand in eqn.~(\ref{eq:sig-part-rint-z}) because the dependence of the HEF coefficient function as well as the dependence of the factor $1/M_T^4$ on ${\bf q}_{T1}^2$ is smooth. This approximation allows one to trivially integrate out ${\bf q}_{T1}^2$ in the region ${\bf q}_{T1}^2<{\bf q}_{T0}^2$ introducing the error which scales as ${\cal O}({\bf q}_{T0}^2/M^2)$ and can be made negligible by choosing a reasonably small value of ${\bf q}_{T0}^2$. In the region ${\bf q}_{T1}^2>{\bf q}_{T0}^2$ no approximations are made. The ${\bf q}_{T2}^2$-integration can be decomposed in the same manner and the region where both ${\bf q}_{T1}^2<{\bf q}_{T0}^2$ and ${\bf q}_{T2}^2<{\bf q}_{T0}^2$ produces the LO CF term in  eqn.~(\ref{eq:sig-check-def}), while the LO-subtracted partonic cross section in the method under consideration is decomposed as follows:
   \begin{equation}
   \check{\sigma}_{gg}^{[m],\ {\rm HEF}} = \check{\sigma}_{\rm 2UI}^{[m]} + \check{\sigma}_{\rm 2F}^{[m]} + 2\left[ \check{\sigma}_{\rm 1UI-C}^{[m]} - \check{\sigma}_{\rm 1UI-CF}^{[m]} - \check{\sigma}_{\rm F}^{[m]} \right], \label{eq:meth1:sig-gg}
   \end{equation}
   where the separate contributions depend on the value of the cut ${\bf q}_{T0}$ but this dependence should cancel in their sum for sufficiently small ${\bf q}_{T0}^2\ll M^2$. For the quark-induced channels, the resummed partonic coefficient can be calculated in terms of the same contributions as:
    \begin{eqnarray}
   \hat{\sigma}_{gq}^{[m],\ {\rm HEF}} &=& \frac{C_F}{C_A}\left[ \check{\sigma}_{\rm 2UI}^{[m]} + \check{\sigma}_{\rm 2F}^{[m]} + \check{\sigma}_{\rm 1UI-C}^{[m]} - 2\check{\sigma}_{\rm 1UI-CF}^{[m]} - \check{\sigma}_{\rm F}^{[m]}  \right], \label{eq:meth1:sig-gq} \\
   \hat{\sigma}_{q\bar{q}}^{[m],\ {\rm HEF}} &=& \left(\frac{C_F}{C_A}\right)^2 \left[ \check{\sigma}_{\rm 2UI}^{[m]} + \check{\sigma}_{\rm 2F}^{[m]} - 2\check{\sigma}_{\rm 1UI-CF}^{[m]}   \right]. \label{eq:meth1:sig-qq} 
   \end{eqnarray} 
 
   The {\it doubly-unintegrated} contribution, $\check{\sigma}_{\rm 2UI}^{[m]}$, entering eqns.~(\ref{eq:meth1:sig-gg}), (\ref{eq:meth1:sig-gq}) and (\ref{eq:meth1:sig-qq}) is just  eqn.~(\ref{eq:sig-part-rint-z}) with both ${\bf q}_{T1,2}^2>{\bf q}_{T0}^2$ and the resummation factor (\ref{eq:Bluemlein}). Other contributions are calculated as follows:
   \begin{eqnarray}
    \check{\sigma}^{[m]}_{\rm 1UI-CF} &=& \int\limits_{-\infty}^{+\infty} d\eta \int\limits_{{\bf q}_{T0}^2}^\infty d{\bf q}_{T1}^2 {\cal C}^{\rm DLA}_{gg}\left( \sqrt{z}\frac{M_{T1}}{M}e^{\eta},{\bf q}_{T1}^2,\mu_F,\mu_R \right)  \nonumber \\
 &\times& {\cal F}\left( \frac{{\bf q}_{T0}^2}{\mu_F^2}, \sqrt{z}\frac{M_{T1}}{M}e^{-\eta} \right)\int\limits_0^{2\pi} \frac{d\phi}{2}\  \frac{H({\bf q}^2_{T1},0,\phi)}{M_{T1}^4}, \label{eq:sig-part-z:1UI-CF} \\
   \check{\sigma}^{[m]}_{\rm 1UI-C} &=& \int\limits_{{\bf q}_{T0}^2}^\infty d{\bf q}_{T1}^2\ {\cal C}^{\rm DLA}_{gg}\left( z\frac{M_{T1}^2}{M^2},{\bf q}_{T1}^2,\mu_F,\mu_R \right)  \int\limits_0^{2\pi} \frac{d\phi}{2}\ \frac{H({\bf q}^2_{T1},0,\phi)}{M_{T1}^4}, \label{eq:sig-part-z:1UI-CF} \\
   \check{\sigma}^{[m]}_{\rm 2F} &=& \int\limits_{-\infty}^{\infty} d\eta\  {\cal F}\left( \frac{{\bf q}_{T0}^2}{\mu_F^2} , \sqrt{z}e^{\eta} \right) {\cal F}\left( \frac{{\bf q}_{T0}^2}{\mu_F^2}, \sqrt{z}e^{-\eta} \right)\int\limits_0^{2\pi} \frac{d\phi}{2}\  \frac{H(0,0,\phi)}{M^4}, \label{eq:sig-part-z:I-2F} \\
    \check{\sigma}^{[m]}_{\rm F} &=&  {\cal F}\left( \frac{{\bf q}_{T0}^2}{\mu_F^2} , z \right) \int\limits_0^{2\pi}\frac{d\phi}{2}\  \frac{H(0,0,\phi)}{M^4}, \label{eq:sig-part-z:I-F} 
   \end{eqnarray}   
   where $M_{T1}^2=M^2+{\bf q}_{T1}^2$.

\subsection{Method 2: small-${\bf q}_T$ subtraction}
\label{sec:method-subtr}

The approach described in this section has been inspired by the treatment of the same problem in the ref.~\cite{Collins:1991ty}. The method relies on the same kind of decomposition of the transverse-momentum dependence of the integrand as in  eqn.~(\ref{eq:f-nT-exp}), but now we perform it to isolate the LO contribution, using the normalisation condition (\ref{eq:Blu-norm}). To save space, let us introduce the short-hand notation for the integrand of eqn.~(\ref{eq:sig-part-rint-z}) :
\begin{eqnarray}
{\cal J}(\xi_1,\xi_2)&=&{\cal C}_{gg}\left(\sqrt{z}\frac{M_T(\xi_1,\xi_2)}{M}e^{\eta},{\bf q}_{T1}^2,\mu_F,\mu_R \right)  {\cal C}_{gg}\left(\sqrt{z}\frac{M_T(\xi_1,\xi_2)}{M}e^{-\eta},{\bf q}_{T2}^2,\mu_F,\mu_R \right) \nonumber \\
&\times& \frac{H(\xi_1^2{\bf q}^2_{T1},\xi_2^2{\bf q}^2_{T2},\phi)}{M_T^4(\xi_1,\xi_2)}, \label{eq:J-int}
\end{eqnarray}
where $M_T^2(\xi_1,\xi_2)=M^2+(\xi_1{\bf q}_{T1}+\xi_2{\bf q}_{T2})^2$ and variables $\xi_{1,2}=\{0,1\}$ are introduced for us to be able to turn on and off the ``smooth part'' of the dependence of the integrand on transverse momenta, while the dependence of ${\cal J}$ on ${\bf q}^2_{T1,2}$, $\phi$ and $\eta$ is implicit. Similarly to eqn.~(\ref{eq:f-nT-exp}) we decompose 
\[
{\cal J}(1,1)={\cal J}_1+{\cal J}_2+{\cal J}_3 + {\cal J}(0,0)\theta(\mu_F^2-{\bf q}_{T1}^2)\theta(\mu_F^2-{\bf q}_{T2}^2),
\]
where the last term of this decomposition, when substituted to  eqn.~(\ref{eq:sig-part-rint-z}) gives us $\sigma_{\rm LO}^{[m]}\delta(1-z)$ due to normalisation condition (\ref{eq:Blu-norm}), while the terms ${\cal J}_{1,2,3}$ are:
\begin{eqnarray*}
{\cal J}_1&=&{\cal J}(1,1) - {\cal J}(0,1)\theta(\mu_F^2-{\bf q}_{T1}^2) \nonumber \\&-& {\cal J}(1,0)\theta(\mu_F^2-{\bf q}_{T2}^2) + {\cal J}(0,0)\theta(\mu_F^2-{\bf q}_{T1}^2)\theta(\mu_F^2-{\bf q}_{T2}^2), \\
{\cal J}_2&=& \left[ {\cal J}(1,0)-{\cal J}(0,0)\theta(\mu_F^2-{\bf q}_{T1}^2) \right]\theta(\mu_F^2-{\bf q}_{T2}^2), \\
{\cal J}_3&=& \left[ {\cal J}(0,1)-{\cal J}(0,0)\theta(\mu_F^2-{\bf q}_{T2}^2) \right]\theta(\mu_F^2-{\bf q}_{T1}^2).
\end{eqnarray*}

Using these expressions, the {\it doubly-unintegrated} contribution $\check{\sigma}_{\rm 2UI-S}^{[m]}$ in this method is defined as eqn.~(\ref{eq:sig-part-rint-z}) with integrand ${\cal J}_1$ and without any low-${\bf q}_{T1,2}$ cuts. The {\it single-unintegrated} contribution is obtained by substituting ${\cal J}_2$ into eqn.~(\ref{eq:sig-part-rint-z}) and integrating out ${\bf q}_{T2}^2$ via eqn.~(\ref{eq:Blu-norm}):
 \begin{eqnarray}
 \check{\sigma}^{[m]}_{\rm 1UI-S} = \int\limits_0^\infty d{\bf q}_{T1}^2\int\limits_0^{2\pi} \frac{d\phi}{2}\ \Bigg[ &&{\cal C}_{gg}\left(z\frac{M_{T1}^2}{M^2},{\bf q}_{T1}^2,\mu_F^2,\mu_R^2 \right) \frac{H({\bf q}_{T1}^2,0,\phi)}{M_{T1}^4}  \nonumber \\
 -  &&{\cal C}_{gg}\left(z,{\bf q}_{T1}^2,\mu_F^2,\mu_R^2 \right) \frac{H(0,0,\phi)}{M^4} \theta(\mu_F^2-{\bf q}_{T1}^2) \Bigg]. 
 \end{eqnarray}
 
 The LO subtracted resummed cross section for the $gg$ channel is calculated as:
\begin{equation}
\check{\sigma}_{gg}^{[m],\ {\rm HEF}} = \check{\sigma}_{\rm 2UI-S}^{[m]} + 2 \check{\sigma}^{[m]}_{\rm 1UI-S},  
\end{equation}
while, for the quark-induced channels, one has:
\begin{eqnarray}
\hat{\sigma}_{gq}^{[m],\ {\rm HEF}} &=& \frac{C_F}{C_A}\left[ \check{\sigma}_{\rm 2UI-S}^{[m]} +  \check{\sigma}^{[m]}_{\rm 1UI-S} \right], \\
\hat{\sigma}_{q\bar{q}}^{[m],\ {\rm HEF}} &=& \left(\frac{C_F}{C_A}\right)^2 \check{\sigma}_{\rm 2UI-S}^{[m]}. 
\end{eqnarray} 

For the DLA resummation factor, the numerical results obtained in both methods agree within the integration accuracy. However, the subtraction method allows one to work with resummation factor given in the numerical form. The only requirement is that it should satisfy the normalisation condition (\ref{eq:Blu-norm}), which is true for example for the resummation factor (\ref{eq:C-full}) with full LL($\ln(1/z)$) anomalous dimension, obtained as numerical solution of  eqn.~(\ref{eq:gamma-eqn}), but with $R(\gamma)=1$, as discussed in Appendix~\ref{sec:gamma-effects}, or for the Gaussian-smeared resummation factor discussed in Appendix~\ref{sec:Gauss}. 

\section{Effects of the anomalous dimension beyond LO}
\label{sec:gamma-effects}

\begin{figure}
  \begin{center}
  \parbox{0.5\textwidth}{\includegraphics[width=0.5\textwidth]{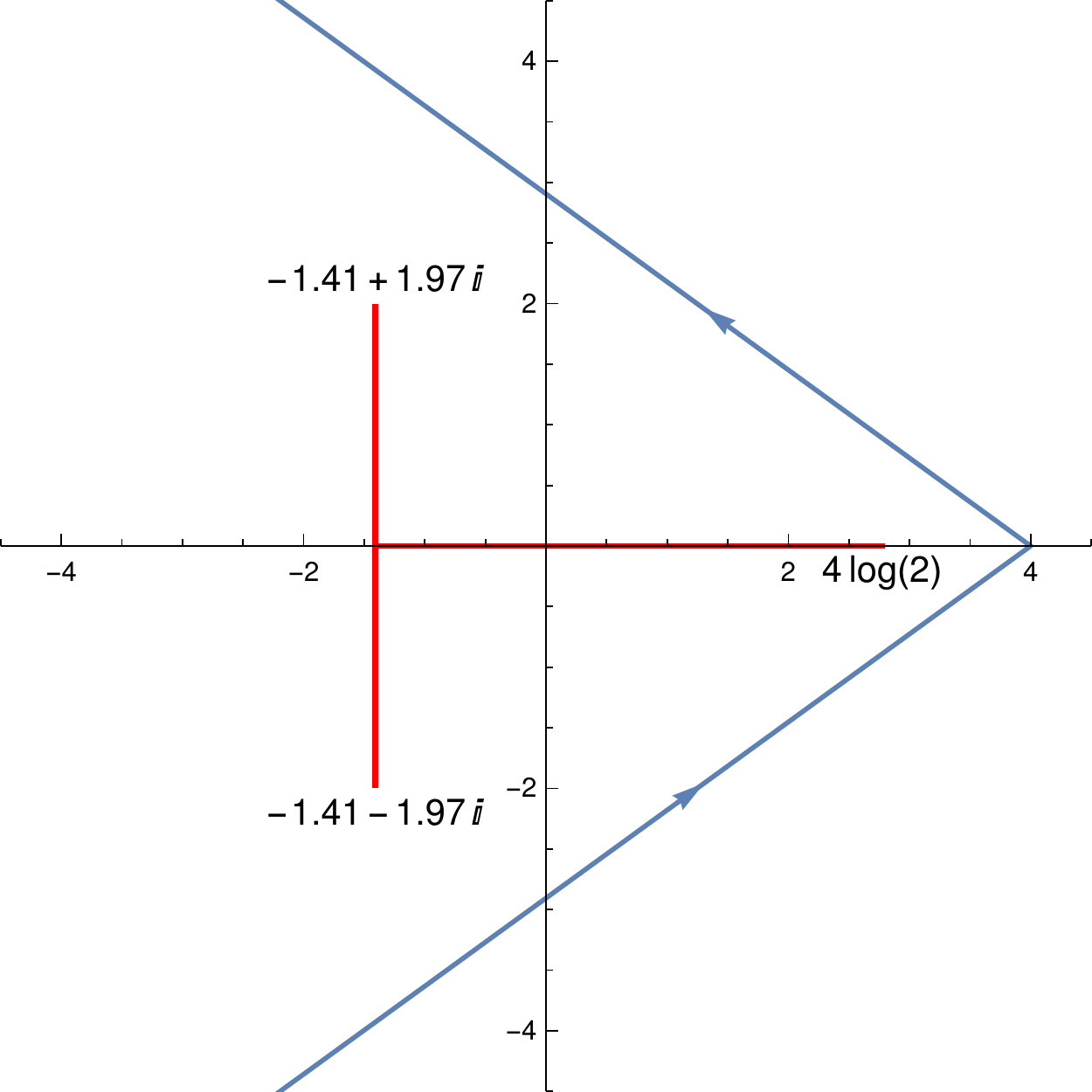}}
  \parbox{0.4\textwidth}{\includegraphics[width=0.4\textwidth]{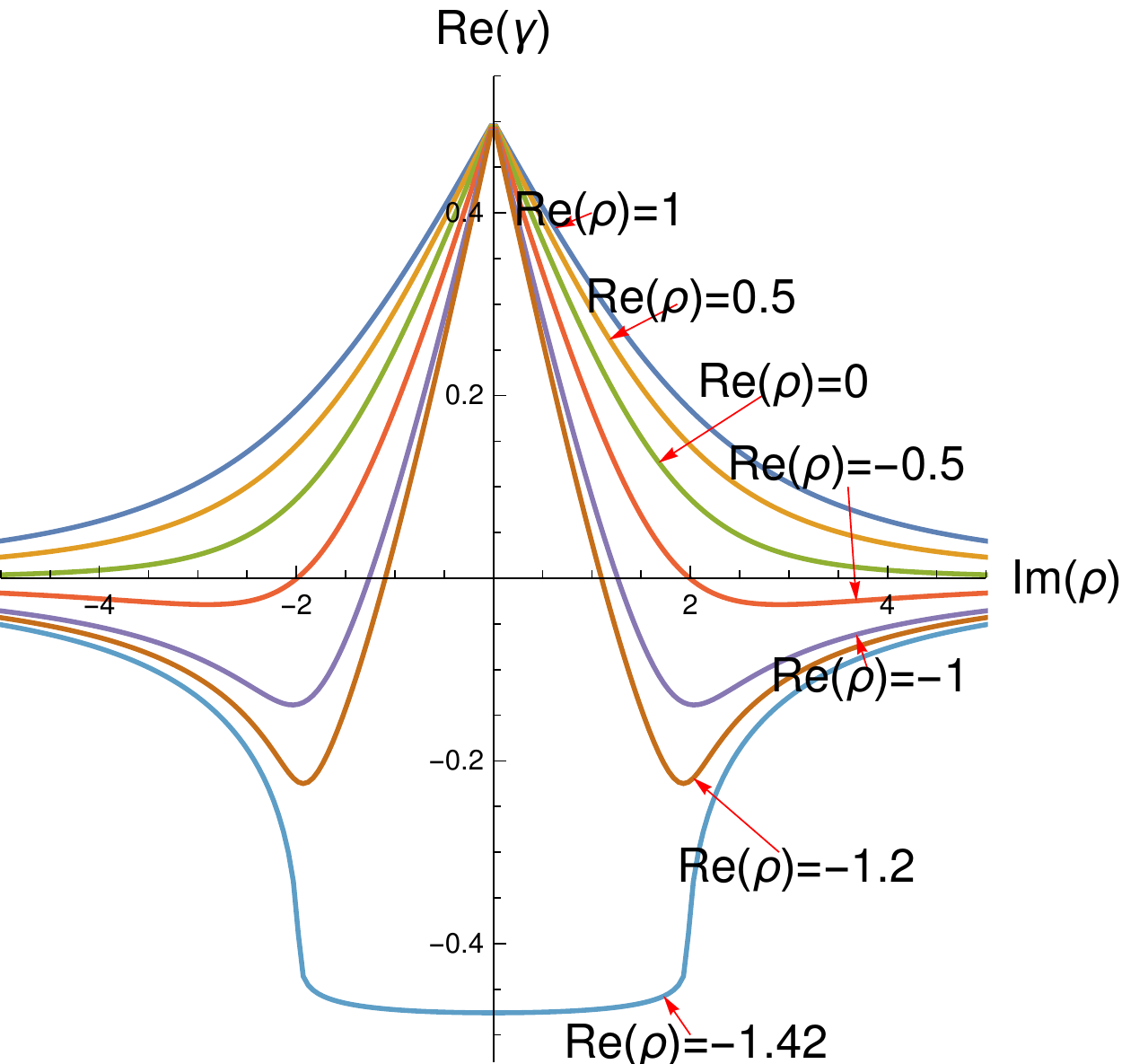} \includegraphics[width=0.4\textwidth]{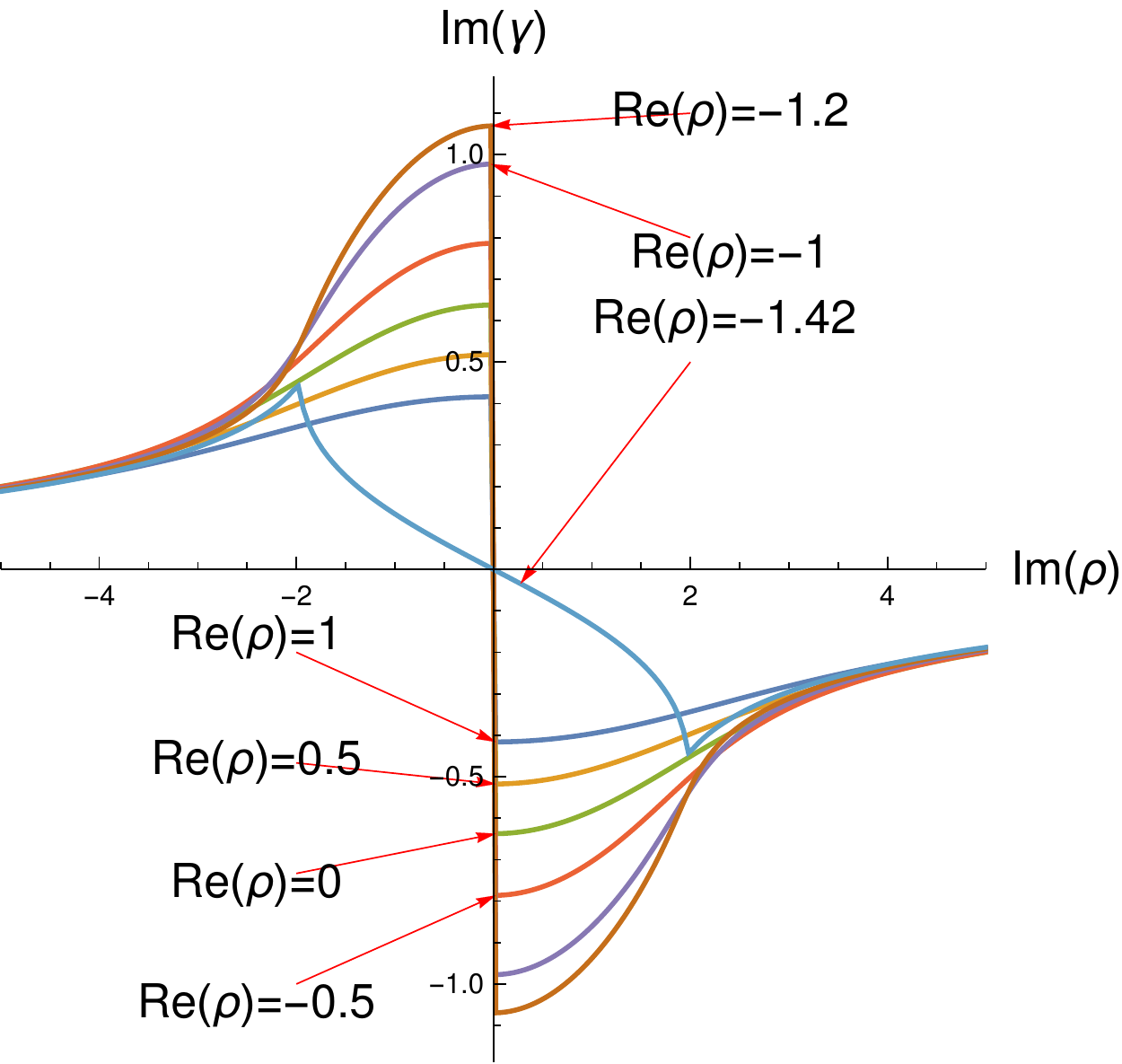}}
  \end{center}  
  \caption{{\bf Left panel:} branch cut discontinuity of the function $\gamma_{gg}(\rho)$ and the integration contour of the inverse Mellin transform in the $\rho$-plane. {\bf Right panel:} real and imaginary parts of $\gamma_{gg}(\rho)$ as functions of $\mathop{\rm Im}\rho$ at fixed $\mathop{\rm Re}\rho=1,0.5,0,-0.5,-1,-1.2,-1.42$. To be compared with the plots in the figure~1 of the ref.~\cite{Blumlein:1995eu}. }\label{fig:gamma-rho}
  \end{figure}

In this appendix, we study the effects of terms beyond LO in $\alpha_s$ in the anomalous dimension (\ref{eq:pert-gamma}) on the resummed cross section, while still omitting the scheme-transformation factor $R(\gamma)$ from the eqn.~(\ref{eq:C-full}). We have obtained the numerical solution of eqn.~(\ref{eq:gamma-eqn}), which in terms of the variable $\rho=N/\hat{\alpha}_s$ has the form:
\begin{equation}
\chi(\gamma_{gg}(\rho))=\rho.
\end{equation}
  This equation has already been studied numerically in refs.~\cite{Ellis:1995gv, Blumlein:1995eu}. The perturbative branch of the solution has a cut-discontinuity depicted in the figure~\ref{fig:gamma-rho}, and inverse Mellin transform integral (\ref{eq:inverse-Mellin}) can be written as an integral over the contour in $\rho$-plane which should avoid crossing this cut. Our numerical results for the real and imaginary parts of the solution agree with results of ref.~\cite{Blumlein:1995eu} and are also shown on figure~\ref{fig:gamma-rho}. 
  
\begin{figure}
  \begin{center}
  \includegraphics[width=0.32\textwidth]{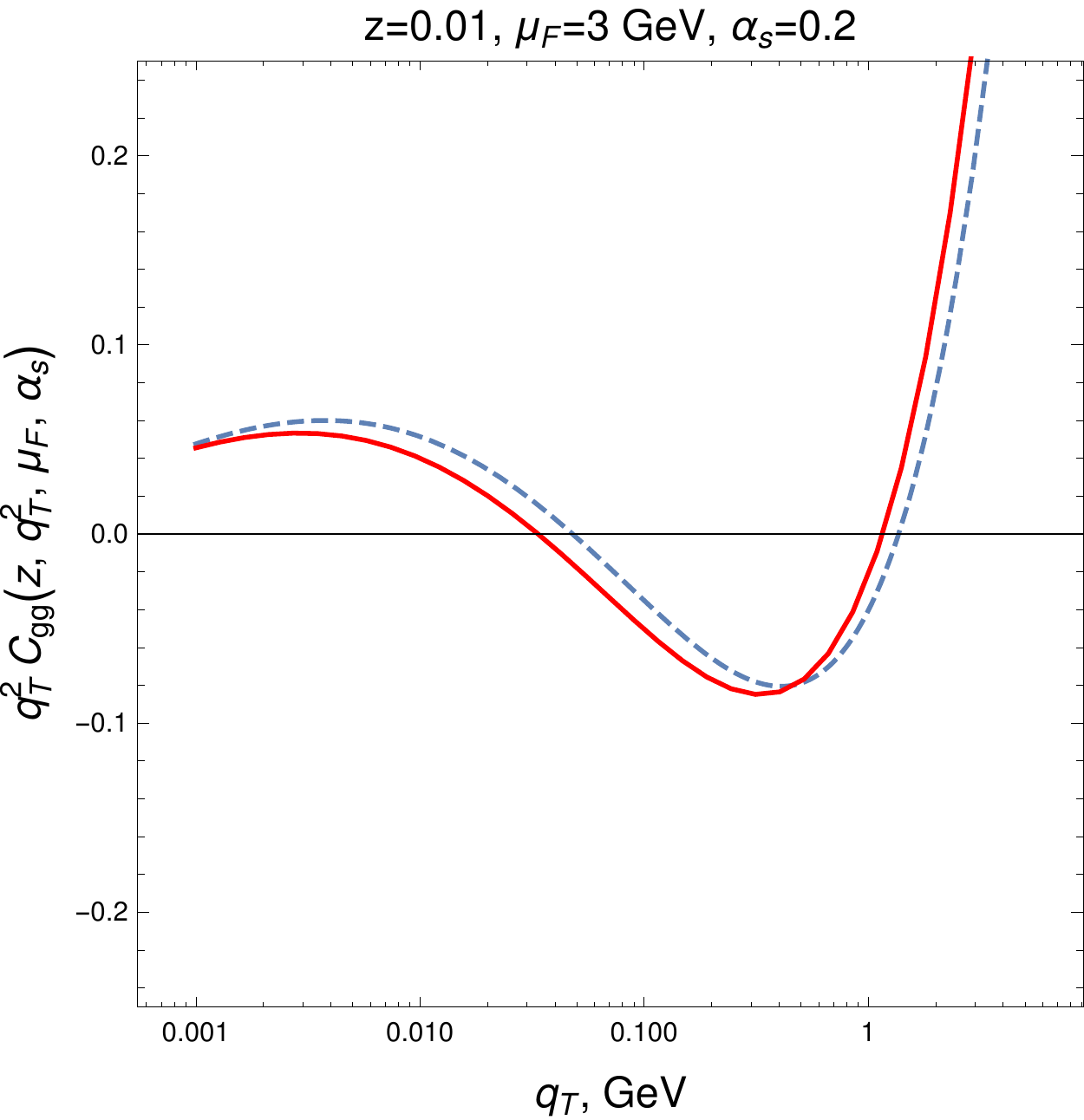}
  \includegraphics[width=0.32\textwidth]{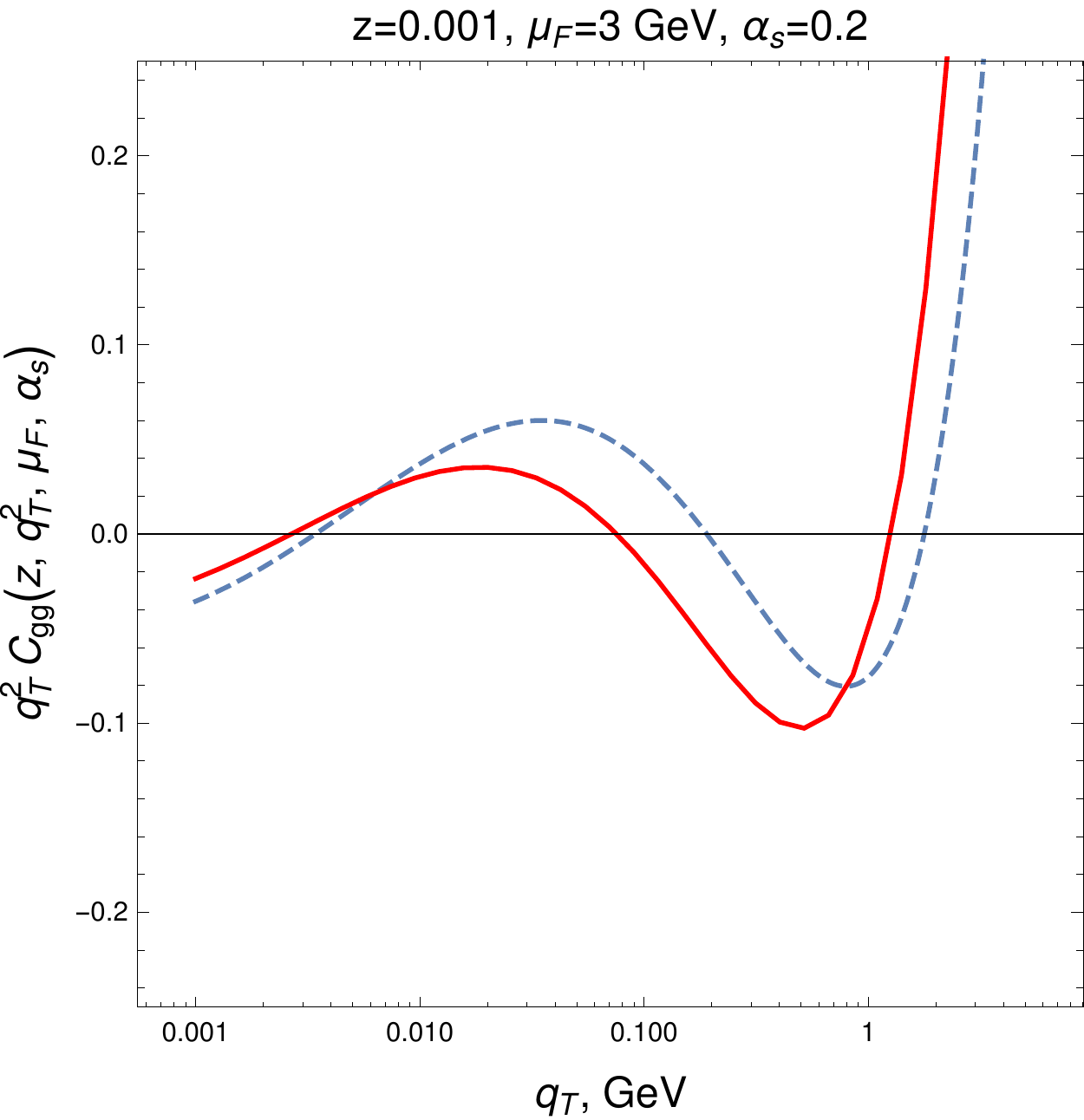}
  \includegraphics[width=0.32\textwidth]{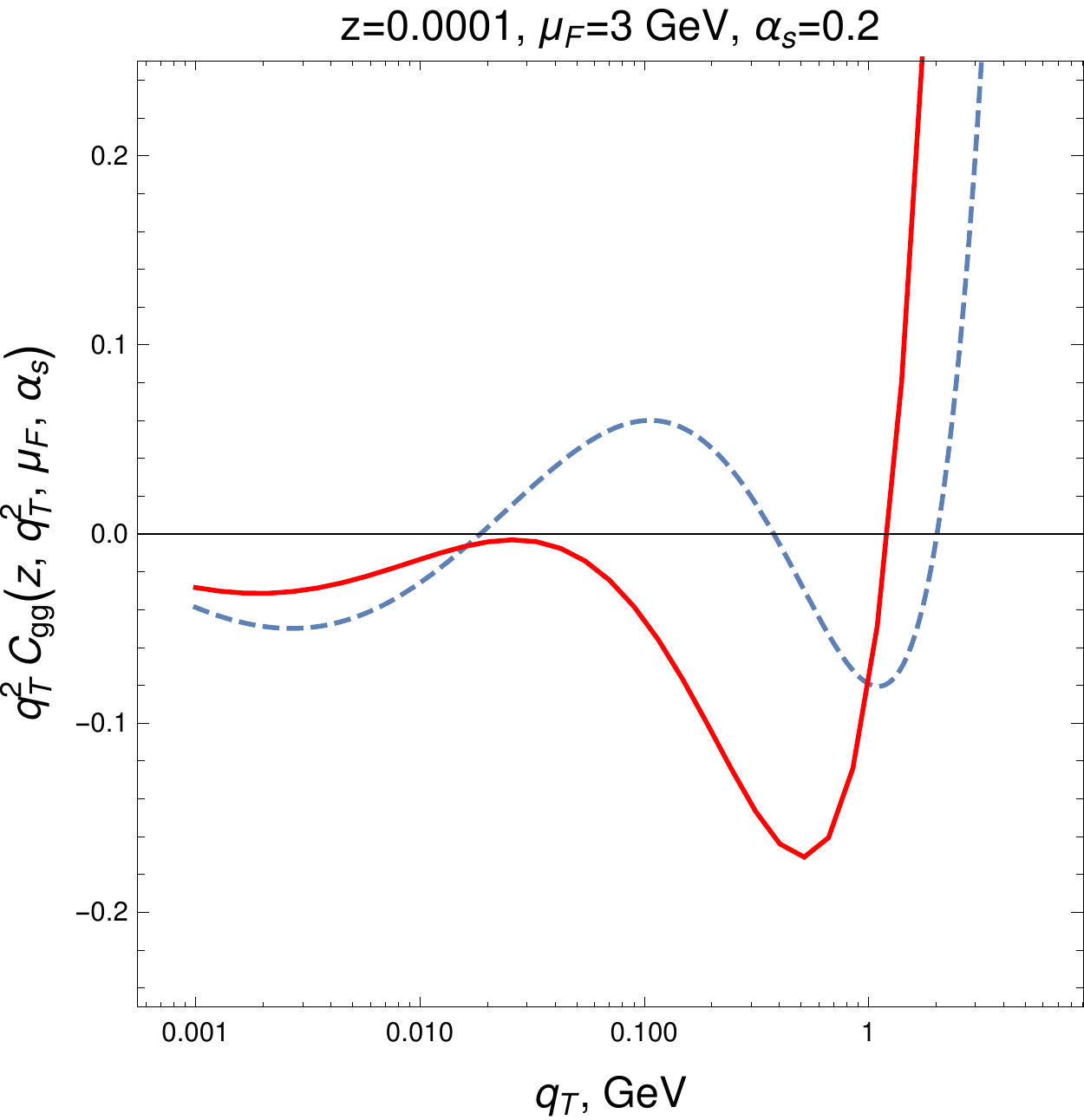}
  \end{center}
  \caption{Resummation factor ${\cal C}_{gg}$ as a function of $|{\bf q}_T|$ for several values of $z$. The DLA result is plotted by the dashed curve and numerical result with LL($\ln(1/z)$) anomalous dimension, obtained as a solution of eqn.~(\ref{eq:gamma-eqn}) is plotted by the solid curve.}\label{fig:C-plots}
  \end{figure}

    Plots of the resummation factor resulting from the numerical inverse Mellin transform are shown on figure~\ref{fig:C-plots}. At moderately small values of $z$, it is close to the DLA, but at smaller values of $z$ it deviates from DLA and the most numerically important region of ${\bf q}_T^2\sim 1$ GeV$^2$ features a significant peak in the negative direction. 
    
\begin{figure}
  \begin{center}
  \includegraphics[width=0.49\textwidth]{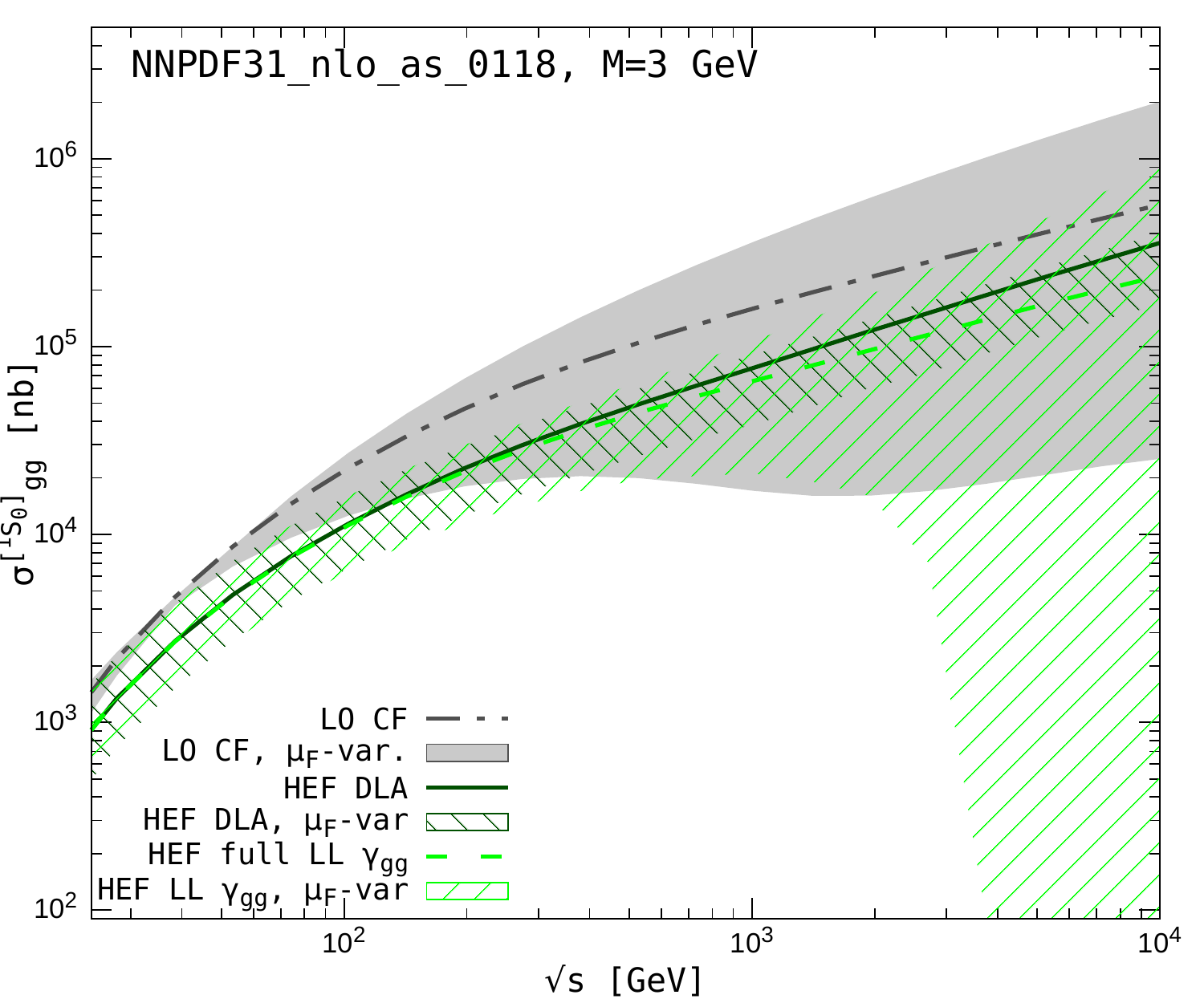}
  \includegraphics[width=0.49\textwidth]{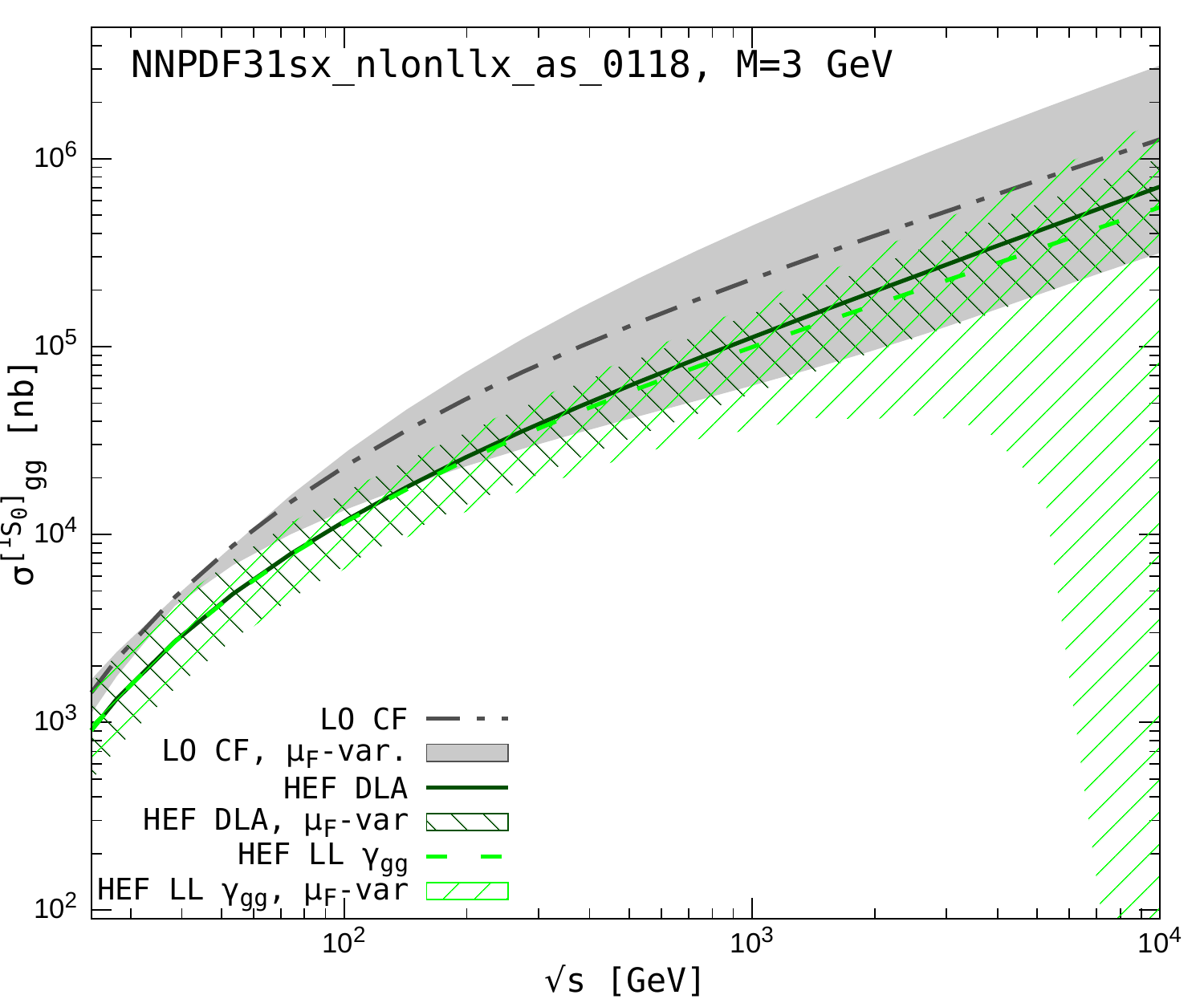}
  \end{center}
  \caption{Energy dependence of the HEF resummed part of the cross section (LO CF included) in the $gg$ channel: the DLA result and it's $\mu_F$-variation are plotted as solid line and left-shaded band, dashed line and right-shaded $\mu_F$-variation band correspond to the calculation using the full numerical solution of eqn.~(\ref{eq:gamma-eqn}) for $\gamma_{gg}(N,\alpha_s)$. For comparison, the LO CF cross section together with its $\mu_F$-variation is shown by the dash-dotted line and the corresponding solid band. The gluon PDFs used are  \texttt{NNPDF31\_nlo\_as\_0118}~\cite{NNPDF:2017mvq}(left panel) and \texttt{NNPDF31sx\_nlonllx\_as\_0118}~\cite{Ball:2017otu}(right panel). The LDME $\langle {\cal O} [ {}^1S_0^{(1)} ] \rangle$ was set to 1~GeV$^3$. }\label{fig:alles-kaput}
  \end{figure}

    On figure~\ref{fig:alles-kaput}, the $\sqrt{s}$-dependence of the HEF-resummed part of the total cross section of production of the $Q\bar{Q}[{}^1S_0^{(1)}]$ state in the $gg$ channel is plotted. The central lines as well as uncertainty bands resulting from the variation of $\mu_F$ by a factor of 2 above and below the default value $\mu_F=M$ are presented. Two approximations for the resummation factor are compared on figure~\ref{fig:alles-kaput}: the DLA, introduced in the section~\ref{sec:C-factors}, and the approximation described above in this appendix, which includes effects beyond LO in $\gamma_{gg}(N,\alpha_s)$. The LO CF central curve and the $\mu_F$ uncertainty is also shown on figure~\ref{fig:alles-kaput} for comparison. One can see that, while DLA significantly reduces the scale uncertainty in comparison to the LO CF result, the inclusion of higher-order corrections to $\gamma_{gg}(N,\alpha_s)$ blows up the scale uncertainty at high energies so much that, for $\sqrt{s}>4$ TeV (left panel of the figure~\ref{fig:alles-kaput}) or $\sqrt{s}>7$ TeV (right panel of the figure~\ref{fig:alles-kaput}), the resummed part of cross section can become negative. 
   
    This catastrophic factorisation-scale uncertainty is a consequence of the mismatch between the $\mu_F$-dependence of the resummation factor and one of the collinear PDFs used. For the left plot on figure~\ref{fig:alles-kaput}, we have employed the central eigenset of the \texttt{NNPDF31\_nlo} \texttt{\_as\_0118}~\cite{NNPDF:2017mvq} PDFs and, in the right panel, the NLL($\ln(1/z)$)-resummed version of the \texttt{NNPDF31} PDFs has been used~\cite{Ball:2017otu}. Figure~\ref{fig:alles-kaput} serves as a numerical illustration of our statement in section~\ref{sec:C-factors} that one should not increase the accuracy of the resummation factors in HEF beyond DLA if collinear PDFs result from a fixed-order DGLAP evolution, as in the left plot of figure~\ref{fig:alles-kaput}. In the case of the NLL($\ln(1/z)$)-resummed PDFs (right panel of figure~\ref{fig:alles-kaput}), the addition of the full LL($\ln(1/z)$) correction to $\gamma_{gg}$ is also not justified because in this case the formal perturbative accuracy of the PDFs is greater that that of the resummation factor and the NLL resummed version $\gamma_{gg}$ is significantly~\cite{Altarelli:1999vw,Ball:2017otu} different from the full LL($\ln(1/z)$)-resummed one, so even in this case we have a mismatch of the $\mu_F$-dependence between the PDFs and the resummation factor.

\section{Higher-twist effects}
\label{sec:Gauss}

\begin{figure}[hbt!]
  \begin{center}
  \parbox{0.49\textwidth}{\includegraphics[width=0.49\textwidth]{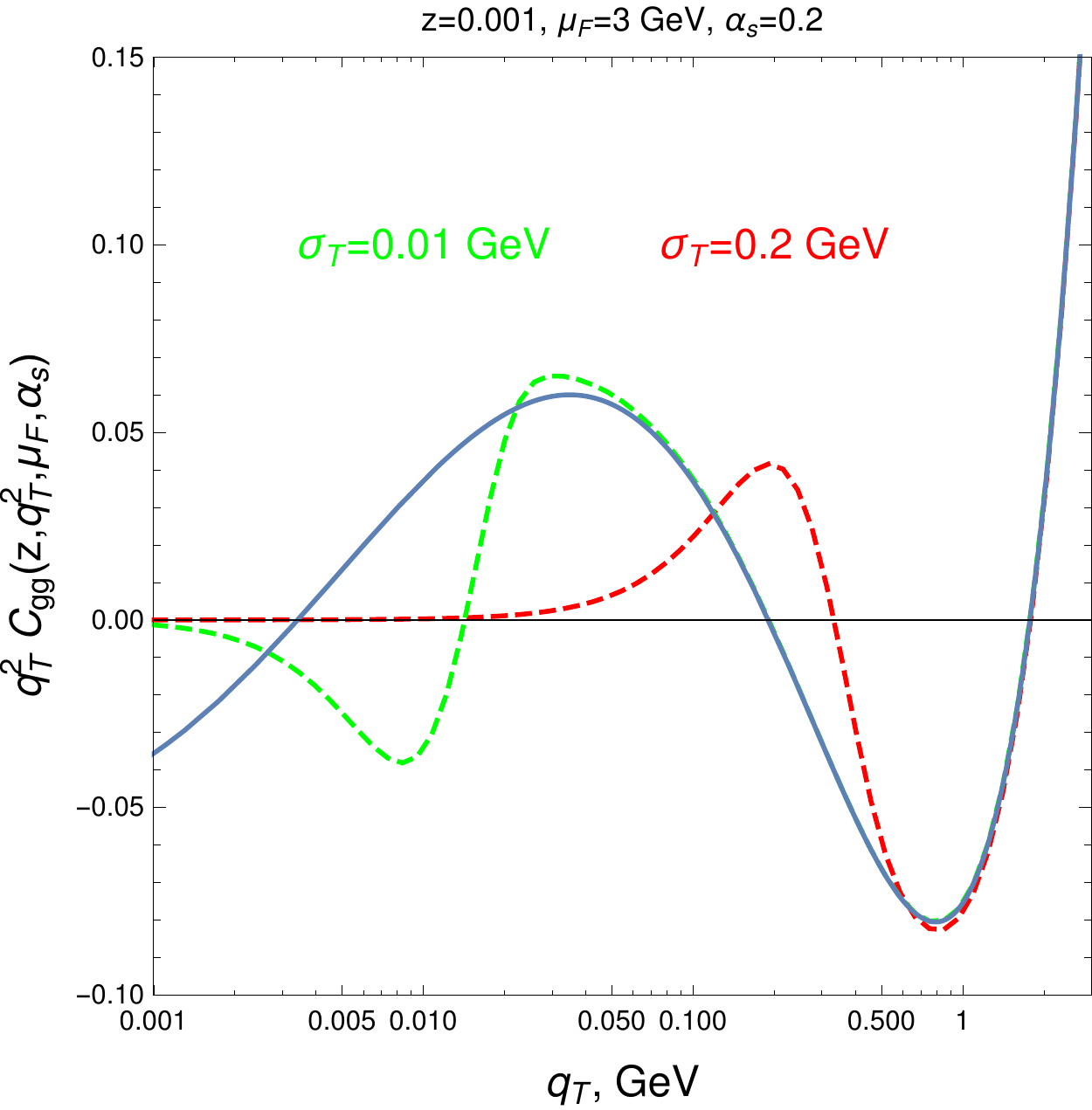}}
  \parbox{0.49\textwidth}{\includegraphics[width=0.49\textwidth]{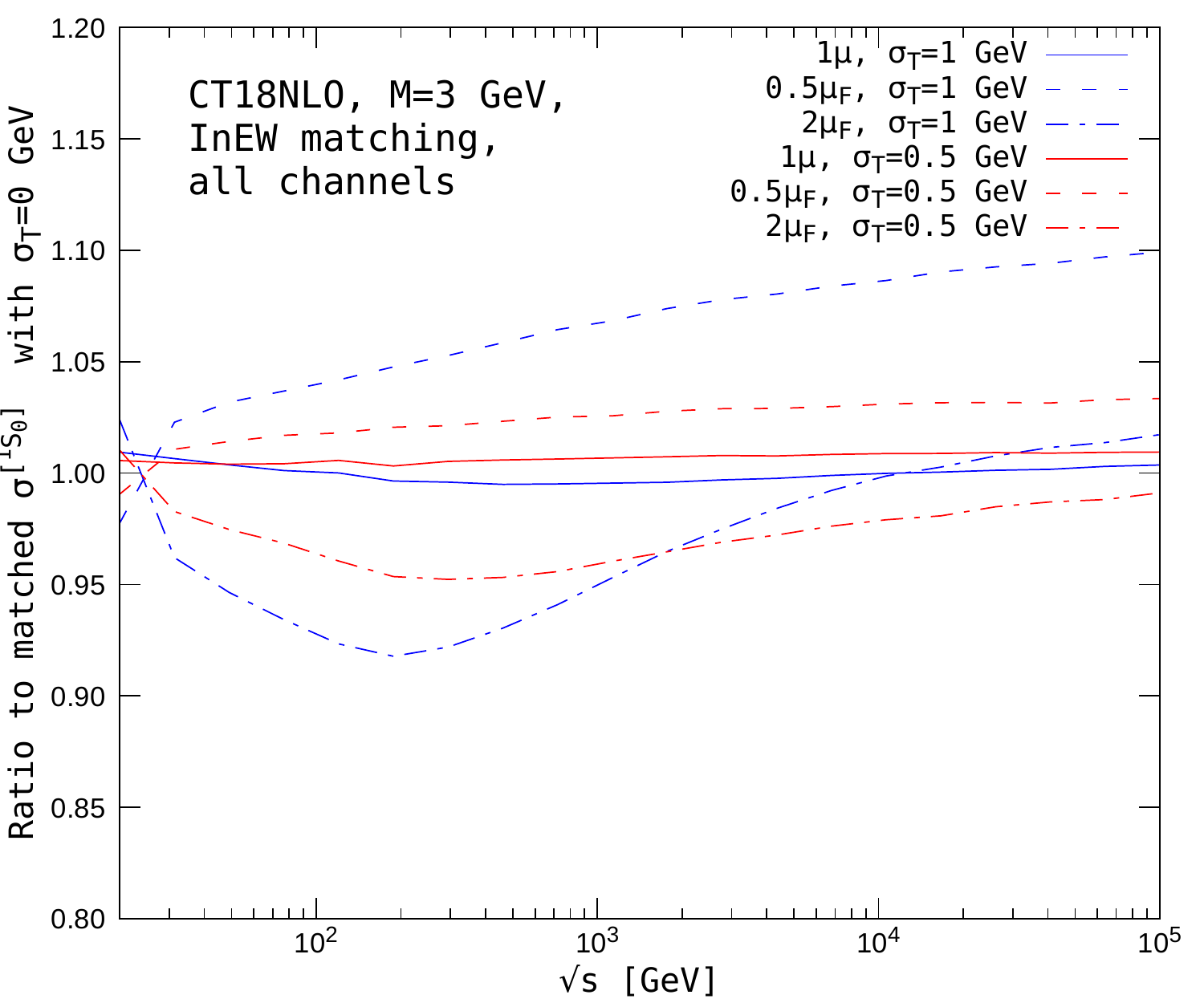}}
  \end{center}
  \caption{{\bf Left panel:} the ${\bf q}_T$-dependence of the Gaussian smeared DLA resummation factor for two values of $\sigma_T$ (dashed curves). The DLA resummation factor without smearing (\ref{eq:Bluemlein}) is shown by solid curve for comparison. {\bf Right panel:} the energy-dependence of the ratio of the matched cross section obtained in the DLA with Gaussian smearing with $\sigma_T=0.5$ and 1 GeV to the result without smearing and its $\mu_F$-scale variation.  }\label{fig:C-smeared}
  \end{figure}

It may be argued that the oscillatory behaviour of the resummation factors at ${\bf q}_T^2\ll 1$~GeV$^2$, depicted on  figure~\ref{fig:C-plots}, is un-physical and that in this domain the resummation factor is dominated by non-perturbative effects. In this context, one would like to understand how much the total cross section depends on the detailed behaviour of the resummation factor at small values of ${\bf q}_T^2$. As a toy model to study this question, we have performed a convolution of the DLA resummation factor (\ref{eq:qT-UPDF}) with Gaussian of the width $\sigma_T$ in transverse momentum which can be done analytically with the Mellin-space result (\ref{eq:qT-UPDF}):
\begin{eqnarray}
  {\cal C}_{gg}^{{\rm DLA},\sigma}(N,{\bf q}_T^2,\mu_F,\mu_R)&=& \int\frac{d^2{\bf k}_T}{\pi\sigma_T^2} \exp\left[-\frac{{\bf k}_T^2}{\sigma_T^2} \right] {\cal C}^{\rm DLA}_{gg}(N,({\bf q}_T+{\bf k}_T)^2,\mu_F,\mu_R) \nonumber \\ &=&  \frac{1}{\sigma_T^2}\left(\frac{\sigma_T^2}{\mu_F^2} \right)^{\gamma_N} e^{-\frac{{\bf q}_T^2}{\sigma_T^2}} {}_1F_1\left(\gamma_N,1,\frac{{\bf q}_T^2}{\sigma_T^2}\right), \label{eq:C-smeared}
  \end{eqnarray} 
  and then converted it to the $z$-space numerically. Eqn.~(\ref{eq:C-smeared}) still satisfies the normalisation condition (\ref{eq:Blu-norm}) up to corrections suppressed as $e^{-\mu_F^2/\sigma_T^2}$, so the method from the section~\ref{sec:method-subtr} can be used to calculate the cross section if $\sigma_T\lesssim 1$ GeV. Several plots of the $|{\bf q}_T|$-dependence of the smeared resummation factor~(\ref{eq:C-smeared}), transformed numerically to the $z$-space are presented in the left panel of  figure~\ref{fig:C-smeared}.
  
  In the right panel of figure~\ref{fig:C-smeared}, the ratio of the matched cross section obtained with the Gaussian-smeared resummation factor to the cross section without ${\bf k}_T$-smearing is plotted as a function of $\sqrt{s}$ for $\sigma_T=0.5$ and $1$ GeV. The results with $\sigma_T=0.5$ and $1$ GeV differ from the matched cross section calculated without a Gaussian by at most ten percent for $\mu_F=0.5M$ and $2M$, while the results at the default scale are even closer. 
  
  From the detailed consideration of the small-${\bf q}_T$ subtraction method, described in section~\ref{sec:method-subtr}, one can see that the effect of the ${\bf k}_T$-smearing on the total cross section is suppressed as ${\cal O}(\sigma_T^2/M^2)\sim O(\Lambda_{\rm QCD}^2/M^2)$ if $\sigma_T\sim\Lambda_{QCD}$. In other words, these effect follow a subleading power in our hard scale $M$; it is a {\it higher-twist effect}. 
  
   The width of the ``intrinsic'' ${\bf k}_T$-distribution is expected to be  increasing with energy, due to increasing saturation scale (see e.g. ref.~\cite{Albacete:2014fwa} and references therein), so that, at high enough energies, the twist expansion will completely break down due to the effect discussed above.  However, as one can see from figure~\ref{fig:C-smeared} this effects are unlikely to be the main source of uncertainty in our calculation at LHC energies and even beyond.
  
\bibliographystyle{JHEP}
\bibliography{mybibfile}

\providecommand{\href}[2]{#2}\begingroup\raggedright\begin{thebibliography}{10}

\bibitem{Chapon:2020heu}
E.~Chapon et~al., \emph{{Perspectives for quarkonium studies at the
  high-luminosity LHC}},  \href{https://arxiv.org/abs/2012.14161}{{\ttfamily
  2012.14161}}.

\bibitem{Arbuzov:2020cqg}
A.~Arbuzov et~al., \emph{{On the physics potential to study the gluon content
  of proton and deuteron at NICA SPD}},
  \href{https://doi.org/10.1016/j.ppnp.2021.103858}{\emph{Prog. Part. Nucl.
  Phys.} {\bfseries 119} (2021) 103858}
  [\href{https://arxiv.org/abs/2011.15005}{{\ttfamily 2011.15005}}].

\bibitem{Lansberg:2019adr}
J.-P.~Lansberg, \emph{{New Observables in Inclusive Production of Quarkonia}},
  \href{https://doi.org/10.1016/j.physrep.2020.08.007}{\emph{Phys. Rept.}
  {\bfseries 889} (2020) 1} [\href{https://arxiv.org/abs/1903.09185}{{\ttfamily
  1903.09185}}].

\bibitem{Brambilla:2014jmp}
N.~Brambilla et~al., \emph{{QCD and Strongly Coupled Gauge Theories: Challenges
  and Perspectives}},
  \href{https://doi.org/10.1140/epjc/s10052-014-2981-5}{\emph{Eur. Phys. J. C}
  {\bfseries 74} (2014) 2981}
  [\href{https://arxiv.org/abs/1404.3723}{{\ttfamily 1404.3723}}].

\bibitem{Brambilla:2010cs}
N.~Brambilla et~al., \emph{{Heavy Quarkonium: Progress, Puzzles, and
  Opportunities}},
  \href{https://doi.org/10.1140/epjc/s10052-010-1534-9}{\emph{Eur. Phys. J. C}
  {\bfseries 71} (2011) 1534}
  [\href{https://arxiv.org/abs/1010.5827}{{\ttfamily 1010.5827}}].

\bibitem{Bodwin:1994jh}
G.T.~Bodwin, E.~Braaten and G.P.~Lepage, \emph{{Rigorous QCD analysis of
  inclusive annihilation and production of heavy quarkonium}},
  \href{https://doi.org/10.1103/PhysRevD.55.5853}{\emph{Phys. Rev. D}
  {\bfseries 51} (1995) 1125}
  [\href{https://arxiv.org/abs/hep-ph/9407339}{{\ttfamily hep-ph/9407339}}].

\bibitem{Butenschoen:2010rq}
M.~Butenschoen and B.A.~Kniehl, \emph{{Reconciling $J/\psi$ production at HERA,
  RHIC, Tevatron, and LHC with NRQCD factorization at next-to-leading order}},
  \href{https://doi.org/10.1103/PhysRevLett.106.022003}{\emph{Phys. Rev. Lett.}
  {\bfseries 106} (2011) 022003}
  [\href{https://arxiv.org/abs/1009.5662}{{\ttfamily 1009.5662}}].

\bibitem{Butenschoen:2011yh}
M.~Butenschoen and B.A.~Kniehl, \emph{{World data of J/psi production
  consolidate NRQCD factorization at NLO}},
  \href{https://doi.org/10.1103/PhysRevD.84.051501}{\emph{Phys. Rev. D}
  {\bfseries 84} (2011) 051501}
  [\href{https://arxiv.org/abs/1105.0820}{{\ttfamily 1105.0820}}].

\bibitem{Butenschoen:2012px}
M.~Butenschoen and B.A.~Kniehl, \emph{{J/psi polarization at Tevatron and LHC:
  Nonrelativistic-QCD factorization at the crossroads}},
  \href{https://doi.org/10.1103/PhysRevLett.108.172002}{\emph{Phys. Rev. Lett.}
  {\bfseries 108} (2012) 172002}
  [\href{https://arxiv.org/abs/1201.1872}{{\ttfamily 1201.1872}}].

\bibitem{Butenschoen:2012qr}
M.~Butenschoen and B.A.~Kniehl, \emph{{Next-to-leading-order tests of NRQCD
  factorization with $J/\psi$ yield and polarization}},
  \href{https://doi.org/10.1142/S0217732313500272}{\emph{Mod. Phys. Lett. A}
  {\bfseries 28} (2013) 1350027}
  [\href{https://arxiv.org/abs/1212.2037}{{\ttfamily 1212.2037}}].

\bibitem{Chao:2012iv}
K.-T.~Chao, Y.-Q.~Ma, H.-S.~Shao, K.~Wang and Y.-J.~Zhang, \emph{{$J/\psi$
  Polarization at Hadron Colliders in Nonrelativistic QCD}},
  \href{https://doi.org/10.1103/PhysRevLett.108.242004}{\emph{Phys. Rev. Lett.}
  {\bfseries 108} (2012) 242004}
  [\href{https://arxiv.org/abs/1201.2675}{{\ttfamily 1201.2675}}].

\bibitem{Gong:2012ug}
B.~Gong, L.-P.~Wan, J.-X.~Wang and H.-F.~Zhang, \emph{{Polarization for Prompt
  J/\ensuremath{\psi} and \ensuremath{\psi}(2s) Production at the Tevatron and
  LHC}}, \href{https://doi.org/10.1103/PhysRevLett.110.042002}{\emph{Phys. Rev.
  Lett.} {\bfseries 110} (2013) 042002}
  [\href{https://arxiv.org/abs/1205.6682}{{\ttfamily 1205.6682}}].

\bibitem{Bodwin:2014gia}
G.T.~Bodwin, H.S.~Chung, U.-R.~Kim and J.~Lee, \emph{{Fragmentation
  contributions to $J/\psi$ production at the Tevatron and the LHC}},
  \href{https://doi.org/10.1103/PhysRevLett.113.022001}{\emph{Phys. Rev. Lett.}
  {\bfseries 113} (2014) 022001}
  [\href{https://arxiv.org/abs/1403.3612}{{\ttfamily 1403.3612}}].

\bibitem{Gong:2010bk}
B.~Gong, J.-X.~Wang and H.-F.~Zhang, \emph{{QCD corrections to $\Upsilon$
  production via color-octet states at the Tevatron and LHC}},
  \href{https://doi.org/10.1103/PhysRevD.83.114021}{\emph{Phys. Rev. D}
  {\bfseries 83} (2011) 114021}
  [\href{https://arxiv.org/abs/1009.3839}{{\ttfamily 1009.3839}}].

\bibitem{Wang:2012is}
K.~Wang, Y.-Q.~Ma and K.-T.~Chao, \emph{{$\Upsilon(1S)$ prompt production at
  the Tevatron and LHC in nonrelativistic QCD}},
  \href{https://doi.org/10.1103/PhysRevD.85.114003}{\emph{Phys. Rev. D}
  {\bfseries 85} (2012) 114003}
  [\href{https://arxiv.org/abs/1202.6012}{{\ttfamily 1202.6012}}].

\bibitem{Gong:2013qka}
B.~Gong, L.-P.~Wan, J.-X.~Wang and H.-F.~Zhang, \emph{{Complete
  next-to-leading-order study on the yield and polarization of
  $\Upsilon(1S,2S,3S)$ at the Tevatron and LHC}},
  \href{https://doi.org/10.1103/PhysRevLett.112.032001}{\emph{Phys. Rev. Lett.}
  {\bfseries 112} (2014) 032001}
  [\href{https://arxiv.org/abs/1305.0748}{{\ttfamily 1305.0748}}].

\bibitem{Butenschoen:2014dra}
M.~Butenschoen, Z.-G.~He and B.A.~Kniehl, \emph{{$\eta_c$ production at the LHC
  challenges nonrelativistic-QCD factorization}},
  \href{https://doi.org/10.1103/PhysRevLett.114.092004}{\emph{Phys. Rev. Lett.}
  {\bfseries 114} (2015) 092004}
  [\href{https://arxiv.org/abs/1411.5287}{{\ttfamily 1411.5287}}].

\bibitem{Barger:1979js}
V.D.~Barger, W.-Y.~Keung and R.J.N.~Phillips, \emph{{On psi and Upsilon
  Production via Gluons}},
  \href{https://doi.org/10.1016/0370-2693(80)90444-X}{\emph{Phys. Lett. B}
  {\bfseries 91} (1980) 253}.

\bibitem{Barger:1980mg}
V.D.~Barger, W.-Y.~Keung and R.J.N.~Phillips, \emph{{Hadroproduction of $\psi$
  and $\Upsilon$}}, \href{https://doi.org/10.1007/BF01588844}{\emph{Z. Phys. C}
  {\bfseries 6} (1980) 169}.

\bibitem{Gavai:1994in}
R.~Gavai, D.~Kharzeev, H.~Satz, G.A.~Schuler, K.~Sridhar and R.~Vogt,
  \emph{{Quarkonium production in hadronic collisions}},
  \href{https://doi.org/10.1142/S0217751X95001443}{\emph{Int. J. Mod. Phys. A}
  {\bfseries 10} (1995) 3043}
  [\href{https://arxiv.org/abs/hep-ph/9502270}{{\ttfamily hep-ph/9502270}}].

\bibitem{Ma:2016exq}
Y.-Q.~Ma and R.~Vogt, \emph{{Quarkonium Production in an Improved Color
  Evaporation Model}},
  \href{https://doi.org/10.1103/PhysRevD.94.114029}{\emph{Phys. Rev. D}
  {\bfseries 94} (2016) 114029}
  [\href{https://arxiv.org/abs/1609.06042}{{\ttfamily 1609.06042}}].

\bibitem{Lansberg:2020rft}
J.-P.~Lansberg, H.-S.~Shao, N.~Yamanaka, Y.-J.~Zhang and C.~No\^us,
  \emph{{Complete NLO QCD study of single- and double-quarkonium
  hadroproduction in the colour-evaporation model at the Tevatron and the
  LHC}}, \href{https://doi.org/10.1016/j.physletb.2020.135559}{\emph{Phys.
  Lett. B} {\bfseries 807} (2020) 135559}
  [\href{https://arxiv.org/abs/2004.14345}{{\ttfamily 2004.14345}}].

\bibitem{Brambilla:2021abf}
N.~Brambilla, H.S.~Chung and A.~Vairo, \emph{{Inclusive production of heavy
  quarkonia in pNRQCD}},
  \href{https://doi.org/10.1007/JHEP09(2021)032}{\emph{JHEP} {\bfseries 09}
  (2021) 032} [\href{https://arxiv.org/abs/2106.09417}{{\ttfamily
  2106.09417}}].

\bibitem{Ma:2017xno}
Y.-Q.~Ma and K.-T.~Chao, \emph{{New factorization theory for heavy quarkonium
  production and decay}},
  \href{https://doi.org/10.1103/PhysRevD.100.094007}{\emph{Phys. Rev. D}
  {\bfseries 100} (2019) 094007}
  [\href{https://arxiv.org/abs/1703.08402}{{\ttfamily 1703.08402}}].

\bibitem{Li:2019ncs}
R.~Li, Y.~Feng and Y.-Q.~Ma, \emph{{Exclusive quarkonium production or decay in
  soft gluon factorization}},
  \href{https://doi.org/10.1007/JHEP05(2020)009}{\emph{JHEP} {\bfseries 05}
  (2020) 009} [\href{https://arxiv.org/abs/1911.05886}{{\ttfamily
  1911.05886}}].

\bibitem{Chen:2020yeg}
A.-P.~Chen and Y.-Q.~Ma, \emph{{Theory for quarkonium: from NRQCD factorization
  to soft gluon factorization}},
  \href{https://doi.org/10.1088/1674-1137/abc683}{\emph{Chin. Phys. C}
  {\bfseries 45} (2021) 013118}
  [\href{https://arxiv.org/abs/2005.08786}{{\ttfamily 2005.08786}}].

\bibitem{Kang:2011mg}
Z.-B.~Kang, J.-W.~Qiu and G.~Sterman, \emph{{Heavy quarkonium production and
  polarization}},
  \href{https://doi.org/10.1103/PhysRevLett.108.102002}{\emph{Phys. Rev. Lett.}
  {\bfseries 108} (2012) 102002}
  [\href{https://arxiv.org/abs/1109.1520}{{\ttfamily 1109.1520}}].

\bibitem{Kuhn:1992qw}
J.H.~Kuhn and E.~Mirkes, \emph{{QCD corrections to toponium production at
  hadron colliders}},
  \href{https://doi.org/10.1103/PhysRevD.48.179}{\emph{Phys. Rev. D} {\bfseries
  48} (1993) 179} [\href{https://arxiv.org/abs/hep-ph/9301204}{{\ttfamily
  hep-ph/9301204}}].

\bibitem{Petrelli:1997ge}
A.~Petrelli, M.~Cacciari, M.~Greco, F.~Maltoni and M.L.~Mangano, \emph{{NLO
  production and decay of quarkonium}},
  \href{https://doi.org/10.1016/S0550-3213(97)00801-8}{\emph{Nucl. Phys. B}
  {\bfseries 514} (1998) 245}
  [\href{https://arxiv.org/abs/hep-ph/9707223}{{\ttfamily hep-ph/9707223}}].

\bibitem{Lansberg:2020ejc}
J.-P.~Lansberg and M.A.~Ozcelik, \emph{{Curing the unphysical behaviour of NLO
  quarkonium production at the LHC and its relevance to constrain the gluon PDF
  at low scales}},
  \href{https://doi.org/10.1140/epjc/s10052-021-09258-7}{\emph{Eur. Phys. J. C}
  {\bfseries 81} (2021) 497}
  [\href{https://arxiv.org/abs/2012.00702}{{\ttfamily 2012.00702}}].

\bibitem{Schuler:1994hy}
G.A.~Schuler, \emph{{Quarkonium production and decays}},  1994.

\bibitem{Mangano:1996kg}
M.L.~Mangano and A.~Petrelli, \emph{{NLO quarkonium production in hadronic
  collisions}}, \href{https://doi.org/10.1142/S0217751X97002048}{\emph{Int. J.
  Mod. Phys. A} {\bfseries 12} (1997) 3887}
  [\href{https://arxiv.org/abs/hep-ph/9610364}{{\ttfamily hep-ph/9610364}}].

\bibitem{Feng:2015cba}
Y.~Feng, J.-P.~Lansberg and J.-X.~Wang, \emph{{Energy dependence of
  direct-quarkonium production in $pp$ collisions from fixed-target to LHC
  energies: complete one-loop analysis}},
  \href{https://doi.org/10.1140/epjc/s10052-015-3527-1}{\emph{Eur. Phys. J. C}
  {\bfseries 75} (2015) 313}
  [\href{https://arxiv.org/abs/1504.00317}{{\ttfamily 1504.00317}}].

\bibitem{DGLAP1}
V.N.~Gribov and L.N.~Lipatov, \emph{$e^+e^-$-annihilation and deep-inelastic
  $ep$-scattering in perturbation theory}, {\emph{Sov. J. Nucl. Phys.}
  {\bfseries 15} (1972) 438}.

\bibitem{DGLAP2}
Y.L.~Dokshitzer, \emph{Calculation of structure functions of deep-inelastic
  scattering and $e^+e^-$-annihilation in perturbation theory of quantum
  chromodynamics}, {\emph{Sov. Phys. JETP} {\bfseries 46} (1977) 641}.

\bibitem{DGLAP3}
G.~Altarelli and G.~Parisi, \emph{Asymptotic freedom in parton language},
  \href{https://doi.org/10.1016/0550-3213(77)90384-4}{\emph{Nucl. Phys.}
  {\bfseries B126} (1977) 298}.

\bibitem{Ellis:1990hw}
R.K.~Ellis and D.A.~Ross, \emph{{The Coupling of the {QCD} Pomeron in Various
  Semihard Processes}},
  \href{https://doi.org/10.1016/0550-3213(90)90609-H}{\emph{Nucl. Phys. B}
  {\bfseries 345} (1990) 79}.

\bibitem{Serri:2021fhn}
A.C.~Serri, Y.~Feng, C.~Flore, J.-P.~Lansberg, M.A.~Ozcelik, H.-S.~Shao et~al.,
  \emph{{Revisiting NLO QCD corrections to total inclusive J/psi and Upsilon
  photoproduction cross sections in lepton-proton collisions}},
  \href{https://arxiv.org/abs/2112.05060}{{\ttfamily 2112.05060}}.

\bibitem{BFKL1}
E.A.~Kuraev, L.N.~Lipatov and V.S.~Fadin, \emph{Multi - {Reggeon} processes in
  the {Yang-Mills} theory}, {\emph{Sov. Phys. JETP} {\bfseries 44} (1976) 443}.

\bibitem{BFKL2}
E.A.~Kuraev, L.N.~Lipatov and V.S.~Fadin, \emph{The {Pomeranchuk} singularity
  in {non-Abelian} gauge theories}, {\emph{Sov. Phys. JETP} {\bfseries 45}
  (1977) 199}.

\bibitem{BFKL3}
Y.Y.~Balitsky and L.N.~Lipatov, \emph{The {Pomeranchuk} singularity in {Quantum
  Chromodynamics}}, {\emph{Sov. J. Nucl. Phys.} {\bfseries 28} (1978) 822}.

\bibitem{Catani:1990xk}
S.~Catani, M.~Ciafaloni and F.~Hautmann, \emph{{GLUON CONTRIBUTIONS TO SMALL x
  HEAVY FLAVOR PRODUCTION}},
  \href{https://doi.org/10.1016/0370-2693(90)91601-7}{\emph{Phys. Lett. B}
  {\bfseries 242} (1990) 97}.

\bibitem{Catani:1990eg}
S.~Catani, M.~Ciafaloni and F.~Hautmann, \emph{{High-energy factorization and
  small x heavy flavor production}},
  \href{https://doi.org/10.1016/0550-3213(91)90055-3}{\emph{Nucl. Phys. B}
  {\bfseries 366} (1991) 135}.

\bibitem{Collins:1991ty}
J.C.~Collins and R.K.~Ellis, \emph{{Heavy quark production in very high-energy
  hadron collisions}},
  \href{https://doi.org/10.1016/0550-3213(91)90288-9}{\emph{Nucl. Phys.}
  {\bfseries B360} (1991) 3}.

\bibitem{Catani:1994sq}
S.~Catani and F.~Hautmann, \emph{{High-energy factorization and small x deep
  inelastic scattering beyond leading order}},
  \href{https://doi.org/10.1016/0550-3213(94)90636-X}{\emph{Nucl. Phys.}
  {\bfseries B427} (1994) 475}
  [\href{https://arxiv.org/abs/hep-ph/9405388}{{\ttfamily hep-ph/9405388}}].

\bibitem{Catani:1992rn}
S.~Catani, M.~Ciafaloni and F.~Hautmann, \emph{{Leptoproduction of heavy flavor
  at high energies}},
  \href{https://doi.org/10.1016/0920-5632(92)90441-T}{\emph{Nucl. Phys. B Proc.
  Suppl.} {\bfseries 29} (1992) 182}.

\bibitem{Hautmann:2002tu}
F.~Hautmann, \emph{{Heavy top limit and double logarithmic contributions to
  Higgs production at m(H)**2 / s much less than 1}},
  \href{https://doi.org/10.1016/S0370-2693(02)01761-6}{\emph{Phys. Lett.}
  {\bfseries B535} (2002) 159}
  [\href{https://arxiv.org/abs/hep-ph/0203140}{{\ttfamily hep-ph/0203140}}].

\bibitem{Harlander:2009my}
R.V.~Harlander, H.~Mantler, S.~Marzani and K.J.~Ozeren, \emph{{Higgs production
  in gluon fusion at next-to-next-to-leading order QCD for finite top mass}},
  \href{https://doi.org/10.1140/epjc/s10052-010-1258-x}{\emph{Eur. Phys. J. C}
  {\bfseries 66} (2010) 359} [\href{https://arxiv.org/abs/0912.2104}{{\ttfamily
  0912.2104}}].

\bibitem{Marzani:2008uh}
S.~Marzani and R.D.~Ball, \emph{{High Energy Resummation of Drell-Yan
  Processes}},
  \href{https://doi.org/10.1016/j.nuclphysb.2009.01.029}{\emph{Nucl. Phys. B}
  {\bfseries 814} (2009) 246}
  [\href{https://arxiv.org/abs/0812.3602}{{\ttfamily 0812.3602}}].

\bibitem{Diana:2009xv}
G.~Diana, \emph{{High-energy resummation in direct photon production}},
  \href{https://doi.org/10.1016/j.nuclphysb.2009.09.001}{\emph{Nucl. Phys. B}
  {\bfseries 824} (2010) 154}
  [\href{https://arxiv.org/abs/0906.4159}{{\ttfamily 0906.4159}}].

\bibitem{Ball:2017otu}
R.D.~Ball, V.~Bertone, M.~Bonvini, S.~Marzani, J.~Rojo and L.~Rottoli,
  \emph{{Parton distributions with small-x resummation: evidence for {BFKL}
  dynamics in {HERA} data}},
  \href{https://doi.org/10.1140/epjc/s10052-018-5774-4}{\emph{Eur. Phys. J.}
  {\bfseries C78} (2018) 321}
  [\href{https://arxiv.org/abs/1710.05935}{{\ttfamily 1710.05935}}].

\bibitem{Abdolmaleki:2018jln}
{\scshape xFitter Developers' Team} collaboration, \emph{{Impact of low-$x$
  resummation on QCD analysis of HERA data}},
  \href{https://doi.org/10.1140/epjc/s10052-018-6090-8}{\emph{Eur. Phys. J.}
  {\bfseries C78} (2018) 621}
  [\href{https://arxiv.org/abs/1802.00064}{{\ttfamily 1802.00064}}].

\bibitem{Echevarria:2018qyi}
M.G.~Echevarria, T.~Kasemets, J.-P.~Lansberg, C.~Pisano and A.~Signori,
  \emph{{Matching factorization theorems with an inverse-error weighting}},
  \href{https://doi.org/10.1016/j.physletb.2018.03.075}{\emph{Phys. Lett. B}
  {\bfseries 781} (2018) 161}
  [\href{https://arxiv.org/abs/1801.01480}{{\ttfamily 1801.01480}}].

\bibitem{Ma:2014mri}
Y.-Q.~Ma and R.~Venugopalan, \emph{{Comprehensive Description of
  J/\ensuremath{\psi} Production in Proton-Proton Collisions at Collider
  Energies}}, \href{https://doi.org/10.1103/PhysRevLett.113.192301}{\emph{Phys.
  Rev. Lett.} {\bfseries 113} (2014) 192301}
  [\href{https://arxiv.org/abs/1408.4075}{{\ttfamily 1408.4075}}].

\bibitem{Kang:2013hta}
Z.-B.~Kang, Y.-Q.~Ma and R.~Venugopalan, \emph{{Quarkonium production in high
  energy proton-nucleus collisions: CGC meets NRQCD}},
  \href{https://doi.org/10.1007/JHEP01(2014)056}{\emph{JHEP} {\bfseries 01}
  (2014) 056} [\href{https://arxiv.org/abs/1309.7337}{{\ttfamily 1309.7337}}].

\bibitem{Balitsky:1995ub}
I.~Balitsky, \emph{{Operator expansion for high-energy scattering}},
  \href{https://doi.org/10.1016/0550-3213(95)00638-9}{\emph{Nucl. Phys. B}
  {\bfseries 463} (1996) 99}
  [\href{https://arxiv.org/abs/hep-ph/9509348}{{\ttfamily hep-ph/9509348}}].

\bibitem{Kovchegov:1999yj}
Y.V.~Kovchegov, \emph{{Small x F(2) structure function of a nucleus including
  multiple pomeron exchanges}},
  \href{https://doi.org/10.1103/PhysRevD.60.034008}{\emph{Phys. Rev. D}
  {\bfseries 60} (1999) 034008}
  [\href{https://arxiv.org/abs/hep-ph/9901281}{{\ttfamily hep-ph/9901281}}].

\bibitem{Balitsky:1998kc}
I.~Balitsky, \emph{{Factorization for high-energy scattering}},
  \href{https://doi.org/10.1103/PhysRevLett.81.2024}{\emph{Phys. Rev. Lett.}
  {\bfseries 81} (1998) 2024}
  [\href{https://arxiv.org/abs/hep-ph/9807434}{{\ttfamily hep-ph/9807434}}].

\bibitem{Jalilian-Marian:1997qno}
J.~Jalilian-Marian, A.~Kovner, A.~Leonidov and H.~Weigert, \emph{{The BFKL
  equation from the Wilson renormalization group}},
  \href{https://doi.org/10.1016/S0550-3213(97)00440-9}{\emph{Nucl. Phys. B}
  {\bfseries 504} (1997) 415}
  [\href{https://arxiv.org/abs/hep-ph/9701284}{{\ttfamily hep-ph/9701284}}].

\bibitem{Jalilian-Marian:1997jhx}
J.~Jalilian-Marian, A.~Kovner, A.~Leonidov and H.~Weigert, \emph{{The Wilson
  renormalization group for low x physics: Towards the high density regime}},
  \href{https://doi.org/10.1103/PhysRevD.59.014014}{\emph{Phys. Rev. D}
  {\bfseries 59} (1998) 014014}
  [\href{https://arxiv.org/abs/hep-ph/9706377}{{\ttfamily hep-ph/9706377}}].

\bibitem{Jalilian-Marian:1997ubg}
J.~Jalilian-Marian, A.~Kovner and H.~Weigert, \emph{{The Wilson renormalization
  group for low x physics: Gluon evolution at finite parton density}},
  \href{https://doi.org/10.1103/PhysRevD.59.014015}{\emph{Phys. Rev. D}
  {\bfseries 59} (1998) 014015}
  [\href{https://arxiv.org/abs/hep-ph/9709432}{{\ttfamily hep-ph/9709432}}].

\bibitem{Iancu:2001ad}
E.~Iancu, A.~Leonidov and L.D.~McLerran, \emph{{The Renormalization group
  equation for the color glass condensate}},
  \href{https://doi.org/10.1016/S0370-2693(01)00524-X}{\emph{Phys. Lett. B}
  {\bfseries 510} (2001) 133}
  [\href{https://arxiv.org/abs/hep-ph/0102009}{{\ttfamily hep-ph/0102009}}].

\bibitem{Iancu:2000hn}
E.~Iancu, A.~Leonidov and L.D.~McLerran, \emph{{Nonlinear gluon evolution in
  the color glass condensate. 1.}},
  \href{https://doi.org/10.1016/S0375-9474(01)00642-X}{\emph{Nucl. Phys. A}
  {\bfseries 692} (2001) 583}
  [\href{https://arxiv.org/abs/hep-ph/0011241}{{\ttfamily hep-ph/0011241}}].

\bibitem{IFLQCD}
B.L.~Ioffe, V.S.~Fadin and L.N.~Lipatov, \emph{Quantum Chromodynamics
  Perturbative and Nonperturbative Aspects}, Cambridge University Press,
  Cambridge (2010).

\bibitem{kovchegov_levin_2012}
Y.V.~Kovchegov and E.~Levin, \emph{Quantum Chromodynamics at High Energy},
  Cambridge Monographs on Particle Physics, Nuclear Physics and Cosmology,
  Cambridge University Press (2012),
  \href{https://doi.org/10.1017/CBO9781139022187}{10.1017/CBO9781139022187}.

\bibitem{RevDelDuca95}
V.~Del~Duca, \emph{An introduction to the perturbative {QCD} pomeron and to jet
  physics at large rapidities}, {\emph{Scientifica Acta} {\bfseries 10} (1995)
  91} [\href{https://arxiv.org/abs/hep-ph/9503226}{{\ttfamily
  hep-ph/9503226}}].

\bibitem{CollinsQCD}
J.C.~Collins, \emph{Foundations of perturbative {QCD}}, Cambridge University
  Press, Cambridge (2011).

\bibitem{Nefedov:2021vvy}
M.~Nefedov, \emph{{Sudakov resummation from the BFKL evolution}},
  \href{https://doi.org/10.1103/PhysRevD.104.054039}{\emph{Phys. Rev. D}
  {\bfseries 104} (2021) 054039}
  [\href{https://arxiv.org/abs/2105.13915}{{\ttfamily 2105.13915}}].

\bibitem{Hentschinski:2021lsh}
M.~Hentschinski, \emph{{Transverse momentum dependent gluon distribution within
  high energy factorization at next-to-leading order}},
  \href{https://doi.org/10.1103/PhysRevD.104.054014}{\emph{Phys. Rev. D}
  {\bfseries 104} (2021) 054014}
  [\href{https://arxiv.org/abs/2107.06203}{{\ttfamily 2107.06203}}].

\bibitem{Lipatov95}
L.N.~Lipatov, \emph{Gauge invariant effective action for high-energy processes
  in {QCD}}, \href{https://doi.org/10.1016/0550-3213(95)00390-E}{\emph{Nucl.
  Phys.} {\bfseries B452} (1995) 369}.

\bibitem{Kniehl:2006vm}
B.A.~Kniehl, V.A.~Saleev and D.V.~Vasin, \emph{{Bottomonium production in the
  Regge limit of QCD}},
  \href{https://doi.org/10.1103/PhysRevD.74.014024}{\emph{Phys. Rev.}
  {\bfseries D74} (2006) 014024}
  [\href{https://arxiv.org/abs/hep-ph/0607254}{{\ttfamily hep-ph/0607254}}].

\bibitem{He:di-Jpsi}
Z.-G.~He, B.A.~Kniehl, M.A.~Nefedov and V.A.~Saleev, \emph{{Double Prompt
  $J/\psi$ Hadroproduction in the Parton Reggeization Approach with High-Energy
  Resummation}},
  \href{https://doi.org/10.1103/PhysRevLett.123.162002}{\emph{Phys. Rev. Lett.}
  {\bfseries 123} (2019) 162002}
  [\href{https://arxiv.org/abs/1906.08979}{{\ttfamily 1906.08979}}].

\bibitem{Hagler:2000dda}
P.~Hagler, R.~Kirschner, A.~Schafer, L.~Szymanowski and O.~Teryaev,
  \emph{{Heavy quark production as sensitive test for an improved description
  of high-energy hadron collisions}},
  \href{https://doi.org/10.1103/PhysRevD.62.071502}{\emph{Phys. Rev. D}
  {\bfseries 62} (2000) 071502}
  [\href{https://arxiv.org/abs/hep-ph/0002077}{{\ttfamily hep-ph/0002077}}].

\bibitem{Hagler:2000dd}
P.~Hagler, R.~Kirschner, A.~Schafer, L.~Szymanowski and O.V.~Teryaev,
  \emph{{Towards a solution of the charmonium production controversy: $k^-$
  perpendicular factorization versus color octet mechanism}},
  \href{https://doi.org/10.1103/PhysRevLett.86.1446}{\emph{Phys. Rev. Lett.}
  {\bfseries 86} (2001) 1446}
  [\href{https://arxiv.org/abs/hep-ph/0004263}{{\ttfamily hep-ph/0004263}}].

\bibitem{Kniehl:2006sk}
B.A.~Kniehl, D.V.~Vasin and V.A.~Saleev, \emph{{Charmonium production at high
  energy in the $k_{T}$ -factorization approach}},
  \href{https://doi.org/10.1103/PhysRevD.73.074022}{\emph{Phys. Rev. D}
  {\bfseries 73} (2006) 074022}
  [\href{https://arxiv.org/abs/hep-ph/0602179}{{\ttfamily hep-ph/0602179}}].

\bibitem{Jaroszewicz:1982gr}
T.~Jaroszewicz, \emph{{Gluonic Regge Singularities and Anomalous Dimensions in
  QCD}}, \href{https://doi.org/10.1016/0370-2693(82)90345-8}{\emph{Phys. Lett.}
  {\bfseries 116B} (1982) 291}.

\bibitem{Kirschner:2009qu}
R.~Kirschner and M.~Segond, \emph{{Small x resummation in collinear
  factorisation}},
  \href{https://doi.org/10.1140/epjc/s10052-010-1363-x}{\emph{Eur. Phys. J. C}
  {\bfseries 68} (2010) 425} [\href{https://arxiv.org/abs/0910.5443}{{\ttfamily
  0910.5443}}].

\bibitem{Vogt:2004mw}
A.~Vogt, S.~Moch and J.A.M.~Vermaseren, \emph{{The Three-loop splitting
  functions in QCD: The Singlet case}},
  \href{https://doi.org/10.1016/j.nuclphysb.2004.04.024}{\emph{Nucl. Phys. B}
  {\bfseries 691} (2004) 129}
  [\href{https://arxiv.org/abs/hep-ph/0404111}{{\ttfamily hep-ph/0404111}}].

\bibitem{Lukowski:2009ce}
T.~Lukowski, A.~Rej and V.N.~Velizhanin, \emph{{Five-Loop Anomalous Dimension
  of Twist-Two Operators}},
  \href{https://doi.org/10.1016/j.nuclphysb.2010.01.008}{\emph{Nucl. Phys. B}
  {\bfseries 831} (2010) 105}
  [\href{https://arxiv.org/abs/0912.1624}{{\ttfamily 0912.1624}}].

\bibitem{Kniehl:2011hc}
B.A.~Kniehl, V.A.~Saleev, A.V.~Shipilova and E.V.~Yatsenko, \emph{{Single jet
  and prompt-photon inclusive production with multi-Regge kinematics: From
  Tevatron to LHC}},
  \href{https://doi.org/10.1103/PhysRevD.84.074017}{\emph{Phys. Rev. D}
  {\bfseries 84} (2011) 074017}
  [\href{https://arxiv.org/abs/1107.1462}{{\ttfamily 1107.1462}}].

\bibitem{Blumlein:1995eu}
J.~Blumlein, \emph{{On the k(T) dependent gluon density of the proton}},  in
  \emph{{Deep inelastic scattering and QCD. Proceedings, Workshop, Paris,
  France, April 24-28, 1995}}, pp.~265--268, 1995,
  \href{http://www-library.desy.de/cgi-bin/showprep.pl?desy95-121}{http://www-library.desy.de/cgi-bin/showprep.pl?desy95-121}
  [\href{https://arxiv.org/abs/hep-ph/9506403}{{\ttfamily hep-ph/9506403}}].

\bibitem{Hou:2019efy}
T.-J.~Hou et~al., \emph{{New CTEQ global analysis of quantum chromodynamics
  with high-precision data from the LHC}},
  \href{https://doi.org/10.1103/PhysRevD.103.014013}{\emph{Phys. Rev. D}
  {\bfseries 103} (2021) 014013}
  [\href{https://arxiv.org/abs/1912.10053}{{\ttfamily 1912.10053}}].

\bibitem{Bailey:2020ooq}
S.~Bailey, T.~Cridge, L.A.~Harland-Lang, A.D.~Martin and R.S.~Thorne,
  \emph{{Parton distributions from LHC, HERA, Tevatron and fixed target data:
  MSHT20 PDFs}},
  \href{https://doi.org/10.1140/epjc/s10052-021-09057-0}{\emph{Eur. Phys. J. C}
  {\bfseries 81} (2021) 341}
  [\href{https://arxiv.org/abs/2012.04684}{{\ttfamily 2012.04684}}].

\bibitem{NNPDF:2017mvq}
{\scshape NNPDF} collaboration, \emph{{Parton distributions from high-precision
  collider data}},
  \href{https://doi.org/10.1140/epjc/s10052-017-5199-5}{\emph{Eur. Phys. J. C}
  {\bfseries 77} (2017) 663}
  [\href{https://arxiv.org/abs/1706.00428}{{\ttfamily 1706.00428}}].

\bibitem{Buckley:2014ana}
A.~Buckley, J.~Ferrando, S.~Lloyd, K.~Nordström, B.~Page, M.~Rüfenacht
  et~al., \emph{{LHAPDF6: parton density access in the LHC precision era}},
  \href{https://doi.org/10.1140/epjc/s10052-015-3318-8}{\emph{Eur. Phys. J.}
  {\bfseries C75} (2015) 132}
  [\href{https://arxiv.org/abs/1412.7420}{{\ttfamily 1412.7420}}].

\bibitem{Ellis:1995gv}
R.K.~Ellis, F.~Hautmann and B.R.~Webber, \emph{{QCD scaling violation at small
  x}}, \href{https://doi.org/10.1016/0370-2693(95)00148-E}{\emph{Phys. Lett. B}
  {\bfseries 348} (1995) 582}
  [\href{https://arxiv.org/abs/hep-ph/9501307}{{\ttfamily hep-ph/9501307}}].

\bibitem{Altarelli:1999vw}
G.~Altarelli, R.D.~Ball and S.~Forte, \emph{{Resummation of singlet parton
  evolution at small x}},
  \href{https://doi.org/10.1016/S0550-3213(00)00032-8}{\emph{Nucl. Phys.}
  {\bfseries B575} (2000) 313}
  [\href{https://arxiv.org/abs/hep-ph/9911273}{{\ttfamily hep-ph/9911273}}].

\bibitem{Albacete:2014fwa}
J.L.~Albacete and C.~Marquet, \emph{{Gluon saturation and initial conditions
  for relativistic heavy ion collisions}},
  \href{https://doi.org/10.1016/j.ppnp.2014.01.004}{\emph{Prog. Part. Nucl.
  Phys.} {\bfseries 76} (2014) 1}
  [\href{https://arxiv.org/abs/1401.4866}{{\ttfamily 1401.4866}}].

\end{thebibliography}\endgroup








\end{document}